\documentclass[iop]{emulateapj}

\newcommand*{\ac}{$A$(C)}
\newcommand*{\teff}{$T_{\rm eff}$}
\newcommand*{\logg}{$\log~g$}
\newcommand*{\feh}{[Fe/H]}
\newcommand*{\afe}{[$\alpha$/Fe]}
\newcommand*{\cfe}{[C/Fe]}
\newcommand*{\kms}{km s$^{-1}$}
\newcommand*{\zmax}{$Z_{\rm max}$}

\newcommand*{\vrad}{$V_{\rm r}$}
\newcommand*{\vthe}{$V_{\rm \theta}$}
\newcommand*{\vphi}{$V_{\rm \phi}$}
\newcommand*{\vRad}{$V_{\rm R}$}
\newcommand*{\vPhi}{$V_{\rm \Phi}$}
\newcommand*{\vZ}{$V_{\rm Z}$}
\newcommand*{\msun}{$M_\odot$}

\newcommand*{\rsun}{$R_\odot$}

\newcommand*{\z}{$|Z|$}

\newcommand*{\cemps}{CEMP-$s$}

\newcommand*{\cfemean}{$\langle$[C/Fe]$\rangle$}

\newcommand*{\gaia}{$\it Gaia$}

\usepackage{color}
\usepackage{epsfig}
\usepackage{graphicx}
\usepackage{natbib}
\usepackage{hyperref}
\hypersetup{
    bookmarks=true,         
    unicode=false,          
    pdftoolbar=true,        
    pdfmenubar=true,        
    pdffitwindow=true,     
    pdfstartview={FitH},    
    pdftitle={},    
    pdfauthor={Lee},     
    pdfsubject={Astronomy},   
    pdfcreator={dvipdf},   
    pdfproducer={dvipdf}, 
    pdfkeywords={metal-poor stars},
    pdfnewwindow=true,      
    colorlinks=true,       
    linkcolor=red,          
    citecolor=blue,        
    filecolor=magenta,      
    urlcolor=cyan,           
    breaklinks=true,
    linktocpage
}

\shorttitle{The Assembly History of the Galactic Stellar Halo}
\shortauthors{Lee et al.}
\slugcomment{Draft version, September 3, 2019}

\begin{document}

\title{Chemical Cartography. II. The Assembly History of the Galactic Stellar Halo Traced by
Carbon-Enhanced Metal-Poor Stars}

\author{Young Sun Lee\altaffilmark{1}, Timothy C. Beers\altaffilmark{2}, and Young Kwang Kim\altaffilmark{1}}
\altaffiltext{1}{Department of Astronomy and Space Science, Chungnam National University,
                 Daejeon 34134, Republic of Korea; youngsun@cnu.ac.kr}
\altaffiltext{2}{Department of Physics and JINA Center for the Evolution of the Elements, University
	         of Notre Dame, Notre Dame, IN 46556, USA}

\begin{abstract}

We present an analysis of the kinematic properties of stellar
populations in the Galactic halo, making use of over 100,000 main
sequence turnoff (MSTO) stars observed in the Sloan Digital Sky Survey.
After dividing the Galactic halo into an inner-halo region (IHR) and
outer-halo region (OHR), based on the spatial variation of
carbon-to-iron ratios in the sample, we find that stars in the OHR
exhibit a clear retrograde motion of --49 $\pm$ 4 \kms\ and a more
spherical distribution of stellar orbits, while stars in the IHR have
zero net rotation (--3 $\pm$ 1 \kms) with a much more radially biased
distribution of stellar orbits. Furthermore, we classify the
carbon-enhanced metal-poor (CEMP) stars among the MSTO sample in each
halo component into CEMP-no and \cemps\ sub-classes, based on their
absolute carbon abundances, \ac, and examine the spatial distributions
and kinematics associated with each sub-class. The CEMP-no stars are the
majority sub-class of CEMP stars in the OHR ($\sim$ 65\%), and the
minority sub-class in the IHR ($\sim$ 44\%), similar to the results of
several previous analyses. The CEMP-no stars in each halo region exhibit
slightly higher counter-rotation than the \cemps\ stars, but within
statistical errors. The CEMP-no stars also show a more spherical
distribution of orbits than the \cemps\ stars in each halo region. These
distinct characteristics provide strong evidence that numerous low-mass
satellite galaxies (similar to the ultra-faint dwarf galaxies) have
donated stars to the OHR, while more-massive dwarf galaxies provided the
dominant contribution to the IHR.
\end{abstract}


\section{Introduction} \label{sec:intro}

Large photometric and spectroscopic surveys such as the Sloan Digital
Sky Survey (SDSS; \citealt{york2000}) have dramatically changed our
understanding of the stellar components of the Milky Way (MW), in
particular for the Galactic halo. For many years, the diffuse stellar
halo of the MW was thought to consist of a single population of stars
with similar kinematics, chemical abundances, and ages. However,
numerous recent studies have demonstrated that
the halo comprises at least two distinct stellar components (e.g.,
\citealt{carollo2007, carollo2010, dejong2010, beers2012, an2013,
an2015, hattori2013, allende2014, chen2014, fernandez2015,
fernandez2016, fernandez2017, das2016a, janesh2016}). In addition, numerical
simulations of the formation of MW-like galaxies (e.g.,
\citealt{zolotov2009, font2011, mccarthy2012, tissera2013, tissera2014,
cooper2015}) indicate that an inner/outer halo dichotomy of the
Galactic halo is indeed expected.

Spatially, stars in the inner halo are predominantly located at
distances up to 10 -- 15 kpc from the Galactic center, whereas stars in
the outer halo are mostly found in the region beyond 15 -- 20 kpc.
Additionally, the stellar density profile of the inner halo is flatter
than that of the outer halo, which is nearly spherical. Chemically, the
metallicity distribution function (MDF) of the inner-halo stars peaks at
[Fe/H] $\sim$ --1.6, in contrast to the peak at [Fe/H] $\sim$ --2.2 for
the outer-halo stars. Kinematically, the inner-halo stars collectively
exhibit essentially zero net rotation around the Galactic center, and
are dominantly on high-eccentricity orbits, in contrast to the
outer-halo stars, which show a net retrograde motion of about --80 \kms\
and more circular orbits (\citealt{carollo2007,carollo2010, kinman2012,
hattori2013}). It has also been shown that the outer-halo population
(OHP) is kinematically hotter than the inner-halo population (IHP), as
expected (\citealt{carollo2007, carollo2010, carollo2014, an2015,
helmi2017}).

However, this view (especially the interpretation of the 
kinematic analysis) has been challenged by \citet{schonrich2011,
schonrich2014}, who claimed that the findings by \citet{carollo2010} may
have resulted from incorrect distance assignments and selection bias of
the sample. Their reanalysis of the sample used by \citet{carollo2010}
shows no retrograde motion for the OHP. \citet{schonrich2014} further
argued that the results of \citet{carollo2007, carollo2010} needed to be
verified after accounting for possible metallicity bias due to target
selection, and to clarify whether or not the claimed retrograde motion
is due to observational errors.

Meanwhile, \citet{beers2012} refuted the arguments made
by \citet{schonrich2011}, and claimed that the absence of the
retrograde signal for the OHP in their analysis is in part due to their
{\it incorrect} adoption of the relation for the distance modulus from
\citet{ivezic2008}. Furthermore, they demonstrated that the 
retrograde signature of the OHP was detected using the proper motions alone.
These dual kinematic features of the Galactic halo have also been
identified based on local K-giant stars selected (in an entirely
different manner from Carollo et al.) from the Large sky Area
Multi-Object Fiber Spectroscopic Telescope (LAMOST; \citealt{cui2012})
in the study of \citet{tian2019}. Additionally, many recent studies,
which used $in~situ$ halo stars up to tens of kpc away from the Galactic
center, consistently support the dichotomy of the Galactic halo (e.g.,
\citealt{dejong2010,deason2011,an2013,an2015,kafle2013, kafle2017,
allende2014,chen2014,fernandez2015,das2016a}).

Since chemical-abundance ratios of individual stars are closely related
to their star-formation history (in particular to that of the galactic fragments
in which they were born), they provide valuable complimentary
information for assessing the assembly history of the
MW. There have been several previous efforts to understand the origin of
the dichotomy of the Galactic halo using the abundances of chemical
elements other than iron, such as Ca and Mg, the so called
$\alpha$-elements (e.g., \citealt{gratton2003,jonsell2005, ishigaki2010,
nissen2010,nissen2011,fernandez2015,fernandez2016, fernandez2017}). These
studies attempt to relate \afe\ to stellar kinematics or stellar
distances, and characterize the observed trends.

Recently, carbon has emerged as valuable element to study the origin of
the duality of the Galactic halo. Especially, the so-called
carbon-enhanced metal-poor (CEMP; \citealt{beers2005}) stars have been
recognized as an important tracer of the assembly history of the MW,
owing to the variety of their sub-classes, which can be associated with
progenitor stars of different masses.

CEMP stars, which are typically defined by [Fe/H] $\le$ --1.0 and [C/Fe]
$\ge$ $+$1.0 (or $+$0.7, depending on the study), can be divided into four
major categories --- CEMP-no, CEMP-$s$, CEMP-$r$, and CEMP-$r/s$,
according to the level of enhancement of their neutron-capture elements
\citep{beers2005}. CEMP-no stars exhibit no over-abundances of heavy
neutron-capture elements, while CEMP-$s$ stars have enhancements of the
$s$-process (slow neutron-capture process) elements, such as Ba. CEMP-$r$ 
stars show strong enhancements of $r$-process (rapid neutron-capture process) 
elements, such as Eu. Carbon-rich stars with enhancements of both the $r$-process 
and the $s$-process are classified as CEMP-$r/s$ stars. Recently, evidence has been 
presented that the CEMP-$r/s$ sub-class is more likely associated with a proposed
intermediate neutron-capture process, the ``$i$-process (intermediate 
neutron-capture process)''; thus CEMP-$r/s$ stars are more appropriately 
called CEMP-$i$ stars (\citealt{hampel2016}).

CEMP-$s$ and CEMP-no stars account for more than 90\% of known CEMP
stars. CEMP-$s$ stars predominantly have [Fe/H] $>$ --2.5, while most
CEMP-no stars have [Fe/H] $<$ --2.5. (e.g., \citealt{aoki2007,
yoon2016}). Long-term radial-velocity monitoring of CEMP stars (e.g.,
\citealt{starkenburg2014, hansen2016a, hansen2016b, jorissen2016})
has shown that the binary fraction of \cemps\ stars is $\sim$ 82\% (including
CEMP-$r/s$ or CEMP-$i$ stars), while the fraction of CEMP-no stars is
only $\sim$ 17\%, indicating that their origin is not causally connected to
binary membership. Rather, their distinctive chemical-abundance patterns
indicate likely pollution of their natal clouds by the first generations
of massive stars.

In addition, several studies report, based on large numbers of CEMP
stars, that the fraction of CEMP stars dramatically increases with
decreasing metallicity (e.g., \citealt{rossi1999, lucatello2006, lee2013, yong2013,
placco2014}), as well as with increasing distance from the Galactic
plane (\citealt{frebel2006, carollo2012, beers2017}). This latter result
is difficult to reconcile with a single-component Galactic halo.
\citet{carollo2012} argued that there is a higher frequency of CEMP
stars associated with the OHP than with the IHP, based on a kinematic
separation of these populations. Furthermore, \citet{carollo2014}
claimed, from an analysis of a small sample of CEMP stars with available
high-resolution spectroscopy, that the OHP exhibits a relatively higher
fraction of CEMP-no stars than the IHP, whereas the IHP has
a similar fraction of the two sub-classes, indicative of differences in
the star-formation histories of the outer-halo and inner-halo
populations.

One drawback of most previous studies is that the stars used
predominantly explore the Solar Neighborhood. To overcome this
limitation, \citet{lee2017} (Paper I hereafter) made use of
main-sequence turnoff (MSTO) stars, with distances extending up to 15
kpc from the Galactic plane, to investigate the chemical nature of the
Galactic halo system $in~situ$. Paper I demonstrated that, based on
[C/Fe] (``carbonicity'') alone, the stellar populations of the MW halo
can be readily distinguished. The component associated with the IHP
shows a metallicity peak at [Fe/H] $\sim$ --1.5, while the metallicity
associated with the OHP peaks at [Fe/H] $\sim$ --2.2, in agreement with
studies by \citet{carollo2007, carollo2010} and \citet{an2013, an2015},
who employed completely different approaches to distinguish the
populations.

Furthermore, Paper I quantitatively estimated the fraction of CEMP-no
and CEMP-$s$ stars in the outer-halo region (OHR) and inner-halo region
(IHR), classified on the basis of their derived \ac\footnote[3]{The
conventional notation is used, \ac\ = log\, $\epsilon$(C) = log\,
($N_{\rm C}$/$N_{\rm H}) + 12$, and is calculated from medium-resolution
spectroscopy using $A$(C) = [C/Fe] $+$ [Fe/H] $+$ $A$(C)$_{\odot}$,
where we adopt the Solar abundance of carbon from
\citet{asplund2009}, $A$(C) $_{\odot}$ = 8.43.}, and reported that the
stars in the OHR exhibit a higher proportion of CEMP-no
stars compared to CEMP-$s$ stars, in contrast to the similar fractions
of CEMP-no and CEMP-$s$ stars in the IHR, consistent with
results from previous, more-local studies.

\citet{yoon2018} carried out a study similar to Paper I.
They constructed a carbonicity map, using $\sim$ 58,000 stars, which 
dominantly consists of MSTO stars in the Southern Hemisphere observed by
the AAOmega Evolution of Galactic Structure (AEGIS) survey, and identify
the inner- and outer-halo regions based on this map.
Their results are commensurate with those from Paper I --
an increasing level of \cfe\ from the IHR to the OHR, and
larger numbers of CEMP-no stars and \cemps\ stars associated
with the OHR and IHR, respectively.

In this study, the second paper in the ``Chemical Cartography'' series,
we explore the kinematics of the Galactic halo populations, separated on
the basis of their different levels of carbonicity, making use of the
$\sim$ 105,700 MSTO stars from SDSS used in Paper I. Note that our
approach differs from many previous studies, which identified halo
populations by their kinematic and orbital properties, in that we use the
distinctive carbonicity signature to carry out the division, then
consider the resulting kinematics.

This paper is arranged as follows. Section 2 briefly summarizes the sample
selection of the MSTO stars, and describes the derivation of the
quantities used for the kinematic analysis, as well as our division of
the Galactic halo regions and classification of \cemps\ and CEMP-no
stars. In Section 3, we discuss the potential impact of
the target-selection bias present in our MSTO sample on the
identification of the Galactic halo regions, and explore the existence
of the systematic offsets in our adopted distance scale. Section 4
presents the distinct chemical and kinematic signatures of the stellar
populations in the Galactic halo. Section 5 reports the spatial and
kinematic properties of the \cemps\ and CEMP-no stars in each Galactic
halo region, providing insight to the origin of each halo component as
well as the assembly history of the MW, as discussed in Section 6. A
brief summary and conclusions are provided in Section 7.

\section{Halo Main-Sequence Turnoff Stars, Kinematic Calculations, and
Classification of CEMP-$s$ and CEMP-no Stars} \label{sec:Sample}
\subsection{Sample Selection}\label{sec:selection}

As described in detail in Paper I, in order to construct a carbonicity
map, we gathered medium-resolution ($R \sim 1800$) spectra of
MSTO stars from the legacy SDSS program, the Sloan Extension for Galactic
Understanding and Exploration (SEGUE-1 and SEGUE-2; \citealt{yanny2009}),
and the Baryon Oscillation Spectroscopic Survey (BOSS; \citealt{dawson2013}).
Stellar atmospheric-parameter (\teff, \logg, and \feh) estimates, and the
carbonicity, \cfe, were obtained from the SEGUE Stellar Parameter Pipeline (SSPP;
\citealt{lee2008a, lee2008b, allende2008, lee2011, lee2013,
smolinski2011}). Paper I also validated our measurement of \cfe, by
performing various calibrations with SDSS stars in common with the
sample of \citet{yoon2016}, which reports high-resolution determinations
of [Fe/H] and [C/Fe] from the literature. As part of this effort, we
adjusted for systematic offsets arising from the difficulty of detecting
the CH $G$-band in low S/N spectra, by carrying out a noise-injection
experiment over a grid of synthetic spectra. Our final sample of MSTO
stars satisfies the following conditions: 15.0 $ \leq g_{0}
\leq 19.4$, 0.22 $\leq (g-r)_{0} \leq 0.38$, 3.5 $\leq$ \logg\ $\leq$
4.8, 5600 K $\leq $ \teff\ $\leq 6700$~K, S/N $\ge$ 12, and equivalent
width of the CH-$G$ band around 4300 \AA\ larger than 0.6 \AA. These
criteria yielded a total of $N \sim 105,700$ stars.

One subtlety to be aware of is that we use the adopted value of \feh\
from the SSPP, not the value determined during the estimation of \cfe,
as it exhibits a smaller offset and scatter when compared to the
high-resolution results. Accordingly, we calculated [C/Fe]$_{\rm
adjusted}$ = [C/H] -- [Fe/H]$_{\rm adopted}$, where [C/H] = [C/Fe] $+$
[Fe/H]. [C/Fe] and [Fe/H] are the estimates from the
carbon-determination routine. Throughout the remainder of this paper, we
refer to [C/Fe]$_{\rm adjusted}$ and [Fe/H]$_{\rm adopted}$ as simply
\cfe\ and \feh, respectively.

\begin{figure}[t]
\centering
\epsscale{1.15}
\plotone{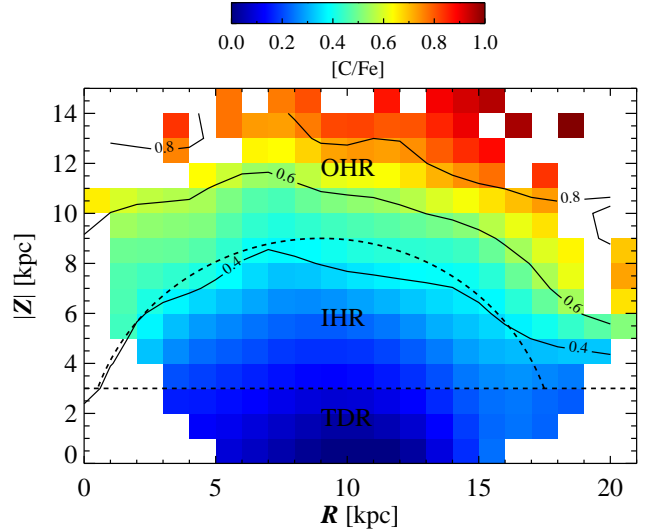}
\caption{Map of carbonicity, [C/Fe], for our MSTO sample in the \z\
versus $R$ plane, as shown in \citet{lee2017}. \z\ is the absolute distance
from the Galactic mid-plane, while $R$ is the projected distance onto the plane
from the Galactic center. The bin size is 1$\times$1 kpc; each pixel
contains at least three stars. Each pixel represents a median value
of \cfe, with the color scale shown in the color bar. Contours of
median [C/Fe] values are overplotted. The dashed line at \z\ = 3 kpc
indicates the approximate upper boundary of a thick-disk region (TDR),
while the area between the thick-disk upper limit and the dashed
circle represents an inner-halo region (IHR). The area above the dashed
circle is assigned to an outer-halo region (OHR). We applied a Gaussian
kernel to the map to obtain a smooth spatial distribution of [C/Fe].
Note that each bin with [C/Fe] $>$ $+$1.0 is forced to a value of [C/Fe]
= $+$1.0, then used to construct our map for stars in the range of
carbonicity [0,+1.0], to better illustrate the subtle contrast in the map.}
\label{fig:cmap}
\end{figure}

\begin{figure*}
\centering
\plotone{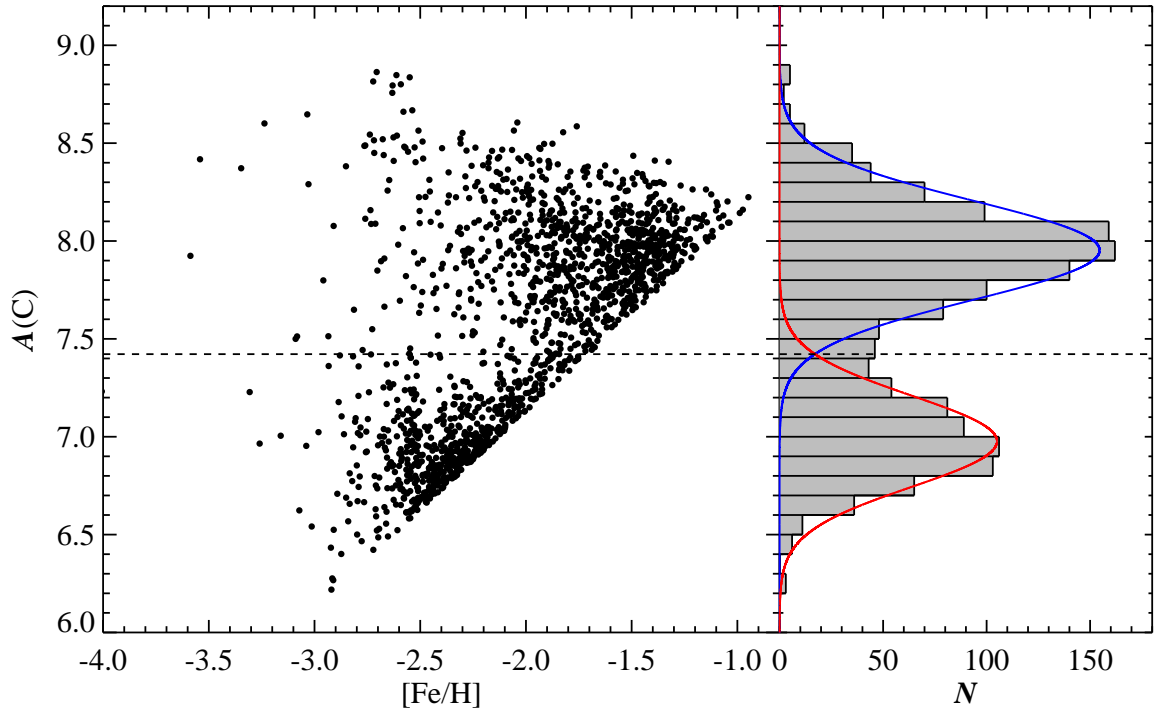}
\caption{$Left~panel$: Plot of \ac\ versus [Fe/H]. $Right~panel$:
Histogram of \ac\ values, which clearly shows a bimodal distribution.
Two Gaussians, shown as red and blue curves, are fitted to the
histogram. The crossing point of the two Gaussians is \ac\ = 7.42, shown
as a black dashed line. This value is used to divide our MSTO stars into
low-\ac\ and high-\ac\ stars. We only considered stars with [C/Fe] $\ge$
+0.7 and S/N $\ge$ 30 in construction of this diagram.}
\label{fig:ac_feh}
\end{figure*}

\subsection{Calculations of Space Velocity Components and Orbital Parameters}\label{sec:velocity}

The primary goal of this study is to carry out a detailed investigation of the
distinct kinematic characteristics of the Galactic halo populations.
Below we describe our procedures for obtaining distances, radial
velocities, and proper motions for the stars in our MSTO sample.

The distance to each star was estimated following the methodology
described by \citet{beers2000, beers2012}, who report a typical
uncertainty on the order of 15 -- 20\%. We computed two other distances
-- one from the Galactic mid-plane, represented by $Z$, and the other
from the Galactic center projected onto the Galactic plane, denoted by
$R$. We assumed that the Sun is located at \rsun\ = 8.0 kpc from the Galactic
center. We adopted the SDSS radial velocity measured by cross-correlated
with the ELODIE spectral library (\citealt{prugniel2001}), with a
typical precision of $\sim 2$ \kms\ (\citealt{allende2008}).

In order to obtain the most accurate space motions for our program
stars, we cross-matched the MSTO stars with \gaia\ Data Release 2 (DR2;
\citealt{gaia2018}) to obtain their proper motions. Even though the
typical uncertainty is less than 3.0 mas yr$^{-1}$ for $G <$ 21,
we excluded from our kinematic analysis stars with errors in
proper motion larger than 1.0 mas yr$^{-1}$.

Based on the above inputs, we first calculated the $U$, $V$, and $W$
velocity components. We adopted ($U$,$V$,$W$)$_{\odot}$ = (--10.1, 4.0, 6.7)
\kms\ (\citealt{hogg2005}) to adjust for the Solar peculiar motions
with respect to the Local Standard of Rest (LSR). For the purpose of our
analysis, we computed three velocity components, $V_{\rm R}$,
$V_{\Phi}$, and $V_{\rm Z}$ in a cylindrical coordinate system centered
on the Galactic center, as well as $V_{\rm r}$, $V_{\theta}$, and
\vphi\ in a spherical coordinate system around the Galactic center. We
assumed in these calculations that the rotation velocity of the LSR is
$V_{\rm LSR}$ = 220 \kms\ (\citealt{kerr1986}).

We adopted an analytic St\"ackel-type gravitational
potential (see \citealt{chiba2000} for details) in order to compute the
apo-Galacticon distance ($r_{\rm apo}$), peri-Galacticon distance
($r_{\rm peri}$), and stellar orbital eccentricity ($e$),
calculated as $e$ = ($r_{\rm apo}-$$r_{\rm peri}$) /($r_{\rm apo}+r_{\rm
peri}$). Uncertainties in the derived kinematics and orbital
parameters were estimated from 100 realizations of a Monte Carlo
simulation, taking into account propagation of the errors in the
observed quantities.

As a check on their possible effect on our results, we 
adopted more recent reported values of $V_{\rm LSR}$ = 236 $\pm$ 3 \kms\
(\citealt{kawata2019}), \rsun\ = 8.2 $\pm$ 0.1 kpc
(\citealt{bland2016}), and $(U,V,W)_\odot$ = (--$11.10,12.24,7.25$)
\kms\ (\citealt{schonrich2010}), and recomputed the
velocity components and orbital parameters. We found that the mean
difference between our adopted velocity components and orbital parameters 
and those derived with the more recent values of $V_{\rm LSR}$, \rsun, and $(U,
V,W)_\odot$ is less than 1 \kms, with a scatter smaller than 5 \kms, 
for \vrad\ and \vthe, and 0.4 kpc, with a scatter less than 1.2 kpc 
for $r_{\rm apo}$. However, as might be expected, we derived a mean
offset of --23 \kms, with a small scatter of 1.8 \kms, for \vphi. This
may affect our rotational motion by about 20 \kms, although the
interpretation of our results does not change much. As we wish to
compare our findings with those of previous studies, which used the
older values, we have retained these for $V_{\rm LSR}$, \rsun, and $(U,
V,W)_\odot$ in our analysis below.

After removing stars with large proper motion errors (or those lacking
proper motion information altogether), non-physically derived orbital
eccentricities, or very high (\vphi\ $>$ 500 \kms)
or low (\vphi\ $<$ --500 \kms) rotation velocities, we obtained a sample
of $N \sim 101,700$ stars for the kinematic analysis.

\subsection{Division of Galactic Halo Regions}\label{sec:division}

In Paper I, we constructed a so-called ``carbonicity map'', as shown
in Figure \ref{fig:cmap}, and divided the map into four primary
regions, based on the level of the carbon enhancement with respect to
iron. However, since the present study aims at inspection of
the kinematics of stars within the chemically divided regions, and
we have a limited number of stars in the outer-halo region (due to
the lack of proper motion information for these more distant stars), we redefined
the map into three regions as follows:
\begin{itemize}

\item Thick-disk region (TDR) -- The region below the straight-dashed line
at \z\ = 3 kpc in Figure \ref{fig:cmap}, where \z\ is the absolute distance from 
the Galactic mid-plane. Stars in this area are dominated by
the thick-disk population (TDP).

\item Inner-halo region (IHR) -- The area surrounded by the line of \z\ = 3 kpc and
the dashed curve, which closely follows the contour line of
[C/Fe] = $+$0.4. Stars in this area are dominated by the
IHP.

\item Outer-halo region (OHR) -- The region above the dashed
curve and \z\ $>$ 6 kpc. The additional constraint by \z\ $>$ 6 kpc is
to minimize overlapping populations from the IHR. Stars
in this region are dominated by the OHP.

\end{itemize}

We cut the sample at \z\ $\leq$ 3 kpc for the TDR after considering the
scale heights of the thick disk and the metal-weak thick disk determined by
\citet{carollo2010}. The (dashed circle) division line between the IHR and OHR is
determined by inspection of the carbonicity map and the map of
fractions of \cemps\ and CEMP-no stars, shown in the left panel of
Figure \ref{fig:sno_map}. Note that, as we removed the transition region in
Figure \ref{fig:cmap}, the dashed curve has a radius of 9 kpc rather than 8.5 kpc
as in Paper I. We stress again that our strategy to separate the stellar
components for the IHR and OHR is $not$ based on metallicity or kinematics, but
solely relies on the level of [C/Fe] at a given location. These regions are shown
with black labels in Figure~\ref{fig:cmap}.

\begin{figure*}[t]
\centering
\epsscale{1.15}
\plotone{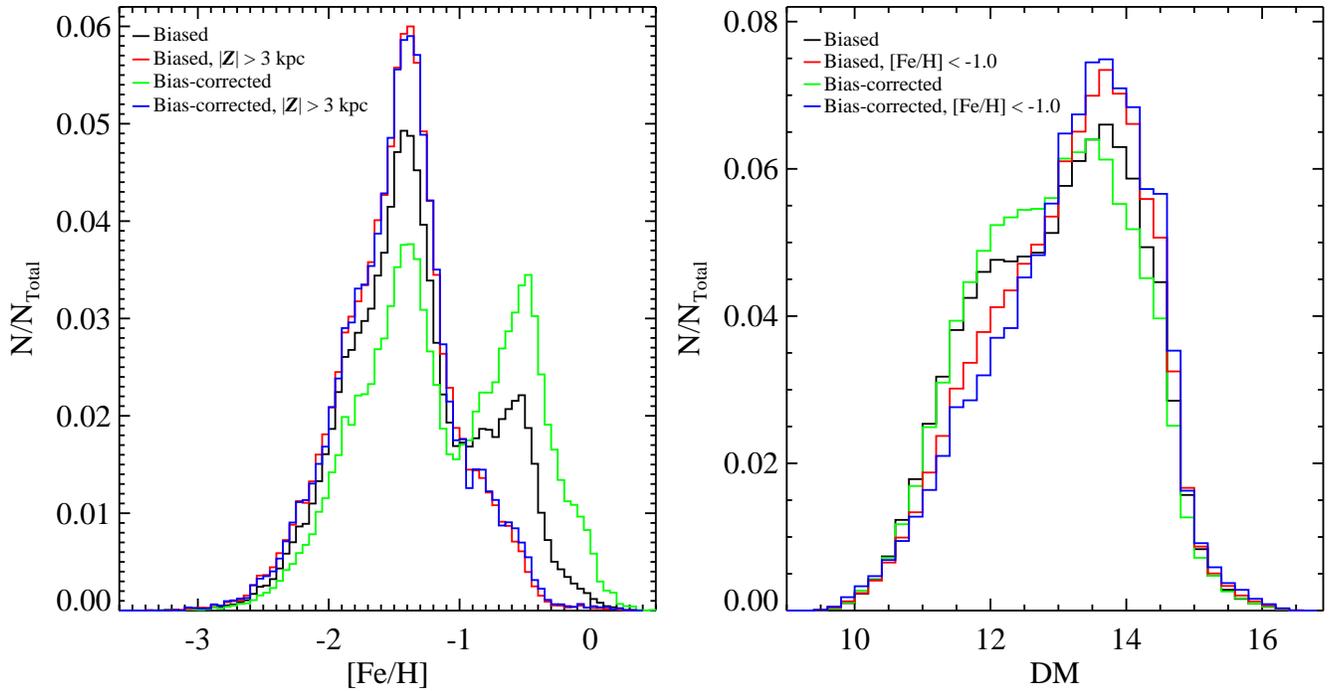}
\caption{$Left~panel$: Metallicity distribution functions for the original, 
target-selection biased (black and red), and bias-corrected (green and blue) samples of our MSTO stars. 
The red and blue histograms are for the stars with \z\ $>$ 3 kpc. $Right~panel$: Same as 
in the left panel, but for the distributions of our adopted distance moduli. The 
red and blue histograms are for the stars with [Fe/H] $<$ --1.0. The two panels indicate 
that the selection bias is very minimal when considering the stars with \z\ $>$ 3 kpc or 
[Fe/H] $<$ --1.0, for which halo stars are dominant.}
\label{fig:feh_hist}
\end{figure*}

\begin{figure*}
\centering
\includegraphics[width=0.32\textwidth]{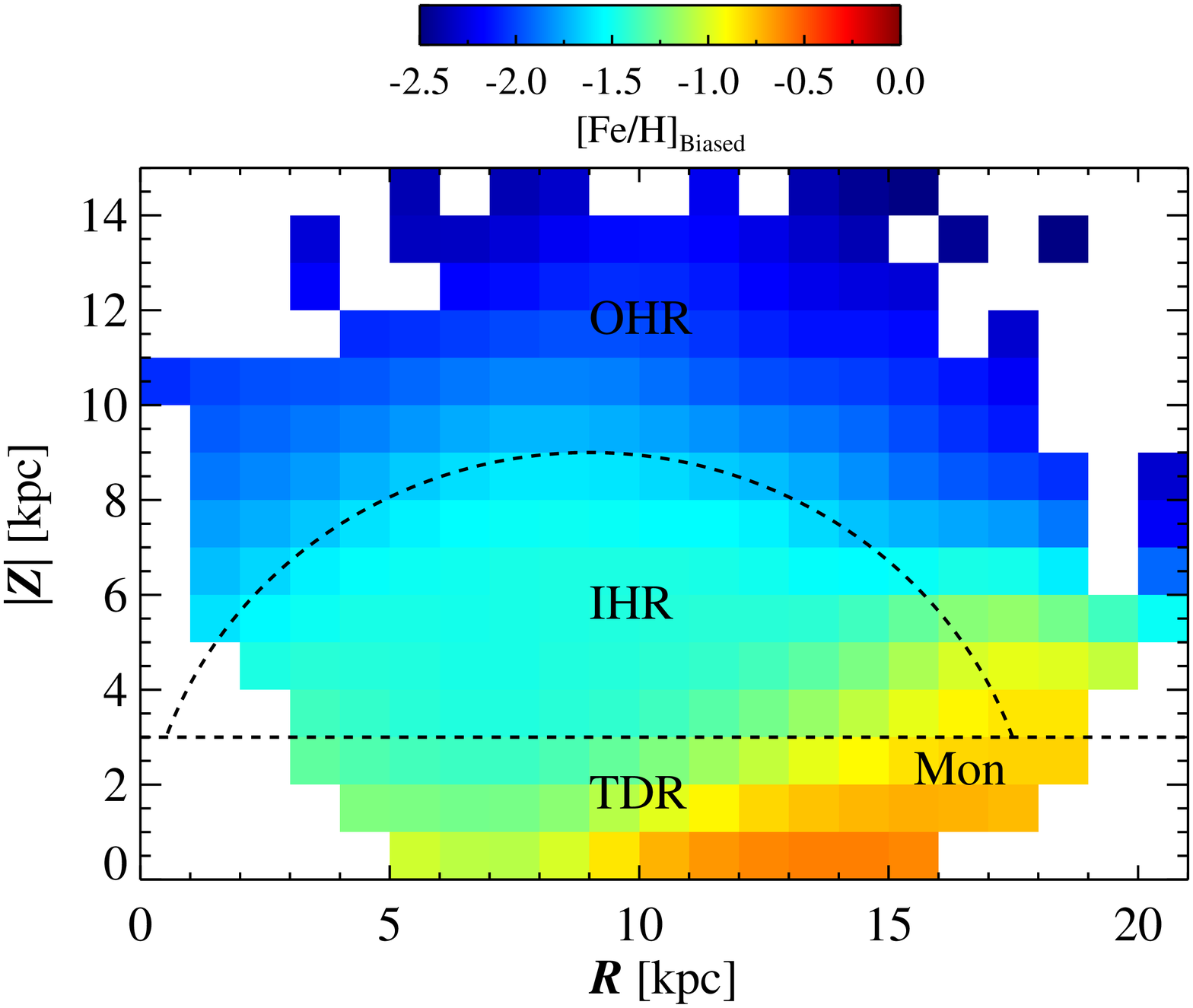}\hspace*{0.2cm}
\includegraphics[width=0.32\textwidth]{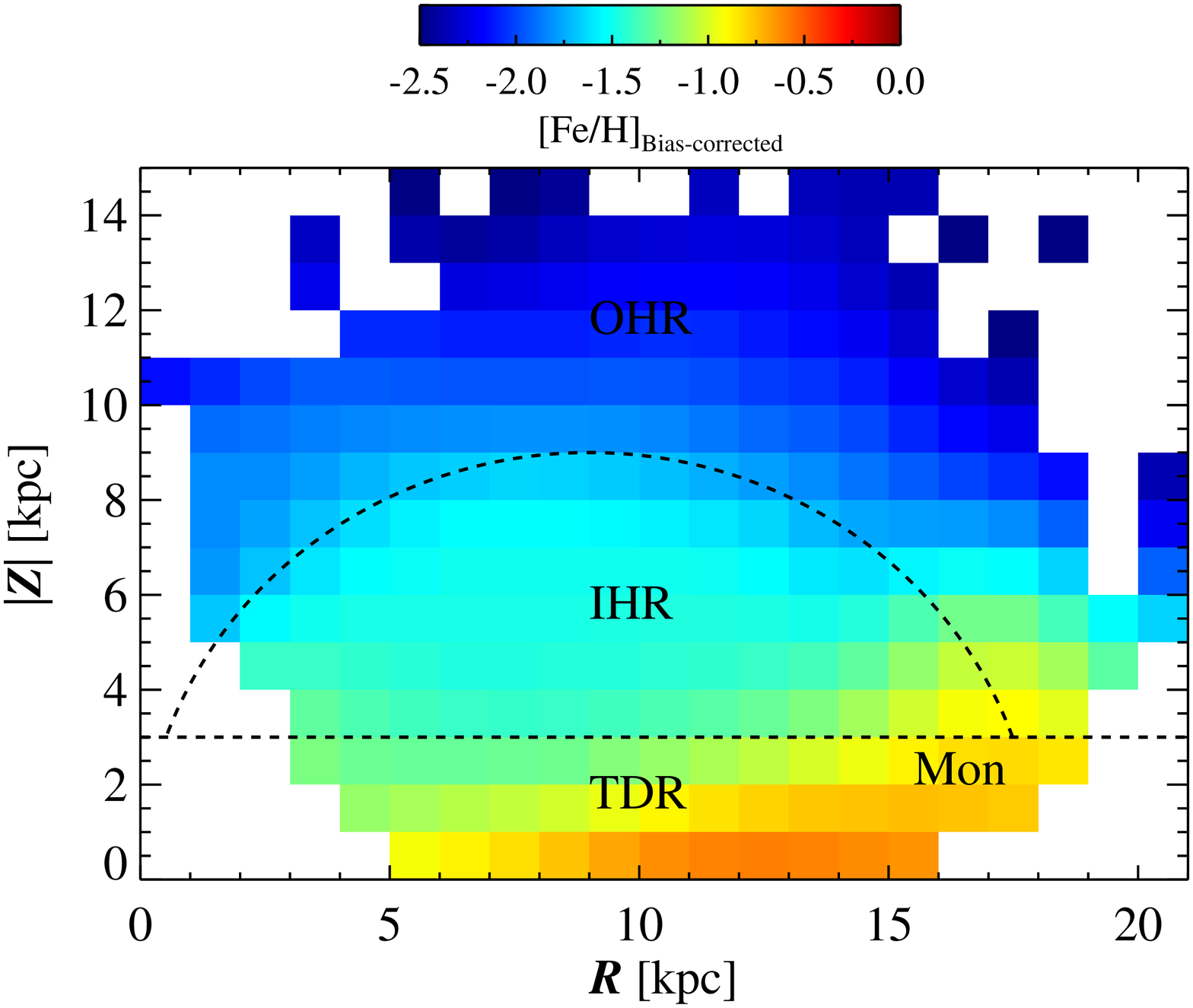}\hspace*{0.2cm}
\includegraphics[width=0.32\textwidth]{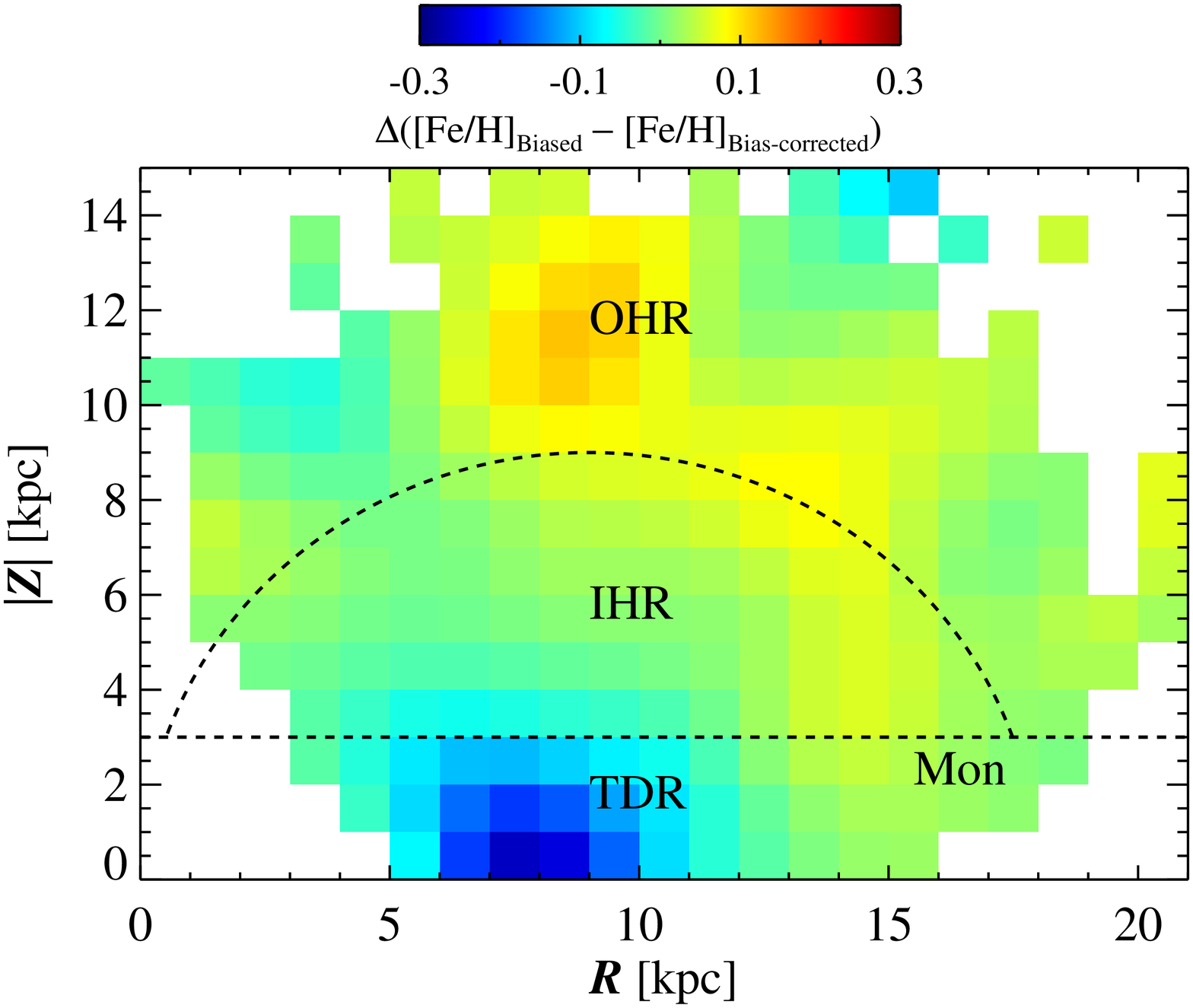}
\caption{$Left~panel$: Same as in Figure \ref{fig:cmap}, but for 
[Fe/H] of our original selection-biased sample. $Middle~panel$: Same as in Figure \ref{fig:cmap}, 
but for [Fe/H] of the bias-corrected sample. The high metallicity 
region with $R$ $>$ 15 kpc and \z\ $<$ 5 kpc indicates the Monoceros Stream 
(\citealt{newberg2002,ivezic2008}). It is marked with ``Mon''. $Right~panel$: Same as in 
Figure \ref{fig:cmap}, but for the difference in 
[Fe/H] between the biased and bias-corrected samples. Note that only
small differences exist, with only regions near the disk approaching 0.3
dex.}
\label{fig:feh_map}
\end{figure*}

\begin{figure*}[t]
\centering
\epsscale{1.15}
\plottwo{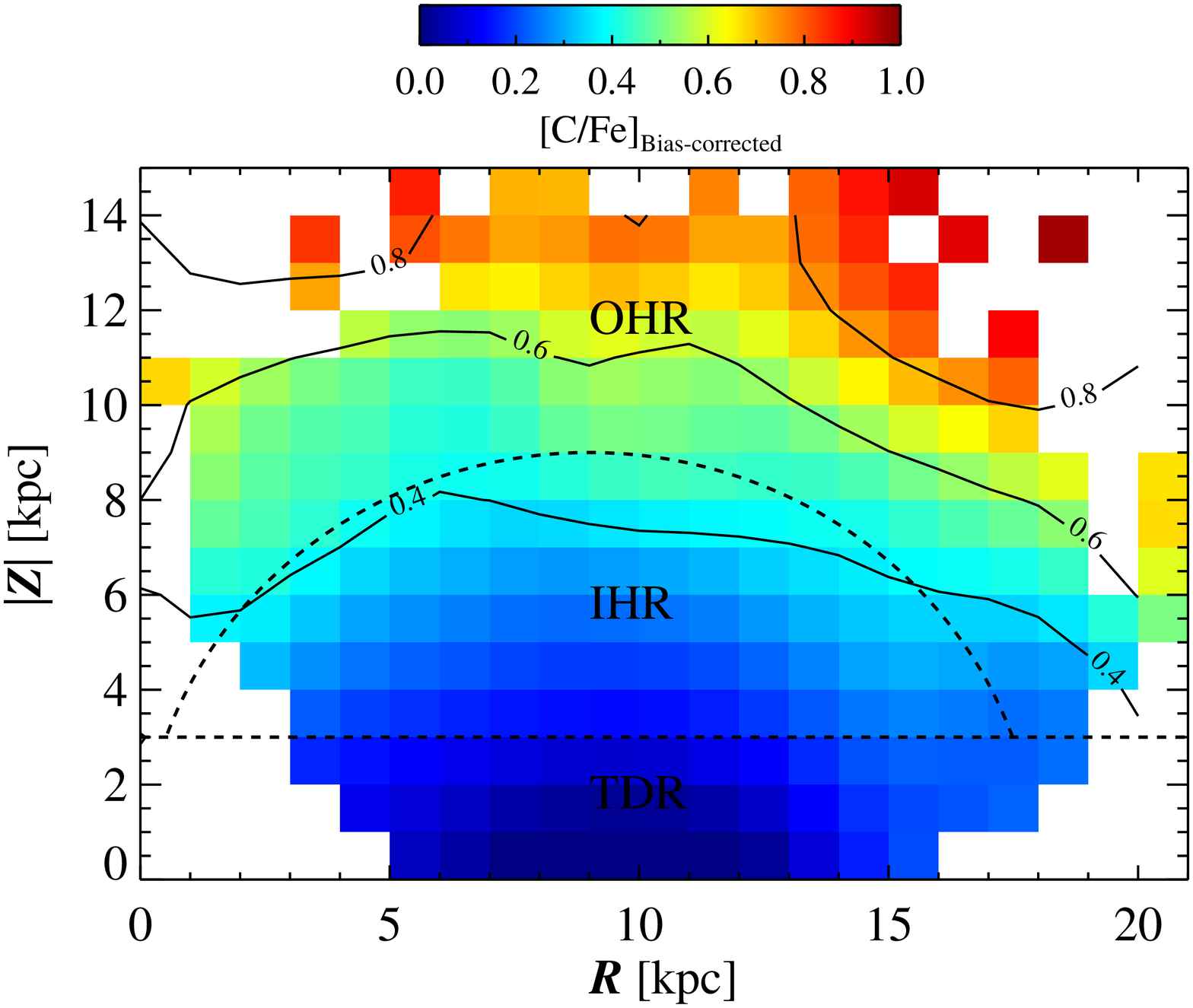}{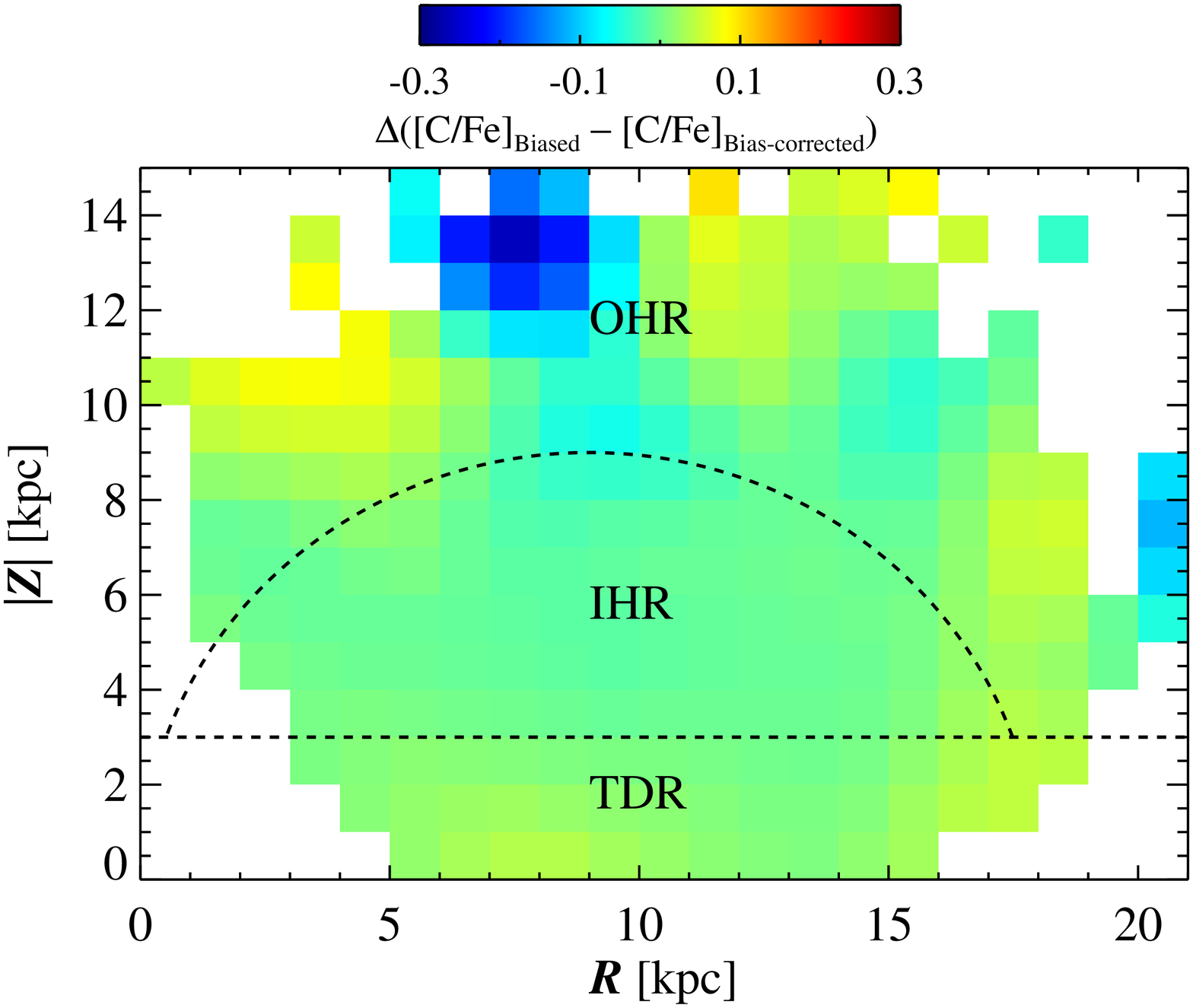}
\caption{$Left~panel$: Same as in Figure \ref{fig:cmap}, but for the
[C/Fe] of the bias-corrected sample. $Right~panel$: Same as in Figure \ref{fig:cmap}, but for the 
difference in [C/Fe] between the biased and bias-corrected samples.}
\label{fig:cmap_diff}
\end{figure*}

\begin{figure*}[t]
\epsscale{1.0}
\plotone{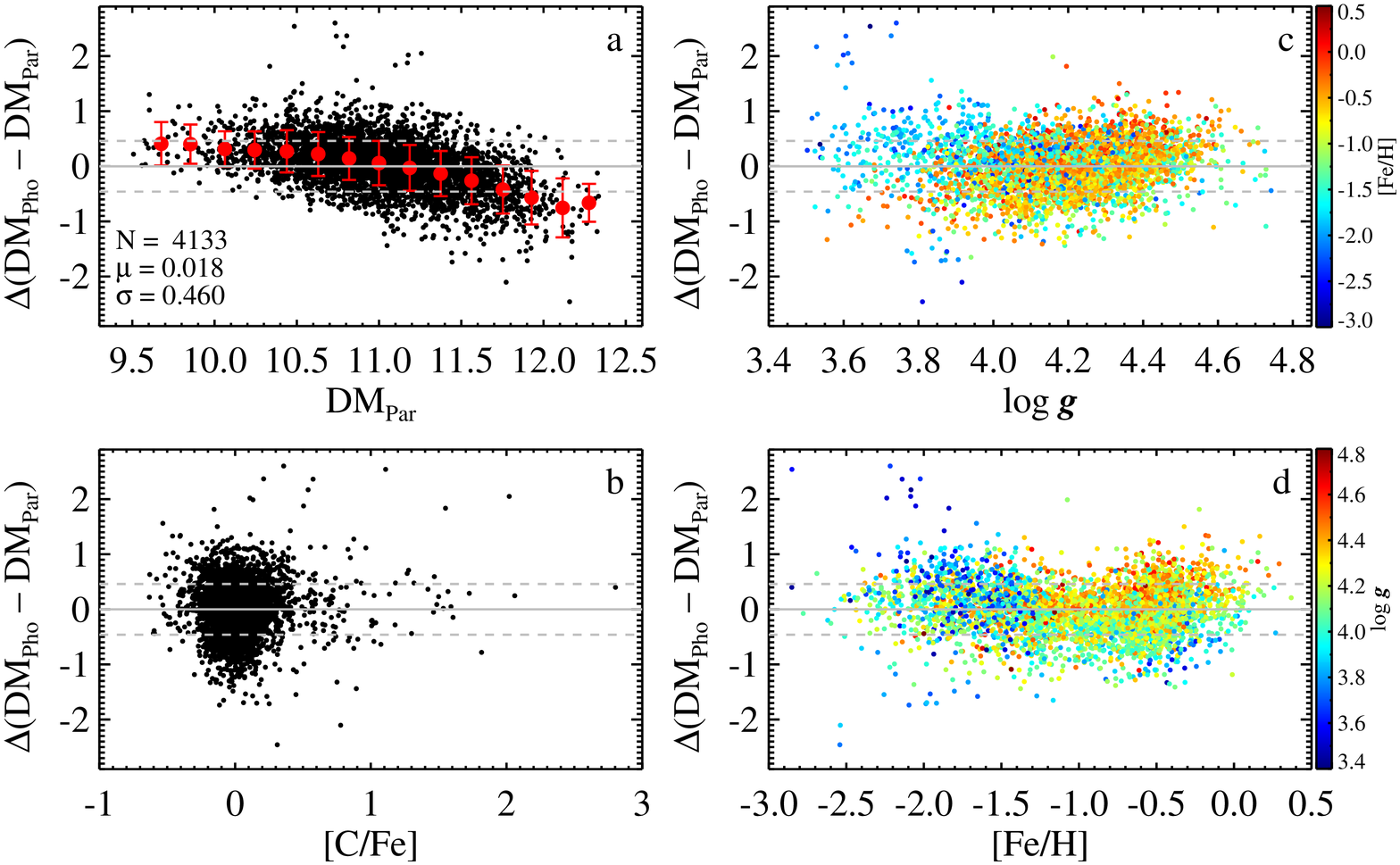}
\caption{Residual plots of distance moduli between our photometric distance
(DM$_{\rm Pho}$) and that (DM$_{\rm Par}$) from \gaia\ DR2, as a
function of DM$_{\rm Par}$ (panel a), [C/Fe] (panel b), \logg\ (panel c), and [Fe/H] (panel d). 
We adjusted the \gaia\ parallax by --0.054 mas, and considered only stars that satisfy
$\sigma_\pi / \pi < 0.1$ and $\sigma_\pi \leq 0.07$ mas, where $\pi$
indicates the parallax. The total number of stars, the mean offset, and
the standard deviation are listed in the left-bottom corner of panel a. 
The gray-dashed lines indicate the 1$\sigma$ regions. In panel a, 
each red dot indicates the mean value of stars in a bin of 0.4 mag in the distance modulus, and 
each bin is overlapped with the next neighboring bin by 0.2 mag. The error bar 
is derived from resampling stars 100 times in each bin. In panels c and d, 
the color-coded dots indicate the scale of metallicty and surface gravity, respectively, 
as shown in the color bar on the right.}
\label{fig:check_dist}
\end{figure*}

\subsection{Sub-classification of the MSTO Sample into CEMP-$s$ and CEMP-no
Stars} \label{sec:classsification}

As we seek to identify possible distinct kinematic differences between
\cemps\ and CEMP-no stars of the IHR and OHR, in order to understand the
origin of the dichotomy of the Galactic halo system, we first need to
establish the criterion for making this separation. The conventional
approach to distinguish CEMP-$s$ from CEMP-no stars is to employ [Ba/Fe]
abundance ratios derived from high-resolution spectroscopy. Recently,
however, \citet{yoon2016} demonstrated that these sub-classes of CEMP
stars can be equally well identified by the level of absolute carbon
abundance, $A$(C), which can be derived from medium-resolution
spectroscopy. As we have no high-resolution measurements of [Ba/Fe] for
the MSTO stars, we applied this latter approach to sub-classify our
program stars.

Figure \ref{fig:ac_feh} illustrates the methodology we employ to
distinguish \cemps\ stars and CEMP-no stars from CEMP stars in our MSTO
sample. The right panel of this figure is the histogram of \ac\ values,
which exhibits a clear bimodal distribution of \ac. To divide the
low-\ac\ stars, which we assign to the CEMP-no sub-class from the
high-\ac\ stars, which we assign to the \cemps\ sub-class, we first fit
two Gaussians to the distribution of \ac. From this exercise, we
identify the point where the two Gaussians cross, \ac\ = 7.42, indicated
as a black dashed line in the figure. We distinguish CEMP-no stars with
\ac\ $\le$ 7.42 from the \cemps\ stars with \ac\ $>$ 7.42. Note that in
this exercise, we only considered stars with [C/Fe] $\ge$ +0.7 with
spectra having S/N $\geq$ 30.

Our adopted value of \ac\ = 7.42 differs somewhat from that
of \citet{yoon2016}, \ac\ = 7.1, which is derived from stars with
available high-resolution spectroscopy. There are several factors that
account for this discrepancy. Our MSTO sample covers substantially
different ranges of metallicity, as can be seen by comparison of
Figure~\ref{fig:ac_feh} with Figure 1 of \citet{yoon2016}. Their CEMP-no
stars are dominated by stars with [Fe/H] $< -3.0$, unlike our sample,
which is dominated by stars with [Fe/H] $> -3.0$. In addition, their
sample includes a large number of sub-giants and giants, whereas our
sample of MSTO excludes these stars by definition. If we only
consider the CEMP stars in their sample with [Fe/H] $> -3.5$ and in
the same temperature and gravity ranges as our sample, application of our
above procedure to separate CEMP-no and \cemps\ stars yields
a division at \ac\ = 7.2, similar to their adopted value.

Our inability to detect low-\ac\ CEMP-no stars among
metal-poor MSTO stars also contributes to the discrepancy. Even though
we restricted our sample to rather narrow ranges of the temperature and
surface gravity, the appropriate $A$(C) division line between CEMP-no
and CEMP-$s$ stars probably still depends, at least weakly, on stellar
temperature, luminosity class, and metallicity. Since a
value of $A$(C) = 7.42 is more suitable for dividing the CEMP-no and
\cemps\ stars in our MSTO sample, we adopted this value for our analysis. As
CEMP-$r$ and CEMP-$r/s$ (or CEMP-$i$) stars generally exhibit relatively 
higher \ac\ values than the CEMP-no stars, we considered all CEMP stars 
with \ac\ $>$ 7.42 as \cemps\ stars. Stars in these sub-classes represent 
a small fraction of CEMP stars, in any case.

Note that, when classifying CEMP stars into 
the \cemps\ and CEMP-no sub-classes following the method described
above, there exists some level of cross-contamination. However,
according to \citet{yoon2016}, the cross-contamination fraction is less
than 10\%, which is very small, compared to the direct use of Ba
abundances derived from high-resolution spectroscopy. In addition, 
there may exists degeneracy between CEMP-no and CEMP-$r$ and 
CEMP-$s$ and CEMP-$i$. However, as they are minority among CEMP stars, 
the contamination by these objects are negligible as well.

\section{Target-Selection Bias and Distance Errors}

Since biases arising from the selection
of spectroscopic targets in SDSS and any systematic errors in derived
distances for our MSTO sample could affect our results and subsequent
interpretation, in this section we examine their possible impact.

\subsection{Target-Selection Bias} \label{sec:selection}

As the target selection for spectroscopy in SDSS 
was mostly carried out by application of cuts in apparent magnitude and
colors, the observed stars can possibly be biased toward the inclusion
of more metal-poor stars, in particular as a function of distance. In
turn, this can affect the carbonicity map (Figure \ref{fig:cmap}), and
identification of the Galactic halo regions, hence their kinematic
properties and interpretation. 

In order to test the severity of this
potential bias, we first obtained a sample of MSTO stars corrected for
this selection bias, in order to evaluate how much it affects the
underlying metallicity distribution of our MSTO sample. We followed the
usual approach for deriving the selection function for our MSTO stars,
as described in other studies (e.g., \citealt{schlesinger2012,nandakumar2017,wojno2017,
chen2018}). The basic idea is to calculate the fraction of the
spectroscopically targeted stars among the photometrically available
targets in a certain range of magnitude and color on a color-magnitude
diagram, individually for each SDSS plug-plate. We adopted a magnitude
and color bin size of 0.2 mag and 0.05 mag, respectively, for a
color-magnitude diagram of $r$ and $g-r$ for this calculation. Here, we
regard the selection function as the ratio of the number of stars
selected for spectroscopic observation to the number of stars present in
the direction of a given plug-plate with available photometry in each 
magnitude and color bin. We corrected for the selection bias of our MSTO 
stars by taking the inverse of the selection function for each object. We then 
cross-checked our estimation of the selection function for a subset of our sample with 
that derived by \citet{mints2019}, and confirmed good agreement.

After obtaining the selection function, we compared the metallicity 
distribution function (MDF) of the as-observed, and potentially biased
sample (black histogram) with that (green histogram) of the bias-corrected
sample, as shown in the left panel of Figure \ref{fig:feh_hist}. In the figure, 
we see clearly that our sample has a relatively greater fraction of metal-poor
stars for [Fe/H] $<$ --1.0, and a smaller number of metal-rich stars for
[Fe/H] $>$ --1.0. However, if we restrict our sample of stars 
to \z\ $>$ 3 kpc, for which the halo stars are dominant and in which we are most 
interested, the selection bias of our sample is minimal as can be 
seen by inspection of the blue and red histograms in the left 
panel of Figure \ref{fig:feh_hist}. 

We recognize as well that the selection bias that we want to remove 
is not only a function of [Fe/H], but also possibly a function of distance and age. 
We investigated the possible distance bias, as shown in 
the right panel of Figure \ref{fig:feh_hist}, which shows the distributions 
of the distance moduli (DM). The distance moduli were calculated 
based on our adopted distance derived by the SSPP. The black histogram 
represents the biased sample of our MSTO stars, while the green histogram 
applies to the unbiased sample. Comparison of the two 
histograms reveals that the distribution of the unbiased sample 
is slightly shifted to shorter distance. However, as in the case 
of the metallicity, by restriction of our sample to [Fe/H] $<$ --1.0, 
which is dominated by halo stars, the distance bias arising from the 
target selection is clearly diminished, as can be seen from comparison of the 
red (biased) and blue (unbiased) histograms. 

Concerning the age bias, since the stars in the halo region (\z\ $<$ 15 kpc) 
under question in this study have, on average, similar ages greater than 12 Gyr 
(\citealt{santucci2015,carollo2016,das2016b}), we can assume that the 
selection function has little dependence on stellar age, and would not 
affect our analysis.

We also investigated how the selection bias
against the metal-rich stars is reflected in the metallicity map, where
we want to delineate the Galactic halo region. Figure \ref{fig:feh_map}
compares the map of our original sample (left panel) with that of the
bias-corrected (middle panel) samples. It appears that there is
little difference between our sample map and unbiased maps. It is also very clear
that the division line between the IHR and OHR (obtained by inspection
of the carbonicity map of Figure 1) is well-established in
both maps. 

The right panel of Figure \ref{fig:feh_map} quantitatively 
shows the difference between the left and the middle panel. According to
the difference map, even though there are small deviations present, they
are less than 0.1 dex over most of the map, except for the disk region
(which is not of interest to this study), for which the discrepancy is
as high as 0.3 dex. 

As a further test, we identified the Galactic halo 
regions on the carbonicity map obtained after correcting for the
target-selection bias, as shown in the left panel of Figure
\ref{fig:cmap_diff}. Compared to Figure \ref{fig:cmap}, even if the
contour lines are slightly changed, we can tell that the boundary for
the OHR established in the original carbonicity map is well-represented
in the bias-corrected map. The right panel of Figure
\ref{fig:cmap_diff}, which shows the difference between the
selection-biased carbonicity map (Figure \ref{fig:cmap}) and the unbiased carbonicity 
map (left panel of Figure \ref{fig:cmap_diff}), quantitatively underscores the 
minor impact of the target-selection bias, as it exhibits very small variations, 
of less than 0.1 dex in [C/Fe], over most of the halo region.

One may naively think that the carbonicity map (Figure \ref{fig:cmap}) 
might be predictable from the metallicity map (left panel 
of Figure \ref{fig:feh_map}), as apparently we can observe a well-behaved trend 
between [C/Fe] and [Fe/H]. However, since the carbon enhancement varies 
with [Fe/H] unpredictably, and the different stellar types and luminosity classes exhibit different 
levels of carbon enhancement (e.g., \citealt{rossi1999, lee2013, yong2013,
placco2014, yoon2016}), it is, in fact,  not possible to predict the carbonicity map from the 
metallicity map. In addition, one can notice a less-smooth distribution of the 
carbonicity map than for the metallicity map, which would not be predicted 
from the metallicity map.  We conclude that the carbonicity map enables us to better 
understand the nature of the Galactic halo, complimenting the metallicity map.

Summarizing, Figures \ref{fig:feh_hist}, \ref{fig:feh_map}, and \ref{fig:cmap_diff} suggest 
that the selection bias present in our MSTO sample does not significantly affect the
metallicity and carbonicity distributions over the Galactic halo region under 
investigation, and might only be a concern if one wants to derive the exact 
shape of the sample MDF. This is primarily due to the fact that the photometric
target selection quickly loses metallicity sensitivity below [Fe/H] =
--2.0. Our examination of large swaths of sky in performing our
analysis also mitigates the potential impact in any given direction. Thus, 
in the following analysis and discussion, we employ the original MSTO halo stars, without 
applying corrections for the selection bias unless otherwise specifically mentioned.

\begin{figure*}
\centering
\includegraphics[width=0.32\textwidth]{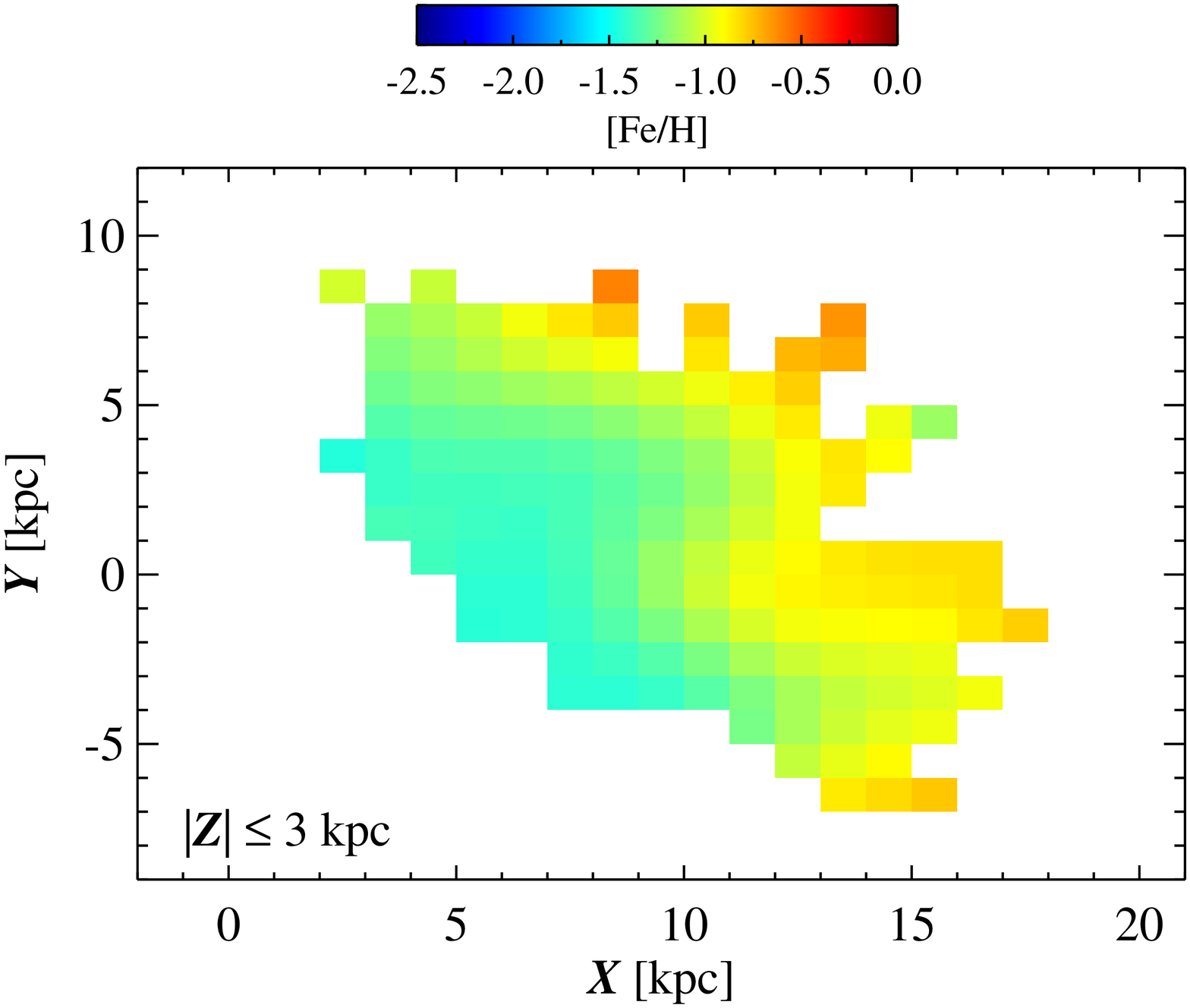}\hspace*{0.2cm}
\includegraphics[width=0.32\textwidth]{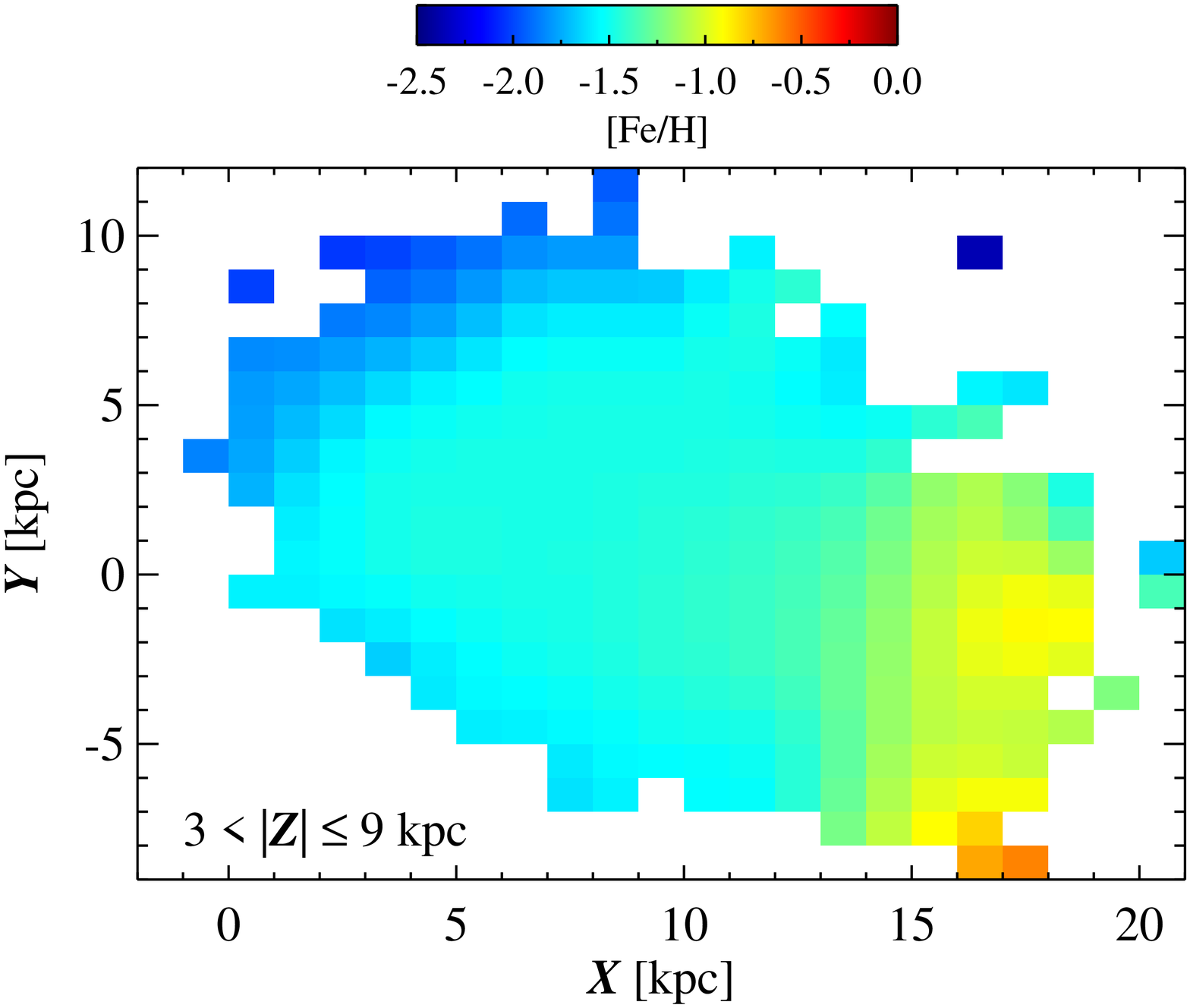}\hspace*{0.2cm}
\includegraphics[width=0.32\textwidth]{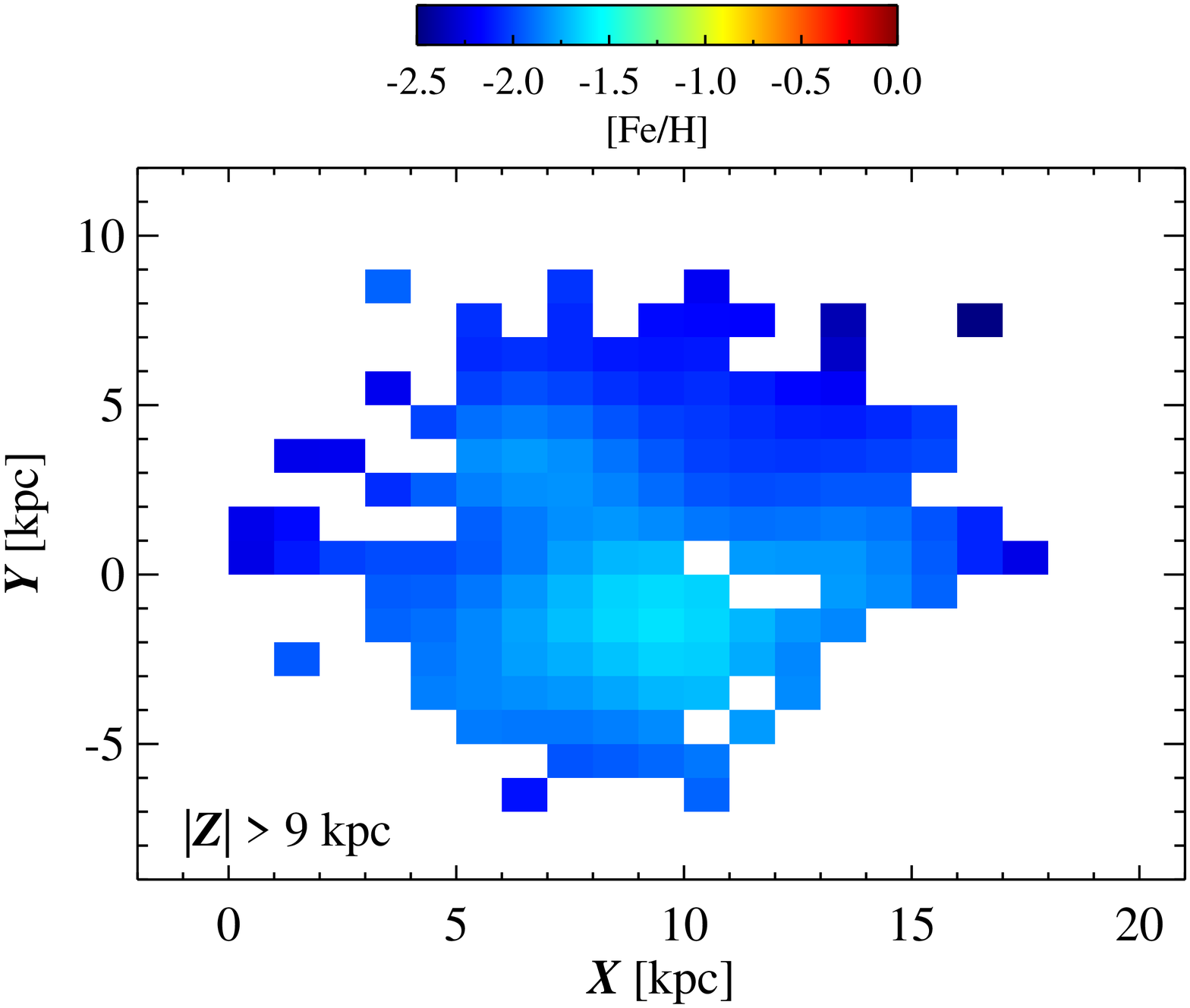}
\caption{Metallicity maps for our MSTO sample in the $X$ and $Y$ plane, 
in the range of \z\ $\leq$ 3 kpc (left panel), 3 $<$ \z\ $\leq$ 
9 kpc (middle panel), and \z\ $>$ 9 kpc (right panel), which approximately correspond to the TDR, 
IHR, and OHR, respectively. In this plane, our Sun is located at ($X$,$Y$)=(8,0) kpc.
The bin size is 1$\times$1 kpc. Each pixel contains at least three stars and 
represents a median value of [Fe/H], with the color scale shown in the color bar.}
\label{fig:feh_zcut}
\end{figure*}

\begin{figure*}
\centering
\includegraphics[width=0.32\textwidth]{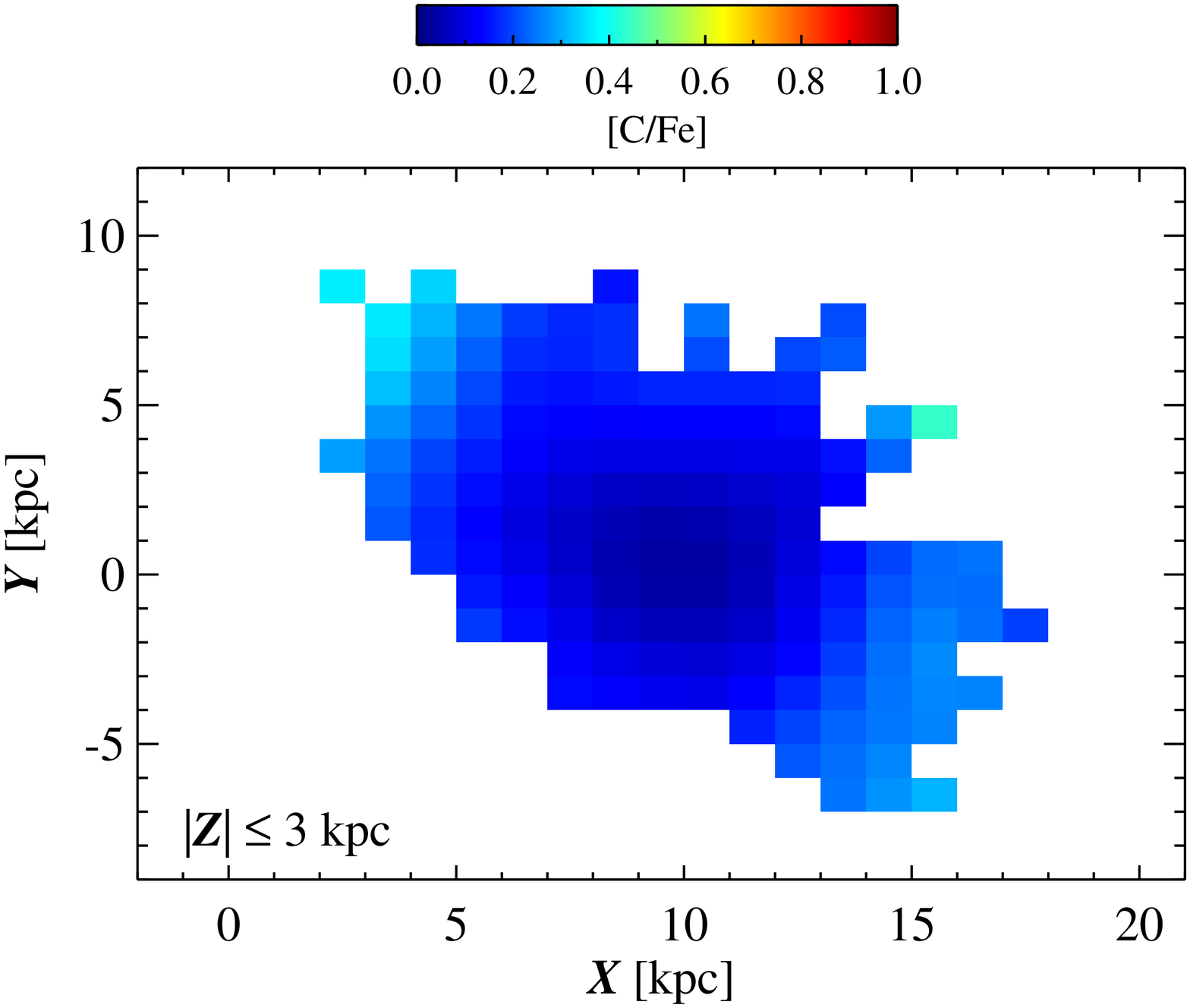}\hspace*{0.2cm}
\includegraphics[width=0.32\textwidth]{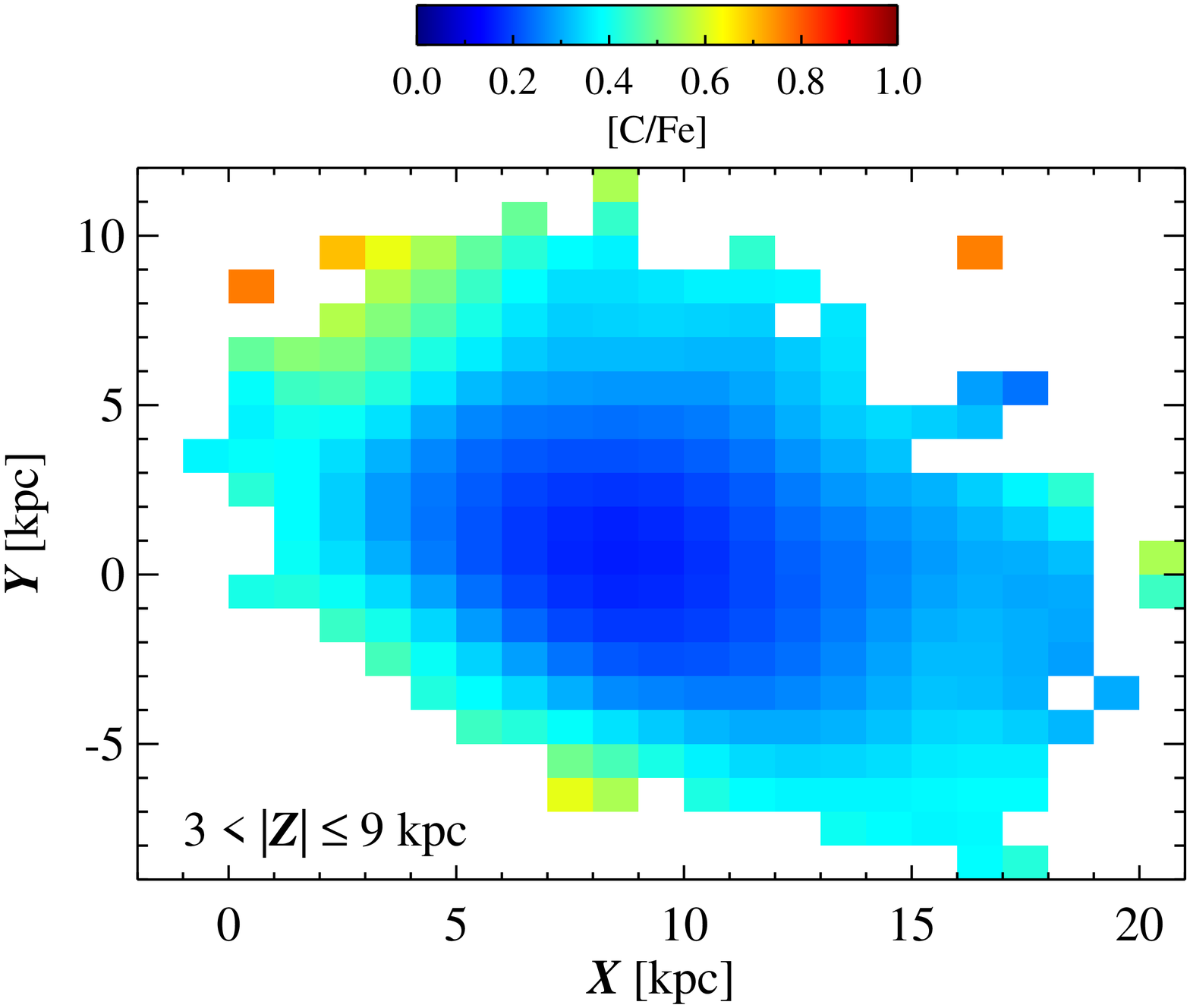}\hspace*{0.2cm}
\includegraphics[width=0.32\textwidth]{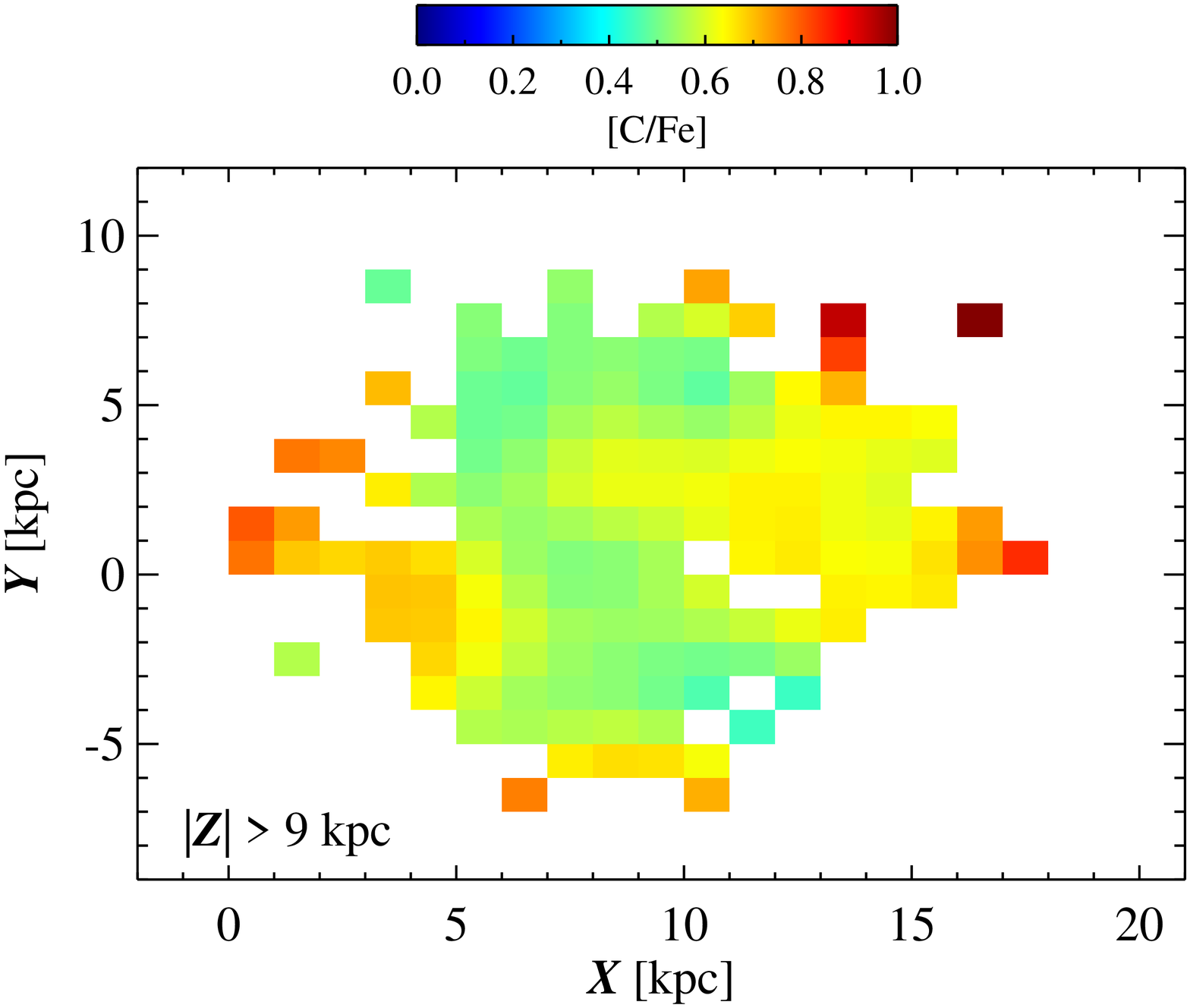}
\caption{Same as in Figure \ref{fig:feh_zcut}, but for [C/Fe].}
\label{fig:cfe_zcut}
\end{figure*}

\subsection{Distance Errors} \label{sec:error}

In order to check on the accuracy of our derived distance scale,
we matched our MSTO sample with stellar sources in \gaia\ DR2, and
selected stars with high-precision parallaxes, by applying
${\sigma}_{\pi}/{\pi} < 0.1$ and ${\sigma}_{\pi} < 0.07$ mas, 
as suggested by \citet{schonrich2019}, where $\pi$
is the parallax and ${\sigma}_{\pi}$ is the uncertainty. After adjusting
for the known zero-point offset of --0.054 $\rm{mas}$
(\citealt{schonrich2019,gracyk2019}), we compared our derived distances with the
\gaia\ DR2 distances. Figure \ref{fig:check_dist} shows the differences in
the distance moduli between our photometric distance (DM$_{\rm pho}$)
and that of \gaia\ DR2 (DM$_{\rm Par}$), as a function of DM$_{\rm Par}$ (panel a), 
[C/Fe] (panel b), \logg\ (panel c), and [Fe/H] (panel d). 
The total number of stars, the mean offset, and the standard deviation are listed in the 
left-bottom corner of panel a. The gray-dashed lines indicate the 1$\sigma$ regions. 
In panel a, each red dot represents a mean value of stars in a bin of 0.4 mag in the 
distance modulus, and each bin is overlapped with the next neighboring bin by 0.2 mag. 
The error bar is derived from resampling stars 100 times in each bin. In panels c and d, 
the color-coded dots indicate the scale of metallicty and surface gravity, respectively, 
as shown in the color bar on the right.

Inspection of panel a of the figure reveals
that, even though there is a tendency such that our distance 
estimate is slightly lower for the most remote stars and slightly higher for closer stars, the overall 
systematic offset in the distance is very small. We note that the distance moduli 
of our MSTO stars mostly agree very well with those having \gaia\ DR2 parallaxes, and are within 
2$\sigma$ for more than 95\% of the stars considered. Moreover, it is noteworthy that the 
trend (red dots) of the mean offset with DM$_{\rm Par}$ is within the 1$\sigma$ 
region, after taking into account the error bar of each red dot. Additionally, in 
panels b, c, and d of Figure \ref{fig:check_dist}, we do not see evidence for any 
trends between our spectroscopic distance moduli and the \gaia\ DR2 moduli over [C/Fe], \logg, 
and [Fe/H], respectively, indicating that there exists no significant systematic errors in 
our derived distance scale due to the presence of strong carbon bands or incorrect assignment
of the luminosity class. We also do not find any complex trends among 
the parameters (panels c and d). These tests indicate the robustness of 
our distance estimates.

\begin{figure*}[!t]
\centering
\includegraphics[width=0.32\textwidth]{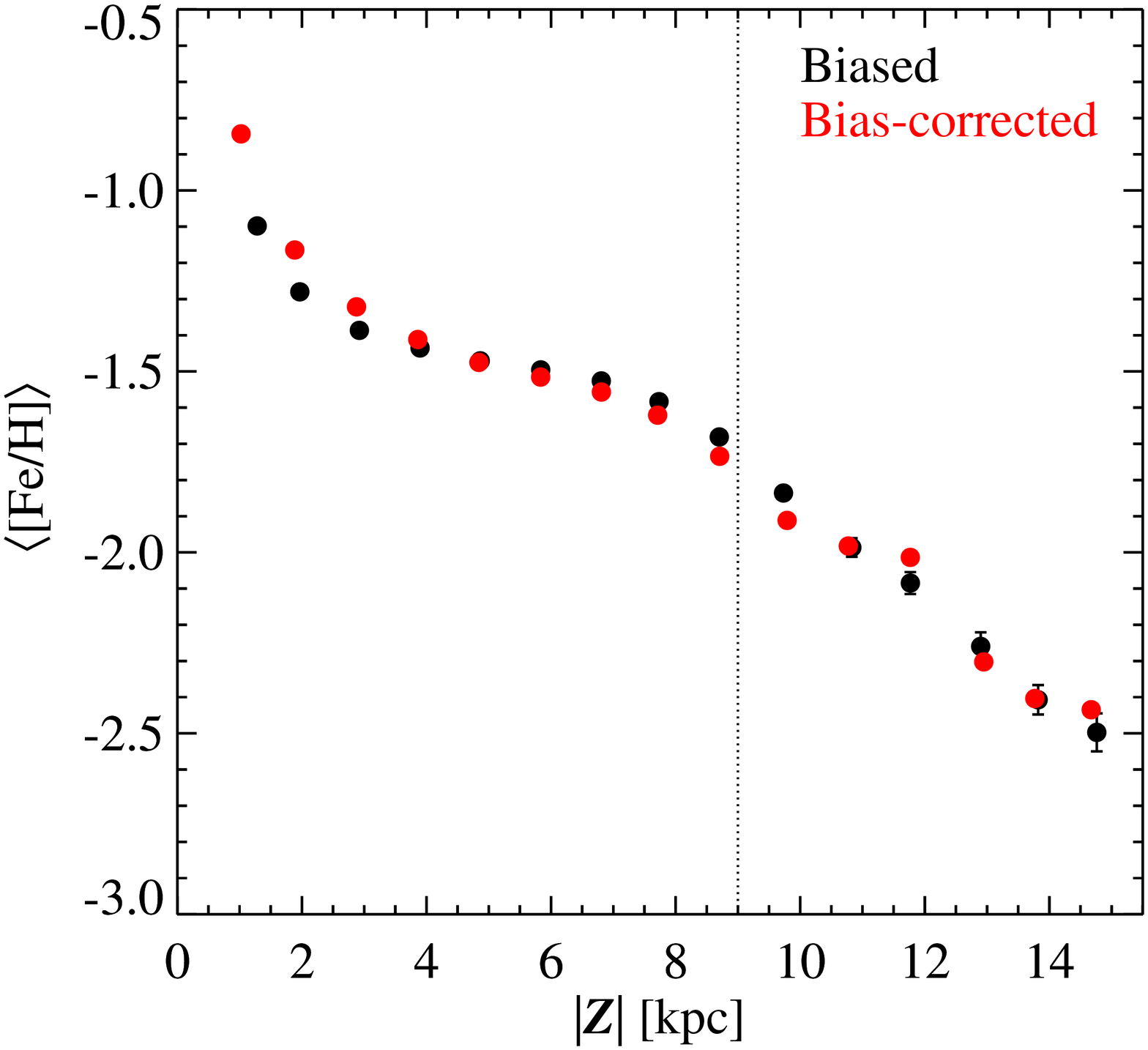}\hspace*{0.2cm}
\includegraphics[width=0.32\textwidth]{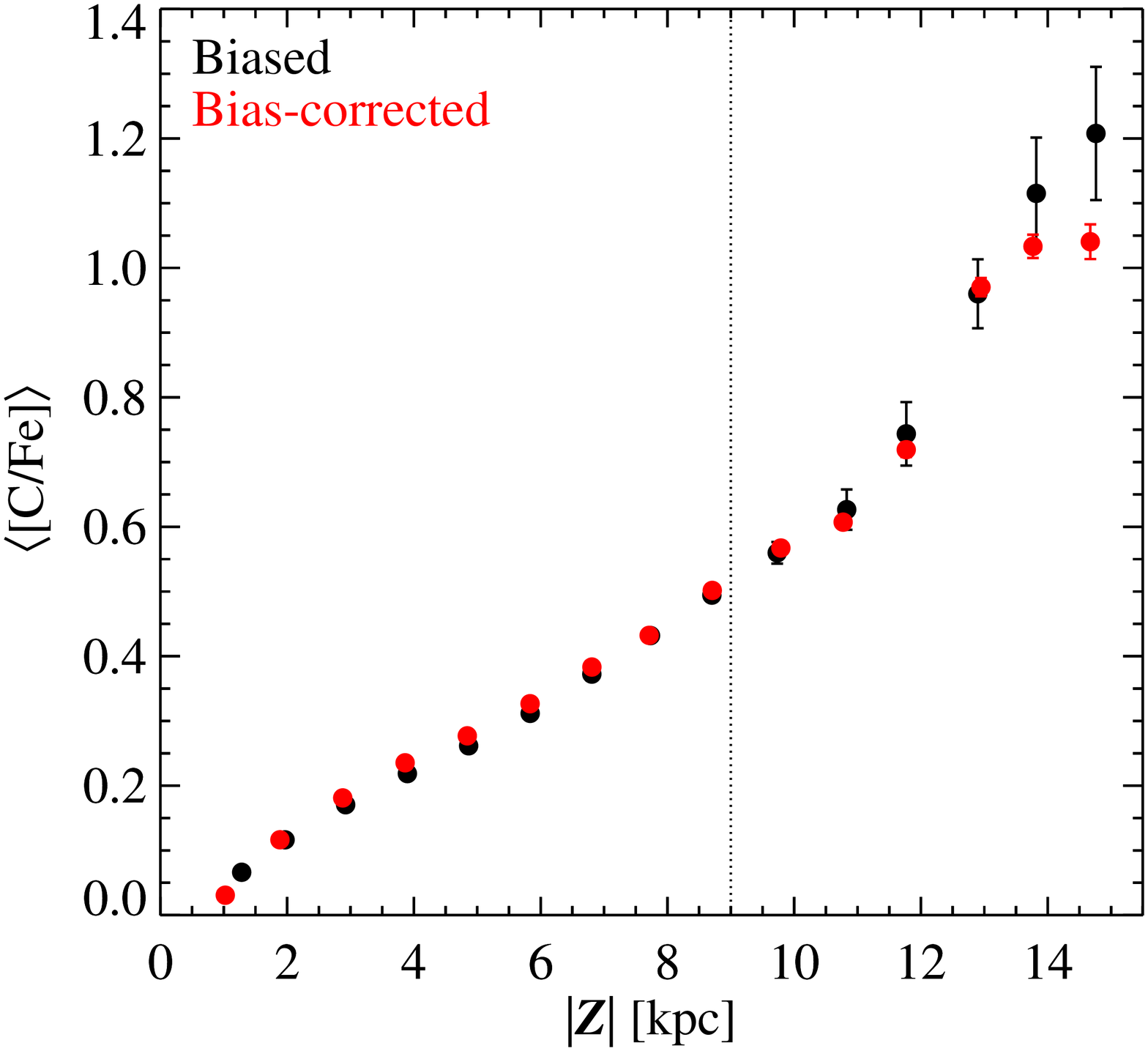}\hspace*{0.2cm}
\includegraphics[width=0.32\textwidth]{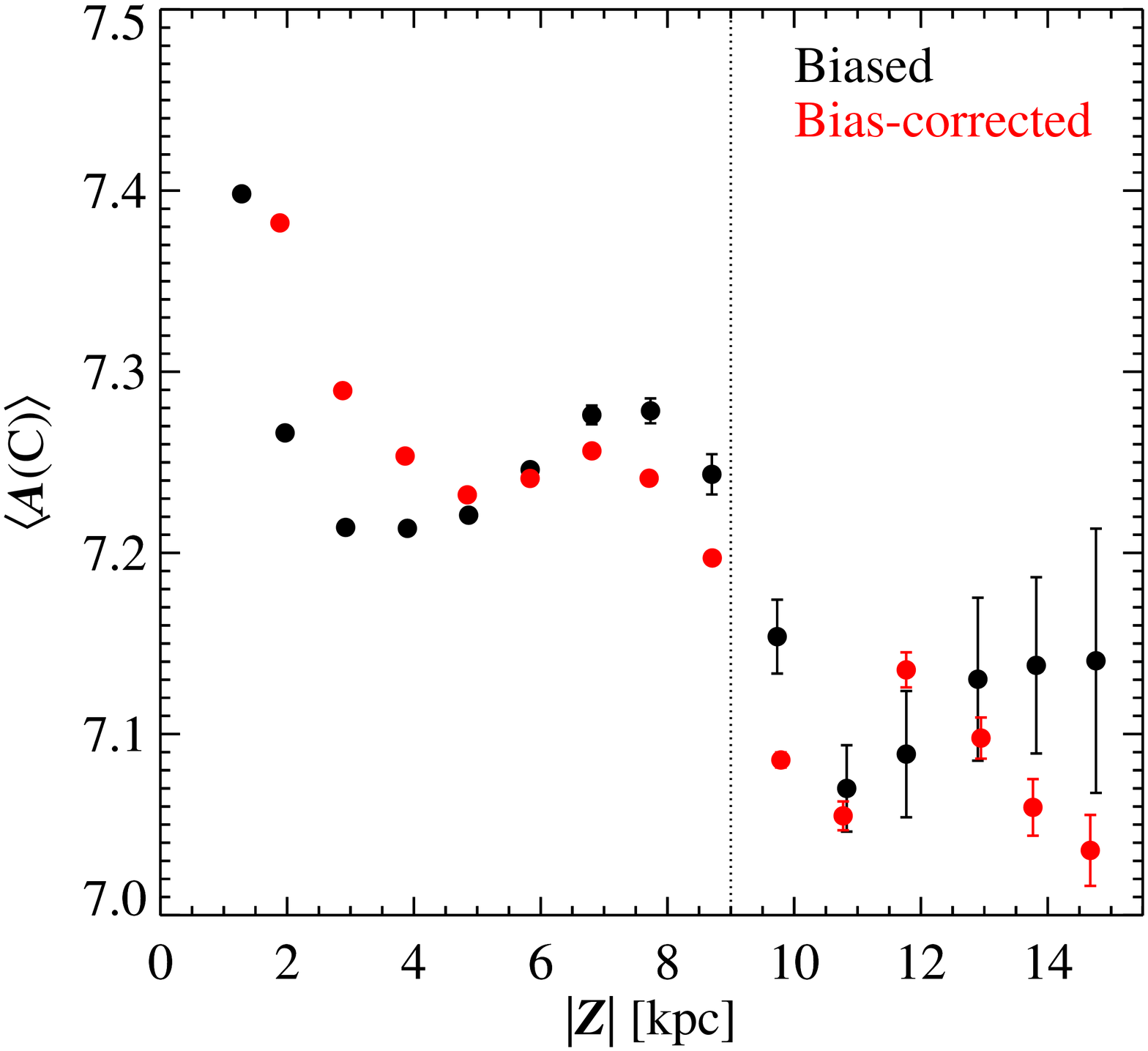}
\caption{Distributions of mean values of [Fe/H] (left panel), [C/Fe] (middle panel), and
\ac\ (right panel), as a function of \z. Each dot represents the mean value for stars within a bin of 2 kpc, and
each bin is overlapped with the next neighboring bin by 1 kpc. The error
is the standard deviation derived by resampling stars 100 times in each
bin. The red dots are derived from the sample corrected for the selection bias. 
The left panel exhibits a slow decrease of $\langle$[Fe/H]$\rangle$
between \z\ = 3 and 8 kpc, and a rather dramatic decrease of [Fe/H]
beyond \z\ = 8 kpc. The bias-corrected sample also shows very similar trend. 
The middle panel shows a steady increase of
$\langle$[C/Fe]$\rangle$ up to \z\ = 11 kpc, and a rapid increase beyond
\z\ = 11 kpc, even though the bias-corrected sample exhibits a somewhat lower increase. 
In the right panel, the mean value of
$\langle$\ac$\rangle$ gradually increases between \z\ = 3 kpc and 8 kpc,
then abruptly drops up to \z\ = 11 kpc, and exhibits a constant, lower 
value of \ac\ for both the biased and bias-corrected samples. 
The vertical dotted line at \z\ = 9 kpc indicates the highest \z\ point in the region 
that separates the IHR from the OHR in Figure \ref{fig:cmap}.}
\label{fig:mean_cfe}
\end{figure*}

\section{Chemical and Kinematic Signatures of Distinct Populations in the Galactic Halo}
\label{sec:distinct}

In this section, we seek to identify differences in stellar populations in the Galactic
halo by searching for distinguishing trends in their chemistry and kinematics.

\subsection{Chemical Signatures}\label{sec:chemical}

We first examine the carbonicity and metallicity distributions 
in the $X$--$Y$ plane in different regions of \z. Figure \ref{fig:feh_zcut} exhibits 
the metallicity maps for our MSTO sample in the $X$ and $Y$ plane for the regions 
of \z\ $\leq$ 3 kpc (left panel), 3 $<$ \z\ $\leq$ 9 kpc (middle panel), and \z\ $>$ 9 kpc (right panel). 
These regions approximately correspond to the TDR, IHR, and OHR, respectively. In 
this plane, our Sun is located at ($X$,$Y$)=(8,0) kpc. Each bin with a size of 1$\times$1 kpc 
contains at least three stars, and represents a median value of \cfe, with the color scale 
shown in the color bar. The general trend noted in the figure is that the overall metallicity 
distribution decreases with increasing distance from the Galactic plane. It is 
interesting to note in the left panel of Figure \ref{fig:feh_zcut} the unexpected 
positive metallicity gradient with the distance from the Galactic center at 
a given $Y$. This may arise from the presence of the Monoceros Stream at $X >$ 11 kpc. 
In the middle panel, we can also see the higher metallicity region at $X >$ 14 kpc, likely 
associated with the Monoceros Stream as well.

Figure \ref{fig:cfe_zcut} shows the carbonicity maps at different heights above the 
Galactic plane. The layout is the same as in Figure \ref{fig:feh_zcut}. In these plots, 
as expected, we notice the higher [C/Fe] at higher \z. Particularly, the two groups 
with relatively larger enhancement of carbon in the right panel imply the presence 
of substructures in the OHR in our sample.

For a more quantitative analysis, we present in Figure \ref{fig:mean_cfe} profiles of 
the mean values of [Fe/H] (left panel), [C/Fe] (middle panel), and \ac\ (right panel), as a
function of \z. Each dot represents an average value for stars within a bin of 2 kpc width,
overlapped with the next neighboring bin by 1 kpc. The error bars are
obtained by bootstrapping the sample 100 times in each bin. The vertical
dotted lines indicate the the radius of the IHR determined from the carbonicity
map shown in Figure \ref{fig:cmap}. Similarly, the red dots are derived from the 
sample corrected for target-selection bias. 

The left panel of Figure~\ref{fig:mean_cfe} reveals three distinct
features in the metallicity profile -- a rapid decrease in
$\langle$[Fe/H]$\rangle$ up to \z\ = 3 kpc, a mild decline between \z\ =
3 and 8 kpc, and a continuous decrease beyond \z\ = 8 kpc. These
transition regions in the [Fe/H] profile correspond well with each
Galactic region assigned in the carbonicity map. We note that 
the bias-corrected sample also exhibits a very similar trend. 

We derived a metallicity gradient of --0.172 $\pm$ 0.043 dex$^{-1}$
kpc$^{-1}$ over \z\ $<$ 3.0 kpc (the location of the TDR),
--0.037 $\pm$ 0.004 dex$^{-1}$ kpc$^{-1}$ over 3 $\le$ \z\ $<$ 8 kpc,
which roughly corresponds to the IHR, and --0.136 $\pm$ 0.003 dex$^{-1}$
kpc$^{-1}$ over \z\ $\ge$ 8 kpc, which we associate with the OHR. The
mild metallicity gradient in the IHR suggests that stars in the IHR
experienced similar chemical-enrichment histories, whereas the
relatively steeper metallicity gradient in the OHR suggests a more
complex star-formation history, possibly involving the accretion of
multiple mini-halos, which may have contributed lower metallicity stars 
to the OHR.

\begin{figure*}
\centering
\includegraphics[width=0.32\textwidth]{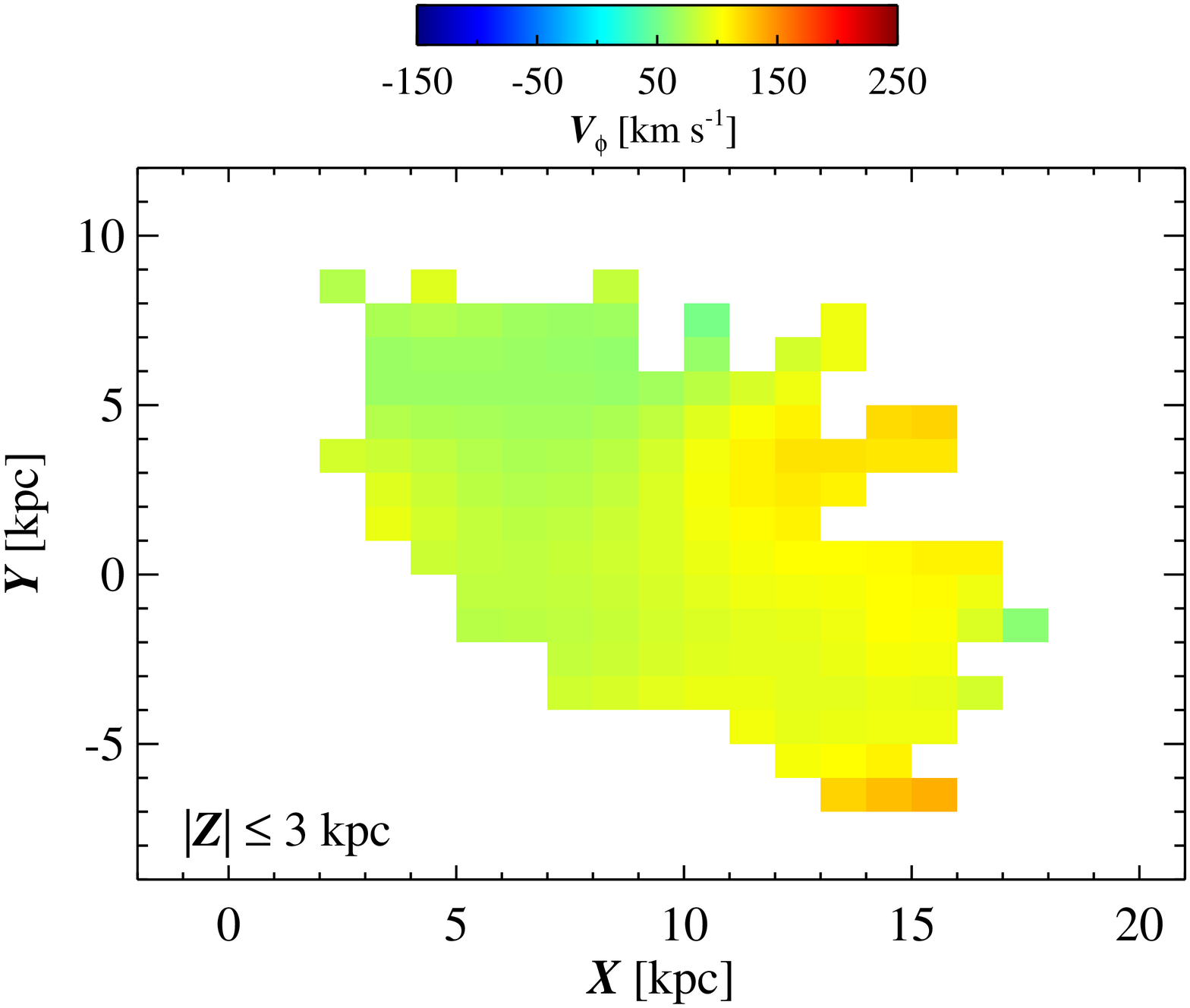}\hspace*{0.2cm}
\includegraphics[width=0.32\textwidth]{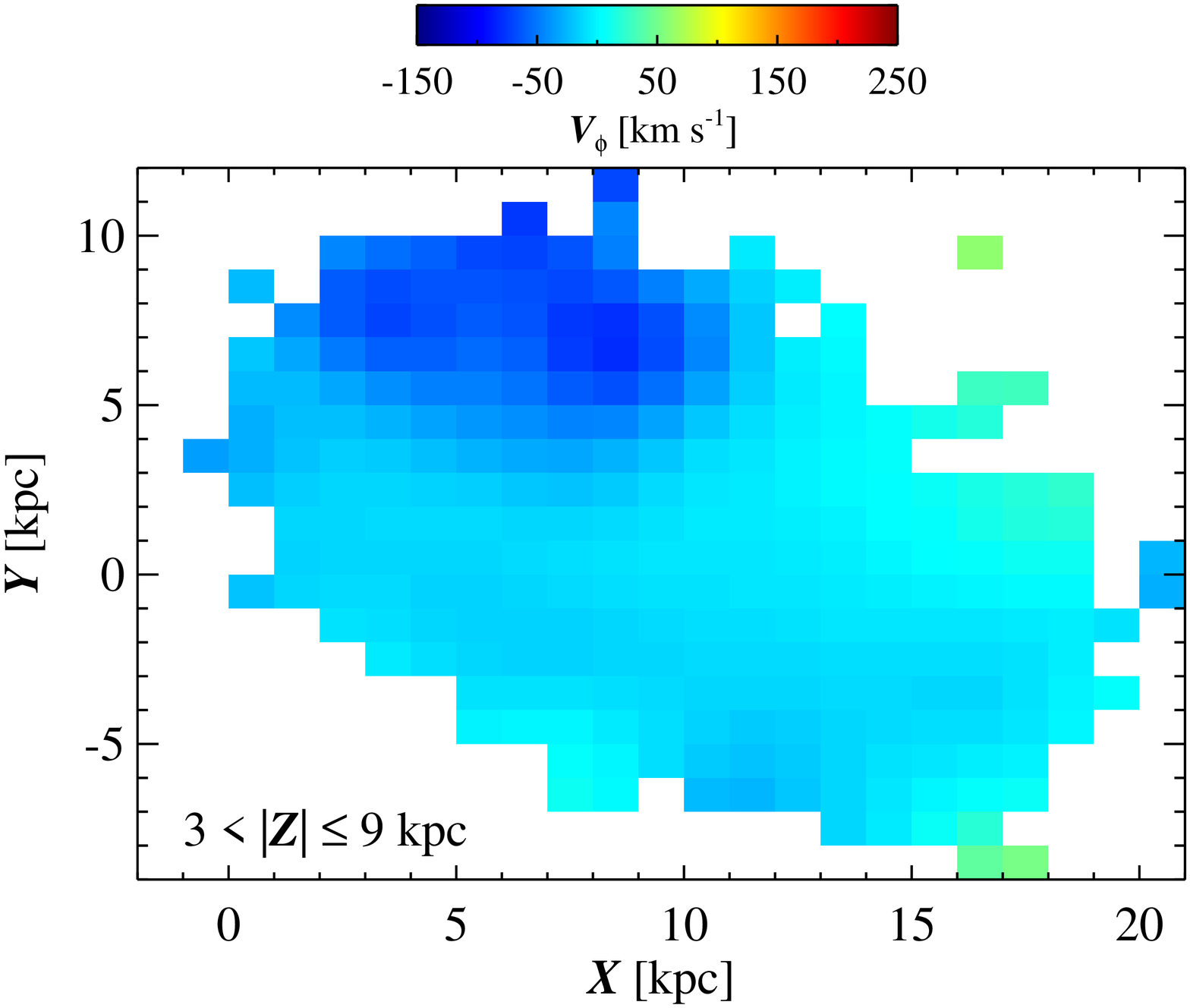}\hspace*{0.1cm}
\includegraphics[width=0.32\textwidth]{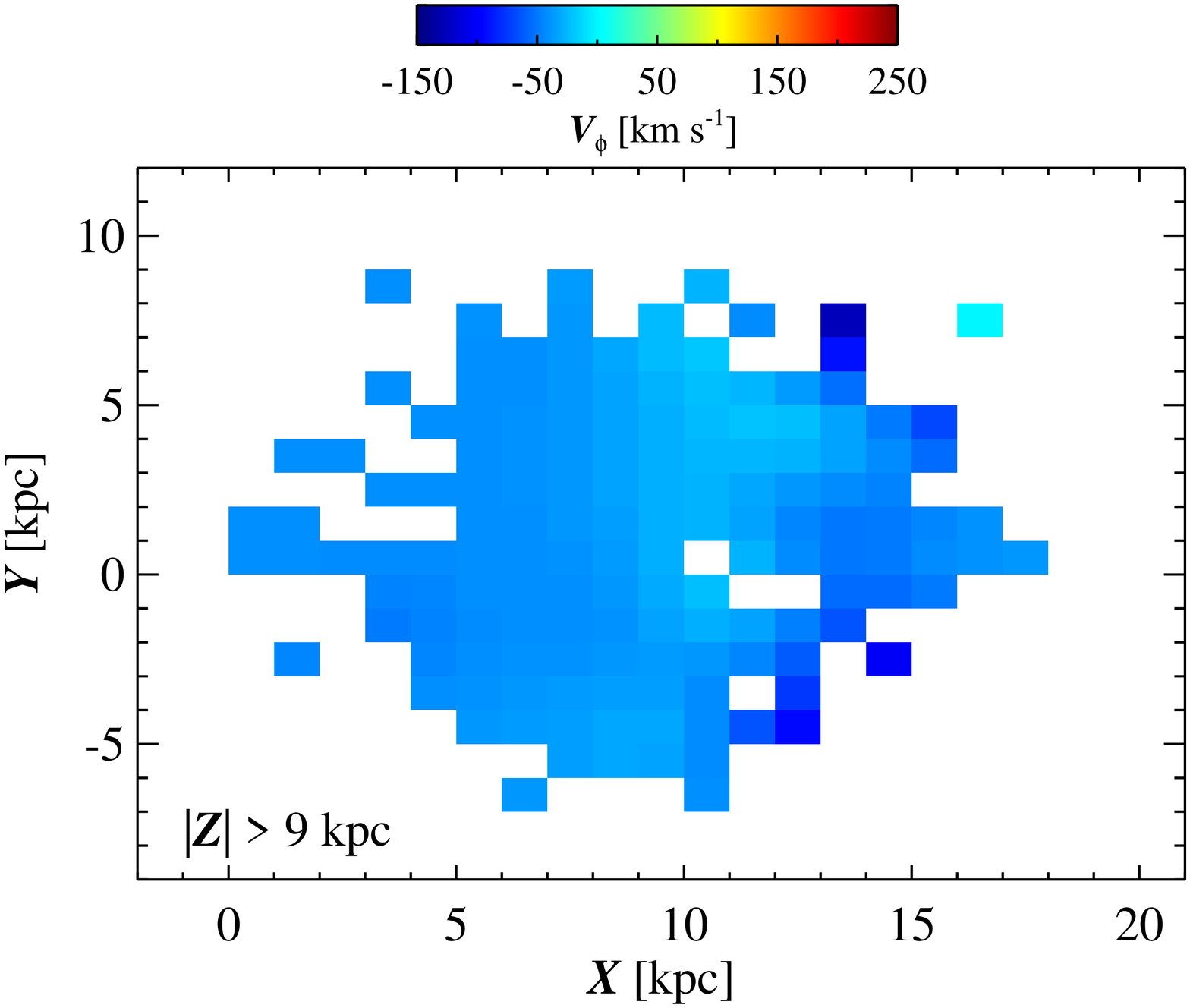}
\caption{Same as in Figure \ref{fig:feh_zcut}, but for \vphi.}
\label{fig:vph_zcut}
\end{figure*}

Our derived metallicity gradient for the OHR is rather steeper
than the results from the simulations of the MW-like galaxies. For
example, \citet{tissera2014} obtain, based on six stellar halos from
Aquarius simulation project (\citealt{scannapieco2009}), a metallicity
gradient between --0.002 and --0.008 dex$^{-1}$ kpc$^{-1}$. However,
the halo region they consider is $r >$ 20 kpc, which is much more
distant than our OHR.

The metallicity gradient of our sample in the OHR is also larger than 
that ($\sim$ --0.001 dex$^{-1}$ kpc$^{-1}$) of \citet{das2016a}, derived 
from an extended distribution function fit to SEGUE K giants with [Fe/H]
$<$ --1.4. Their K giants are also mostly located at distance greater
than 10 kpc from the Galactic center. Consequently, as their stellar
species and distance coverage are different from our sample, it is
difficult to directly compare their result and ours. Nonetheless, if we
only consider the stars with [Fe/H] $<$ --1.4 and \z\ $>$ 8 kpc, we
obtain slightly smaller gradient of $\sim$ --0.1 dex$^{-1}$ kpc$^{-1}$.


The above behavior stands in contrast to that observed in the
the middle panel of the figure, the average carbonicity profile. The
$\langle$[C/Fe]$\rangle$ value steadily increases up to \z\ = 11 kpc,
with a more rapid increase beyond \z\ = 11 kpc, which may be a signature
of the existence of chemical substructure in the OHR. Even though the increasing 
trend at higher \z\ is somewhat weak, we notice the similar trend 
from the bias-corrected sample (red dots).

Differences in the nature of the stellar populations between the IHR and
OHR become more clear in the distribution of $\langle$\ac$\rangle$ in
the right panel of Figure~\ref{fig:mean_cfe}. The mean value of \ac\
drops rapidly over \z\ $<$ 3.0 kpc (the TDR), then slightly increases
between \z\ = 3 kpc and 8 kpc (the IHR), and then abruptly drops up to 
\z\ $>$ 11 kpc (the OHR), and remains at a lower value of \ac\ for 
both the biased and bias-corrected samples. These trends suggest that
the progenitors that contributed to the formation of IHR differ from
those that contributed to the OHR, a clear indication of the duality (at
least) of the Galactic halo.

\begin{figure}
\centering
\epsscale{1.15}
\plotone{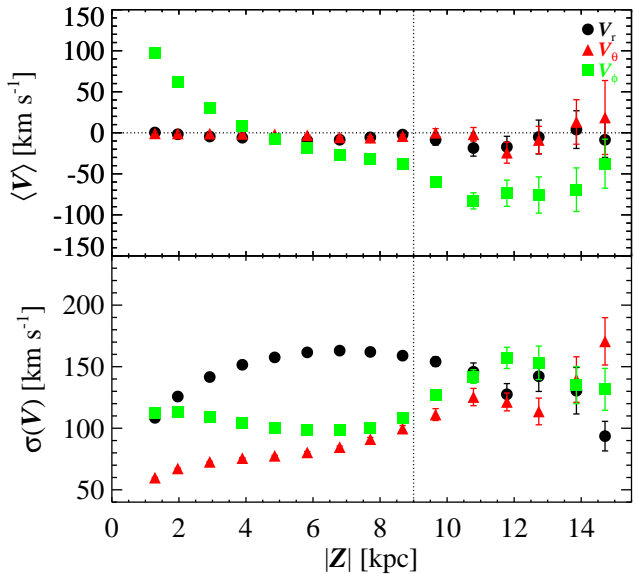}
\caption{Trends of mean velocities (top panel) of $V_{\rm r}$ (black
symbols), $V_{\theta}$ (red symbols), and \vphi\ (green symbols) and their
dispersions (bottom panel) in spherical coordinates, as a function of
\z. Each dot covers a bin size of 2 kpc, and each bin is overlapped with
the next neighboring bin by 1 kpc. The error bar is the standard
deviation of 100 realizations obtained by bootstrapping. The vertical
dotted line at 9 kpc indicates the highest point in the region that
separates the IHR from the OHR in Figure \ref{fig:cmap}, whereas the
horizontal dotted line provides a reference at 0 \kms. In the top
panel, the mean rotational velocity (green symbols) steeply decreases in
the disk region (\z\ $<$ 4 kpc), declines very slowly up to 9 kpc, and
then decreases again beyond \z\ = 9 kpc, becoming retrograde. The other
two velocity components exhibit almost zero velocity throughout the
region considered. Interestingly, we note that the dispersion in $V_r$
crosses over the $V_{\theta}$ and $V_{\rm \phi}$ dispersions at about 11 kpc,
consistent with a transition from the IHP to the OHP in our sample.}
\label{fig:rvrot}
\end{figure}

\begin{figure*}
\centering
\includegraphics[width=0.32\textwidth]{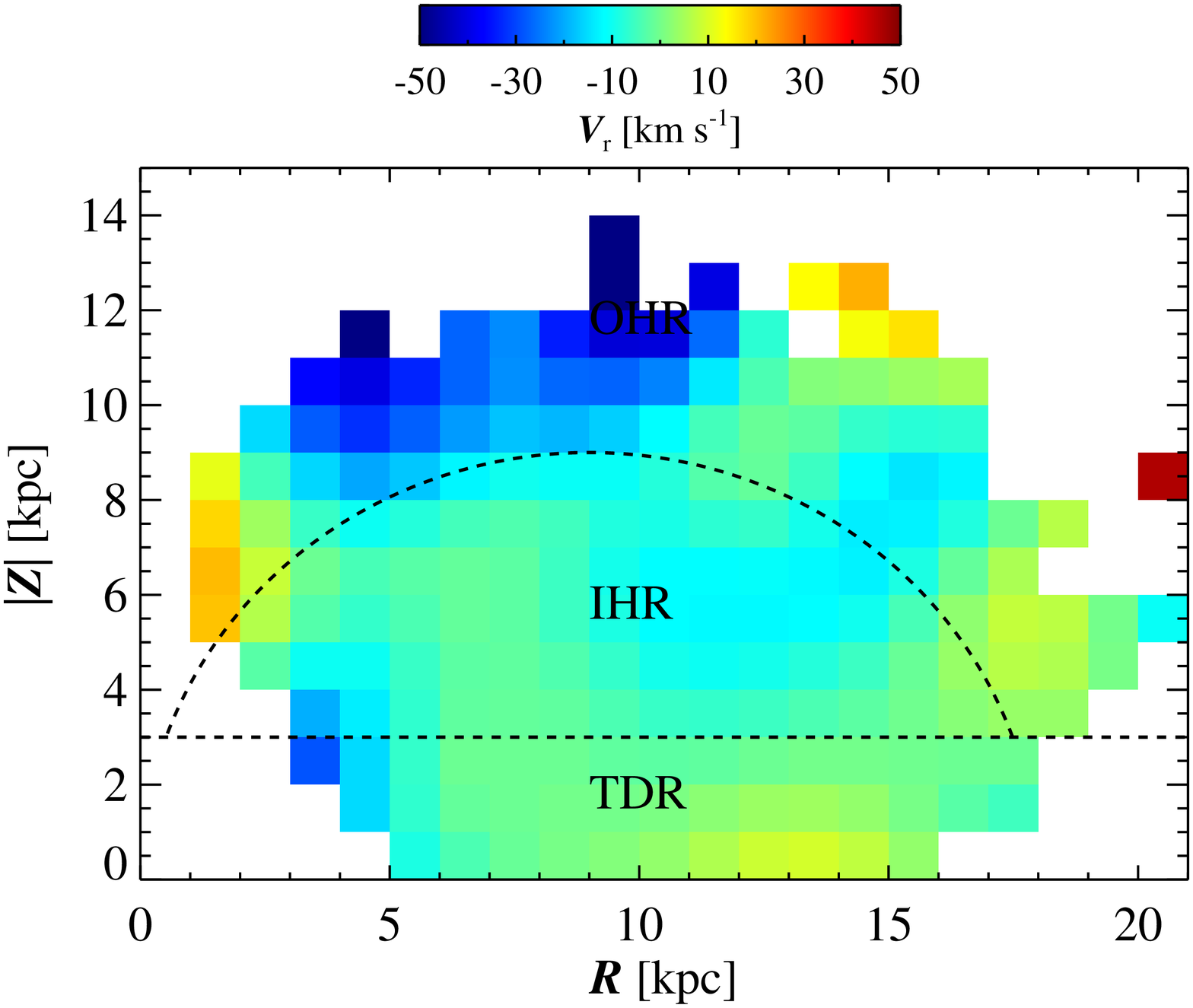}\hspace*{0.2cm}
\includegraphics[width=0.32\textwidth]{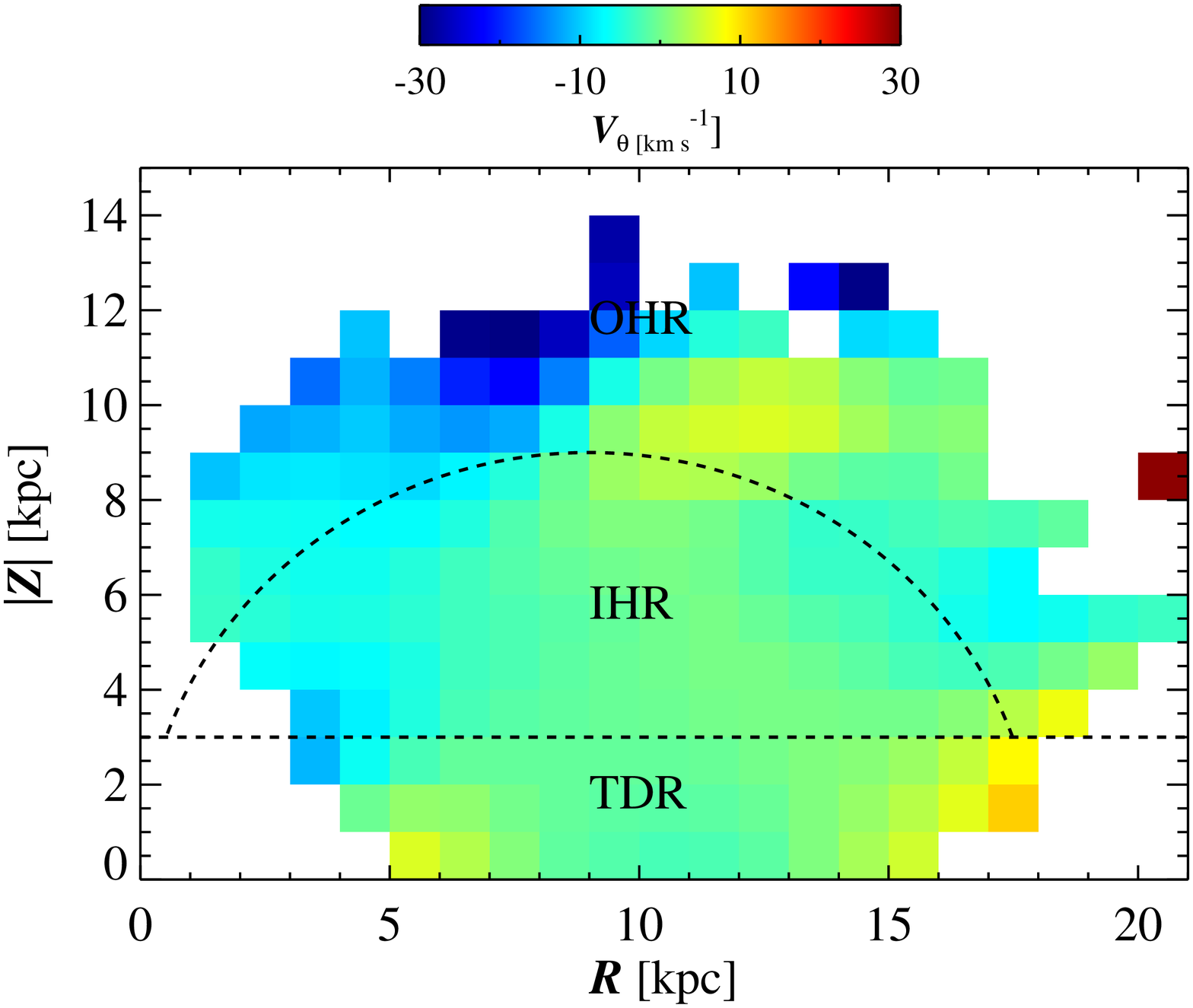}\hspace*{0.1cm}
\includegraphics[width=0.32\textwidth]{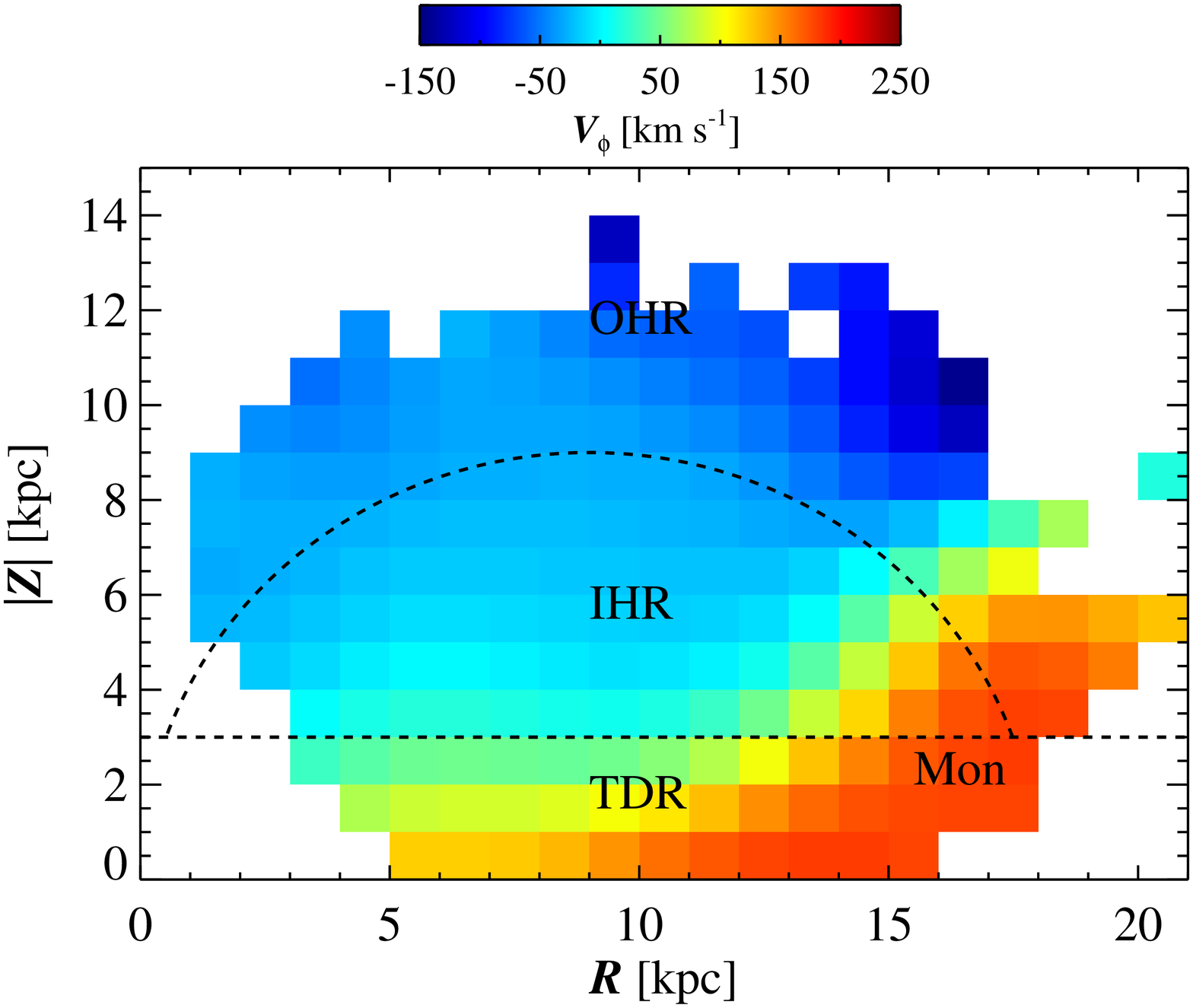}
\caption{Maps of the medians of $V_{r}$ (left), $V_{\theta}$
(middle), and \vphi\ (right) components in the \z\ and $R$ plane. Due to 
the small number of stars in some bins, we considered the median instead
of the mean. 
Each bin has a size of 1$\times$1 kpc and at least three stars. The dashed lines are
the boundaries for the Galactic components defined in Figure \ref{fig:cmap}.
We note that the IHR and OHR are well-distinguished in the three velocity 
components. ``Mon'' in the right panel indicates the Monoceros Stream.}
\label{fig:vel_mean}
\end{figure*}

\begin{figure*}
\centering
\includegraphics[width=0.32\textwidth]{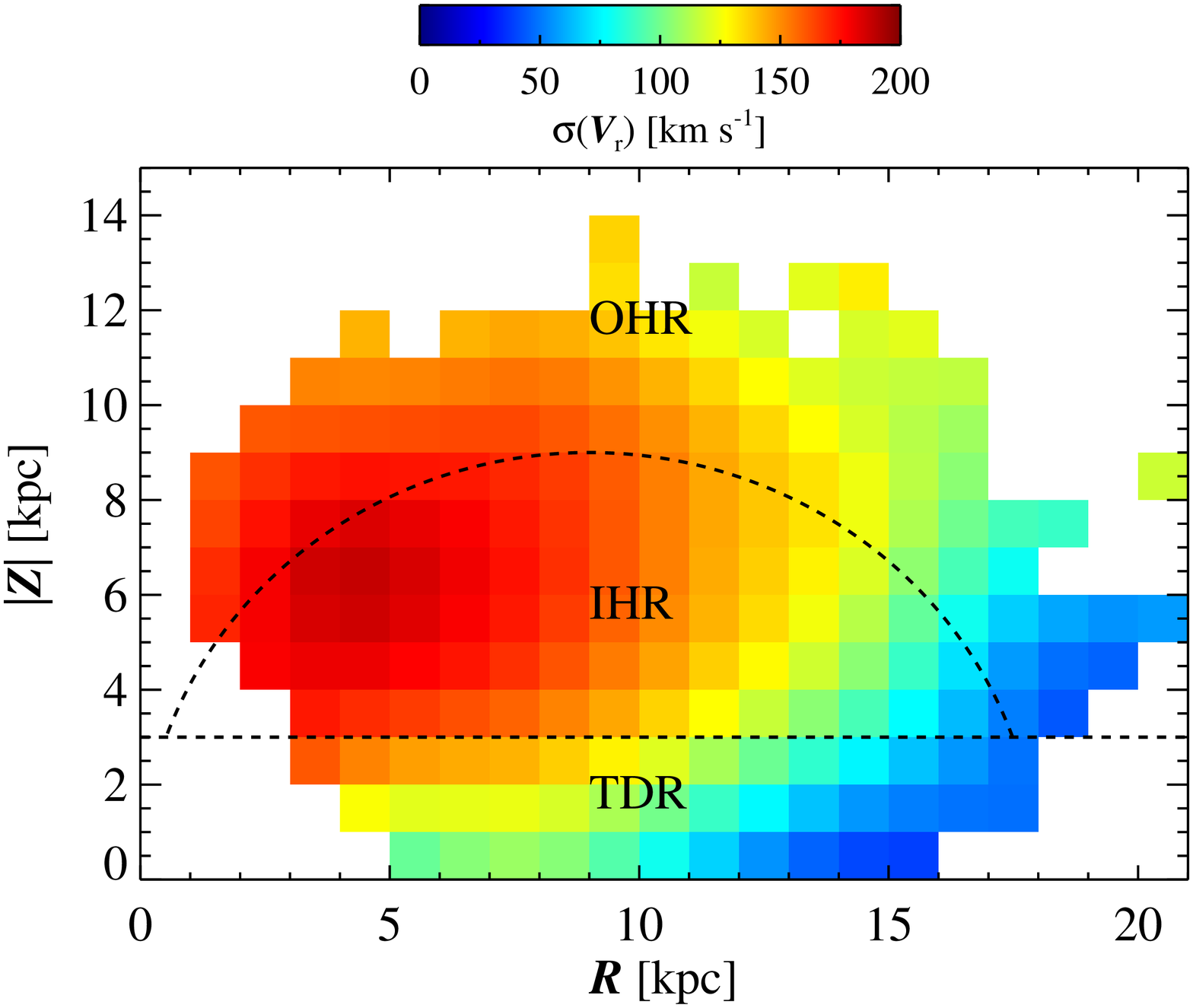}\hspace*{0.2cm}
\includegraphics[width=0.32\textwidth]{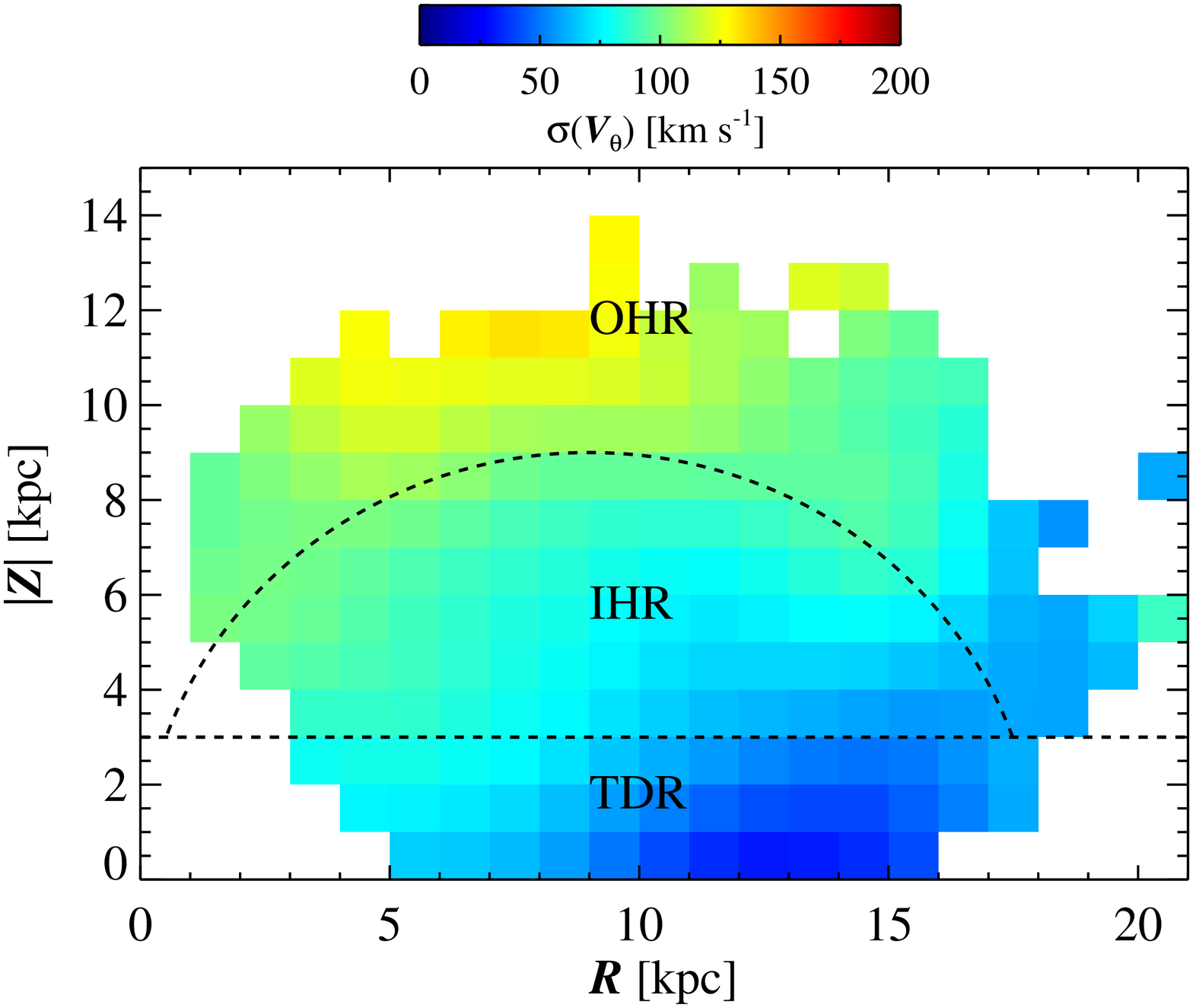}\hspace*{0.1cm}
\includegraphics[width=0.32\textwidth]{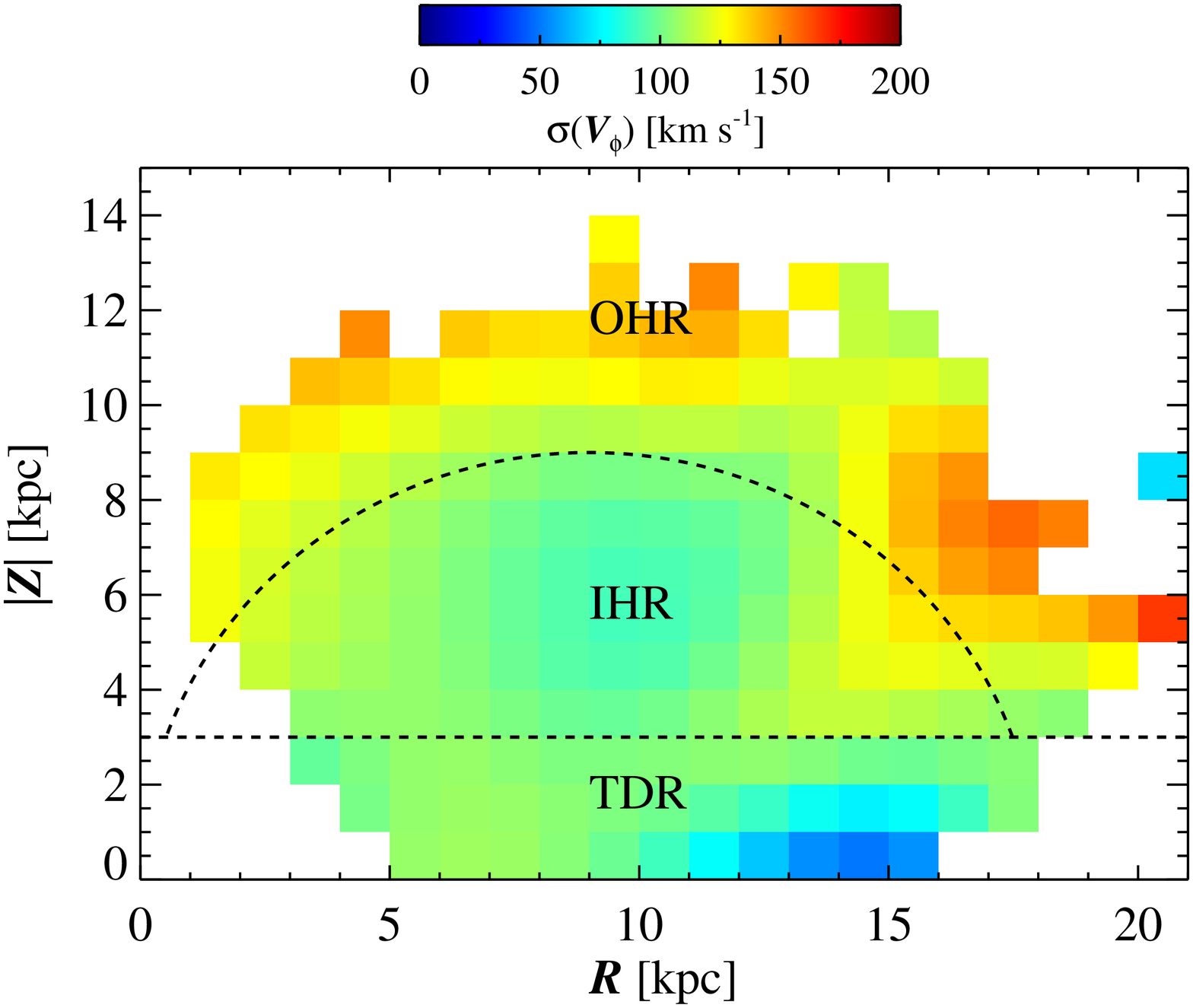}
\caption{Same as in Figure \ref{fig:vel_mean}, but for the velocity
dispersions of the three components.}
\label{fig:vel_disp}
\end{figure*}

\subsection{Kinematic Signatures}\label{sec:kinematic}

As the kinematics associated with different stellar populations in the halo
can provide valuable clues to its assembly history, we now examine the
nature of the velocity structure associated with the components
identified in the carbonicity map, beginning with the spatial distribution 
of the rotational velocity and the mean velocity and
velocity dispersion profiles, followed by consideration of the
differences revealed in their velocity ellipsoids and anisotropies.

\subsubsection{Velocity Structure}

We first investigated the distribution of rotational velocities
for 
our sample in the $X$--$Y$ plane in different bins of \z, as shown in 
Figure \ref{fig:vph_zcut}. The layout of the figure is the same as 
in Figure \ref{fig:feh_zcut}. Generally, we observe a more 
retrograde motion at larger distances from the Galactic plane. One 
interesting aspect is a small, patchy area with relatively stronger 
counter-rotation in the IHR (middle panel), indicative 
of substructures in velocity space. 

Figure \ref{fig:rvrot} shows the profiles of mean velocities (top panel)
and dispersions (bottom panel) in spherical coordinates, as a function
of \z. The black dots correspond to \vrad, while the red triangles
represent \vthe, and the green squares indicate \vphi. Each symbol
represents an average value in a bin size of 2 kpc in \z, with each bin
overlapped with the next neighboring bin by 1 kpc. The error bar is the
bootstrap estimate of the standard deviation based on 100 realizations.
The vertical dotted line marks the highest point in the region that
separates the IHR from the OHR.

Inspection of the top panel of Figure~\ref{fig:rvrot} reveals that the
rotation velocity indicated by the green squares rapidly decreases from
$\sim 100$ \kms\ at \z\ = 1 kpc to almost zero at \z\ = 5 kpc, then
continues to decline, exhibiting a small retrograde motion. Above \z\ =
9 kpc, stars show significant retrograde motions. In contrast, the
other two velocity components (\vrad\ and \vthe) do not vary around
$\langle$$V$$\rangle$ = 0 \kms\ over the vertical distance considered.
The \vphi\ behavior in the region of the TDR is expected from an overlap
between stars belonging to the metal-weak thick disk and of the IHP over
1 $<$ \z\ $< 5$ kpc. Beyond 5 kpc from the plane, the observed behavior of
\vphi\ can be accounted for by the gradual transition from the dominant
contribution by stars of the IHP to the OHP; even though there
are fewer stars in our MSTO sample far from the plane, they display
significantly different \vphi.

Similar distinct behaviors can be also found from inspection of
the bottom panel of Figure~\ref{fig:rvrot}, which presents the
velocity dispersion profiles. The dispersion in \vphi\ does not change
dramatically between 90 and 110 \kms\ over \z\ $<$ 9 kpc, while it
increases to between 130 and 160 \kms\ above \z\ = 9 kpc. The
dispersion of \vrad\ increases with \z\ distance in the TDR,
does not change much in the IHR, then declines to $\sim$ 90 \kms\ in
the OHR. The dispersion of \vthe\ generally exhibits a continuously
increasing trend. In the figure, we note that the dispersion of \vrad\
is much larger than either \vthe\ and \vphi\ in the IHR (i.e., the
velocity ellipsoid of the IHR is dominated by large radial motions, as
shown by many previous studies), but decreases in the OHR, again
indicating contrasting behavior relative to the IHR.

Taken as a whole, Figure~\ref{fig:rvrot} indicates that the transition
regions identified in the trends of the mean velocity and velocity
dispersion correspond (at least qualitatively) to the
differences found in the metallicity and carbonicity profiles shown in
Figure~\ref{fig:mean_cfe}. This strongly suggests that differences in the
stellar populations of the halo are revealed by our MSTO stellar sample.

\begin{table*}
\renewcommand{\tabcolsep}{2pt}
\begin{center}
\caption{Kinematic Properties in Spherical and Cylindrical Coordinates for each Galactic Region}
\label{tab:kin} 
\begin{tabular}{lcrrrrrrrrr}
\hline
\hline
       &               &             &                  & Spherical&                      &                            &                         &              \\
\hline
Region & $N_{\rm tot}$ & $V_{\rm r}$~~~~ & $V_{\theta}$~~~~ & $V_{\phi}$~~~~ & $\sigma_{V_{\rm r}}$~~~ & $\sigma_{V_{\theta}}$~~~ & $\sigma_{V_{\phi}}$~~~ & $\beta$~~~~~~~~ \\
       &               & (\kms)      & (\kms)           & (\kms)         & (\kms)               & (\kms)                     & (\kms)                  &        \\
\hline
  TDR  &  55769 &   --1.1 $\pm$    0.5 &   --0.9 $\pm$    0.3 &    75.7 $\pm$    0.5 &  119.8 $\pm$    0.5 &   64.3 $\pm$    0.4 &  114.6 $\pm$    0.3 &    0.399 $\pm$   0.006 \\
  IHR  &  44546 &   --7.0 $\pm$    0.8 &   --2.1 $\pm$    0.3 &   --2.6 $\pm$    0.5 &  155.6 $\pm$    0.5 &   78.1 $\pm$    0.3 &  102.7 $\pm$    0.4 &    0.656 $\pm$   0.003 \\
  OHR  &   1169 &   --5.8 $\pm$    4.5 &   --6.7 $\pm$    3.1 &  --49.3 $\pm$    3.7 &  152.6 $\pm$    2.6 &  108.8 $\pm$    2.9 &  130.0 $\pm$    2.8 &    0.383 $\pm$   0.030 \\
\hline
        &               &             &                &Cylindrical&                      &                         &                      &  \\
\hline
Reigion & $N_{\rm tot}$ & $V_{\rm R}$~~~~ & $V_{\Phi}$~~~~ & $V_{\rm Z}$~~~~ & $\sigma_{V_{\rm R}}$~~~ & $\sigma_{V_{\Phi}}$~~~ & $\sigma_{V_{\rm Z}}$~~~ &  \\
        &               & (\kms)      & (\kms)         & (\kms)      & (\kms)               & (\kms)                  & (\kms)               &  \\
\hline
TDR   & 55769   &  --1.3 $\pm$ 0.5  &   75.7 $\pm$ 0.5  &    0.7 $\pm$ 0.3  & 117.5 $\pm$ 0.4  & 114.6 $\pm$ 0.3  &  68.4 $\pm$ 0.3  \\
IHR   & 44546   &  --8.6 $\pm$ 0.7  &  --2.6 $\pm$ 0.5  &  --0.7 $\pm$ 0.5  & 144.9 $\pm$ 0.4  & 102.7 $\pm$ 0.4  &  96.5 $\pm$ 0.4  \\
OHR   &  1169   & --12.7 $\pm$ 4.5  & --49.3 $\pm$ 3.7  &  --3.1 $\pm$ 4.0  & 140.7 $\pm$ 2.6  & 130.0 $\pm$ 2.8  & 123.4 $\pm$ 2.4  \\
\hline
\end{tabular}
\tablecomments{$N_{\rm tot}$ is the total number of stars in each Galactic region. The listed uncertainties are derived from 100 bootstrapped resamples.
By coordinate configuration, $V_{\phi}$ is equal to $V_{\Phi}$. $\beta$ is the anisotropy parameter calculated by Equation (1).}
\end{center}
\end{table*}

We now consider the distribution of the median and dispersion 
of the velocity components in the spatial domain. Figure~\ref{fig:vel_mean} 
presents maps of the medians in $V_{r}$ (left panel), $V_{\theta}$ (middle panel), 
and \vphi\ (right panel) components in the \z\ and $R$ plane. Each bin has a
dimension of 1$\times$1 kpc and has at least three stars. The dashed
lines are the boundaries for the Galactic components defined in Figure
\ref{fig:cmap}. Note that, owing to the absence of the proper motions of 
some stars, the shape of the map is slightly different from 
Figure \ref{fig:cmap}. In the map, due to the small number of stars in 
some bins, we considered the median instead of the mean for a more
robust estimate.

The figure indicates that the medians of
\vrad\ and \vthe\ in the IHR are mostly between --10 \kms and $+$10 \kms, and 
there exists no radial or vertical gradient in \vrad\ and \vthe, as 
already seen in Figure \ref{fig:rvrot}. We can also notice almost no 
rotation (\vphi) of the IHR, but there is weak vertical and radial 
gradient in \vphi. We can see the Monoceros Stream (\citealt{newberg2002,ivezic2008}), 
with a high rotation velocity, as well. 

By contrast, in the OHR we see several clumpy structures for 
\vrad\ and \vthe, as well as relatively milder retrograde and stronger counter 
rotation in some regions. In these maps, we notice 
that the boundary defined in Figure~\ref{fig:cmap} between the IHR and 
OHR corresponds well with structures in the three velocity 
components, which suggests that the chemical division of the stellar 
populations can identify distinct kinematic properties as well.

Figure~\ref{fig:vel_disp} shows maps of the velocity 
dispersions. The layout of the figure is the same as in Figure \ref{fig:vel_mean}. 
Inspection of this figure reveals that the dispersion of \vrad\ becomes higher towards 
the bulge and Galactic North Pole. At a given \z, it exhibits a clear radial 
gradient, while the \vthe\ dispersion exhibits a moderate vertical gradient. 
The obvious offset of the high radial dispersion population in the direction 
toward the Galactic center supports the identification by numerous recent 
studies of the ``\gaia\ Sausage'' (\citealt{belokurov2018}) or \gaia-Enceladus 
(\citealt{helmi2018}) structure. There exists no strong dispersion gradient in \vphi\ 
within the IHR, although the dispersion increases in the OHR. 
Generally, the IHR-OHR boundary is well-delineated in the dispersion maps of 
\vthe\ and \vphi. The clumpy structures in the OHR seen in 
Figure \ref{fig:vel_mean} and the relatively higher velocity dispersions shown 
in Figure \ref{fig:vel_disp} strongly suggest that the stars in the IHR have 
experienced different assembly histories from those in the OHR.

The derived mean velocities and dispersions of each velocity component
(in both spherical and cylindrical systems) for each Galactic region
are listed in Table \ref{tab:kin}, along with their bootstrapped errors
based on 100 resamples. We first consider the detailed behavior of the
rotational component \vPhi\ (in the cylindrical system, which is the
same as \vphi\ in the spherical system). We obtained on average
$\langle$\vPhi$\rangle$ = --2.6 $\pm$ 0.5 \kms\ for the IHR 
and --49.3 $\pm$ 3.7 \kms\ for the OHR, clear evidence for its retrograde motion.

Since it is possible for disk stars to reach distances
from the plane (for example, due to disk heating) that are sufficiently
large to be confused with halo stars, we have checked how potential
contamination from disk stars could affect the derived rotation
velocities by restricting our MSTO sample to stars with [Fe/H] $<$
--1.0, as most of disk stars have [Fe/H] $>$ --1.0. After making this
restriction, we obtained $\langle$\vPhi$\rangle$ = --17.5 $\pm$ 0.5
\kms\ for the stars in the IHR, and $\langle$\vPhi$\rangle$ = --60.5
$\pm$ 3.9 \kms\ for the stars in the OHR, indicating a larger retrograde
motion for each halo. However, as we want to study the $in~situ$ halo,
we consider all of our MSTO stars in our analysis, without removing the
possible disk stars.  

\begin{figure*}[t]
\centering
\plottwo{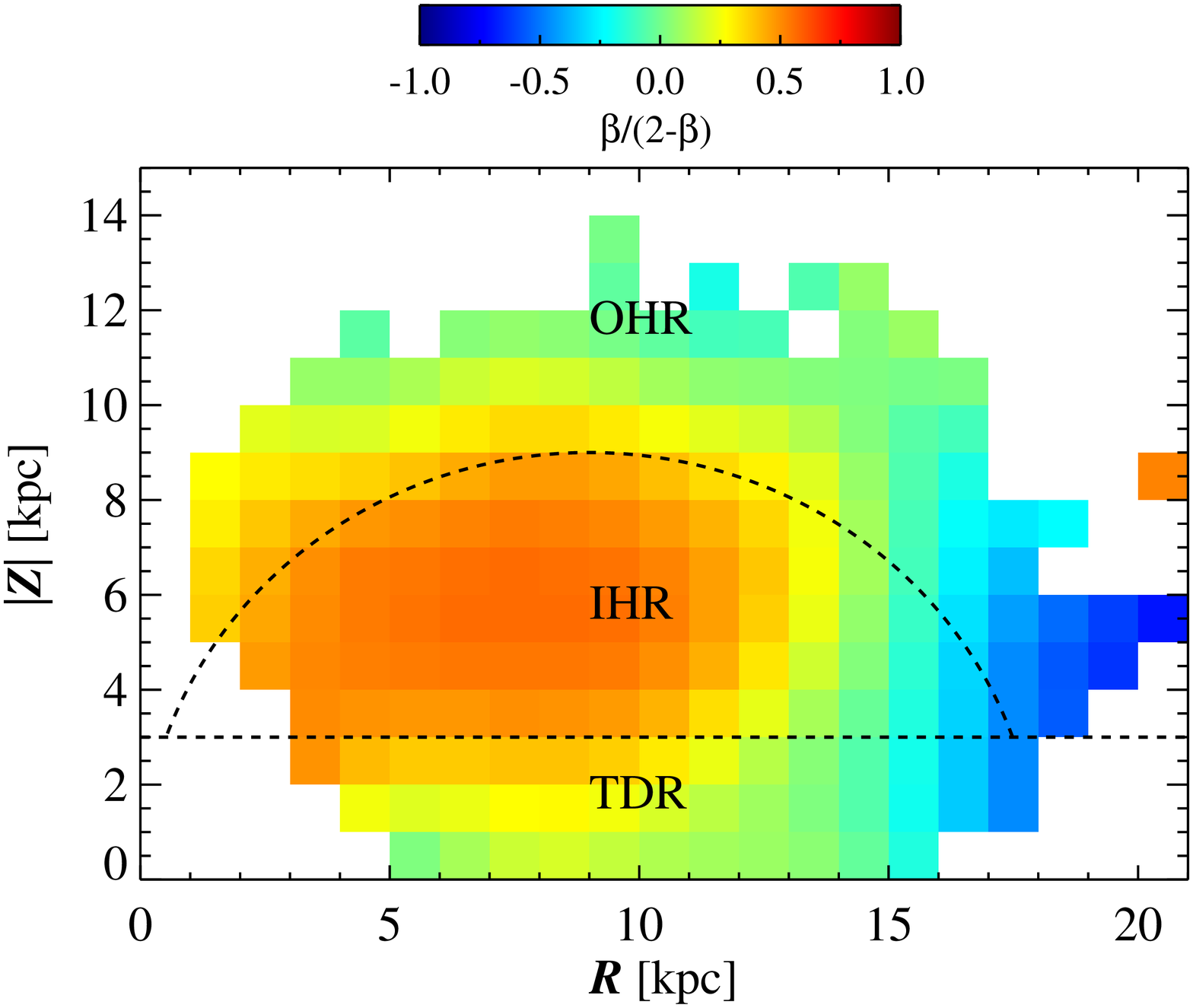}{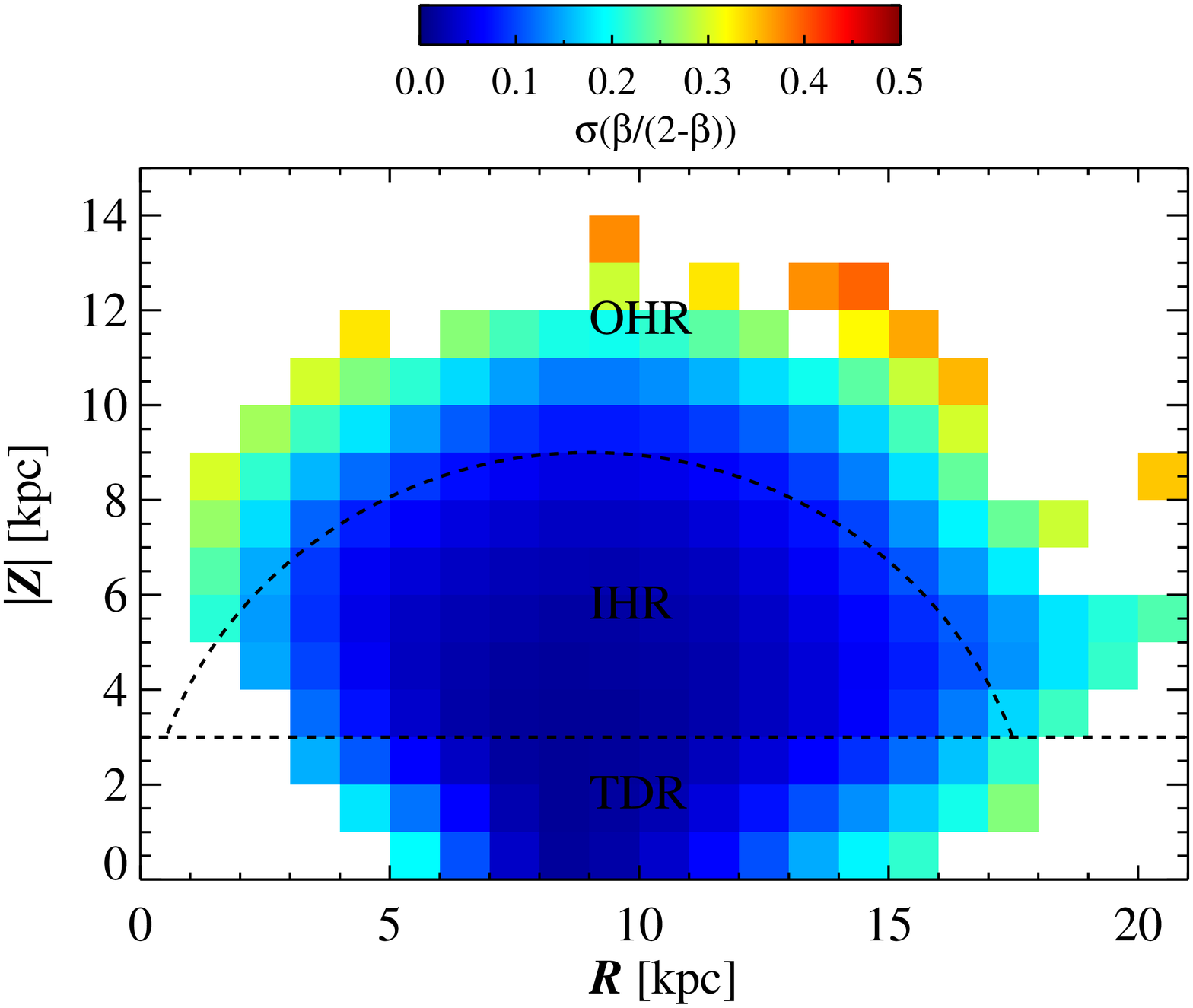}
\caption{Same as in Figure \ref{fig:vel_mean}, but for the rescaled anisotropy parameter,
$\beta^{*} = \beta/(2-\beta)$ (left panel) and its scatter (right
panel), based on 100 samples generated 
from a Monte Carlo simulation.}
\label{fig:beta}
\end{figure*}

\subsubsection{Comparisons with Other Studies}

Our average value of \vPhi\ for the IHR stars is not far from 
that (7 $\pm$ 4 \kms) derived by \citet{carollo2010}, while our 
derived $\langle$\vPhi$\rangle$ of --49 \kms\ suggests that the 
counter-rotating signal is slightly weaker than that (--80 $\pm$ 13 \kms) 
reported by \citet{carollo2010}, even taking the errors into account. 
We found a dispersion for \vPhi\ of 103 and 130 \kms\ for the IHR and OHR, 
respectively, compared to 95 $\pm$ 2 and 165 $\pm$ 9 \kms\ derived by 
Carollo et al.

The above discrepancies may arise in part from the fact that
\citet{carollo2010} employed a local sample of stars, with distances less than 4 kpc
from the Sun, and separated the various halo populations based on the
kinematic information itself, while our separation is purely based on
the spatial distribution of \cfe. Thus, the two methods are describing
two different aspects of the data -- one a population-based method, and
the other an in-situ method (wherein overlapping inner- and outer-halo
populations can contribute, in particular for the IHR).

Comparing with the other two cylindrical velocity components for the IHP
and OHP given in Table 5 of \citet{carollo2010}, our value of \vRad\ =
--8.6 \kms\ for the IHR is slightly lower than their value of 3
\kms\ for the IHP, while our value of --12.7 \kms\ for the OHR does not
differ much from their value of --9 \kms. Our derived
dispersions for \vRad\ are 145 and 141 \kms\ for the
IHR and OHR, respectively, while theirs are 150 and 159 \kms\ for the
IHP and OHP, respectively. For the \vZ\ component, we derived average
values of --1 (97) and --3 (123) \kms\ (where the number listed in
parentheses is the dispersion), compared to their values of 3 (85) and 2
(116) \kms\ for the IHP and OHP, respectively. Once again, the
small differences in these quantities are likely due to their different
definitions and techniques for separation of the sample stars.

We now consider the results obtained by a number of other recent
studies. \citet{kafle2017} also use SDSS MSTO stars to investigate the
kinematic properties of stars in the region $r \leq 15$ kpc from the
Galactic center and \z\ $>$ 4 kpc from the Galactic plane.
Interestingly, they find for the mean and dispersion of the rotation
velocity component values of --16 (72) \kms\ for a metal-rich group
([Fe/H] $>$ --1.4) and 26 (82) \kms\ for a metal-poor group ([Fe/H] $<$
--1.4). Apparently, for the OHR our mean value has opposite sign to theirs, 
while our derived dispersion is higher than theirs. If we consider our sample
separated by their criteria, we obtained $\langle$\vphi$\rangle$ = 2 (87)
\kms\ for the metal-rich component and --31 (102) \kms\ for the
metal-poor component.

\citet{deason2017} make use of the SDSS data and proper motions from the first data release
of \gaia\ (\citealt{gaia2016}) to derive $\langle$\vphi$\rangle$ = 12 \kms\ from RR
Lyrae, 6.0 \kms\ from blue horizontal branch (BHB) stars, and 23 \kms\ from K giants.
These stars reach up to 50 kpc from the Galactic center and do not include stars
with \z\ $<$ 4 kpc, which correspond to the disk populations. Although
they find that metal-poor population ([Fe/H] $<$ --1.5) exhibits a slightly
lower rotation velocity than the metal-rich population for all three
stellar types, their metal-poor population does not show retrograde
motion, contrary to our metal-poor counterpart ([Fe/H] $<$ --1.4), which
has $\langle$\vphi$\rangle$ = --31 \kms.

On the other hand, \citet{belokurov2018} analyze MS stars in the SDSS
to compute $\langle$\vphi$\rangle$ = 20 -- 30 \kms\ for stars with --1.0
$<$ [Fe/H] $<$ --1.7 and 1 $<$ \z\ $<$ 9 kpc. They also find that the metal-poor
stars with [Fe/H] $<$ --1.7 exhibit various values of $\langle$\vphi$\rangle$
from 15 to 50 \kms, depending on the range of \z\ investigated. In
contrast, we found between --20 and 20 \kms\ for the metal-rich
component, and between --46 and --4 \kms\ for the metal-poor component
in the three regions (1 $<$ \z\ $<$ 3 kpc, 3 $<$ \z\ $<$ 5 kpc, and 5
$<$ \z\ $<$ 9 kpc) of \z\ they considered.

Summing up the aforementioned studies, although we found that there does
exist some correlation between the stellar rotation velocity and metallicity,
the reviewed literature does not report a significant counter-rotating
signature.

\subsubsection{Orbital Properties}

We now examine the distribution of stellar orbits in the halo populations
quantified by the velocity anisotropy parameter (\citealt{binney2008}), given by

\begin{equation}
\beta = 1- \frac{\sigma^2_{\theta}+\sigma^2_{\phi}}{2\sigma^2_{r}}
\end{equation}

\noindent where $\sigma^2_{r}$, $\sigma^2_{\theta}$, and $\sigma^2_{\phi}$ are the velocity
dispersions in the spherical coordinate system. A value of $\beta$ = 0
means that the two tangential components are equal to the radial
component, suggesting the distribution of stellar orbits is isotropic.
When $\beta > 0$, the radial component is larger than the tangential
components, corresponding to more radially elongated stellar orbits. On
the other hand, values of $\beta < 0$ indicates a tangentially biased
distribution of the stellar orbits.

Because the anisotropy parameter has a negatively skewed distribution, it is often rescaled by
$\beta^{*} = \beta/(2-\beta)$, so that the value of $\beta^{*}$ has the
range from --1 to 1 when investigating the overall description of
orbital distributions. In the rescaled $\beta^{*}$ case, the
interpretation of a stellar system is the same as for $\beta$, such that
if $\beta^{*}$ = 0, a stellar system is isotropic, while for cases of
$\beta^{*} <$ 0 and $\beta^{*} >$ 0, the system is tangentially and
radially biased, respectively.

The left panel of Figure \ref{fig:beta} is a map of the rescaled anisotropy parameter,
$\beta^{*} = \beta/(2-\beta)$, constructed in a similar manner as
Figure \ref{fig:vel_mean}. Inspection of the figure clearly shows the
radially biased orbits of the stars in the IHR, while the stellar orbits
in the OHR become more isotropic. It is also interesting to note that
the stars in the region with $R >$ 15 kpc and \z\ $<$ 6 kpc have
tangentially biased orbits. This is partly due to the presence of the
Monoceros stellar stream \citep{newberg2002,ivezic2008}, which is
included in the region.

Quantitatively, we obtained $\beta$ = 0.656 and 0.383 for the IHR and
OHR, respectively, as listed in the last column of Table \ref{tab:kin},
implying more spherical distribution of stellar orbits in the OHR. Our
derived value of $\beta$ = 0.656 for the IHR is mostly in good agreement
with other studies. For example, previous studies of sub-dwarfs and MS stars
from SDSS report $\beta \sim 0.68$ for the local halo,
consistent with a radially biased distribution (\citealt{smith2009,
bond2010}). \citet{kafle2012} derive $\beta \sim$ 0.5 from BHB stars
observed by SDSS/SEGUE, covering the region $r$ = 9 -- 12 kpc from the
Galactic center, which corresponds to our IHR. \citet{king2015} use a
sample of stars observed with the MMT along with F-type and BHB stars
from SDSS to derive the anisotropy parameter based on radial velocities
alone. They find $\beta$ = 0.15 -- 0.54 in the range $r$ = 9 -- 12 kpc.
\citet{kafle2017} also obtain a value of 0.58 for metal-rich ([Fe/H] $>$
--1.4) MSTO stars with \z\ $>$ 4 kpc and $r \leq$ 15 kpc from radial
velocities alone. On the other hand, using SDSS-\gaia\ proper motions,
\citet{belokurov2018} report a higher value of $\beta$ $\sim$ 0.9
for MS stars with [Fe/H] $>$ --1.7, located within $\sim$ 10 kpc of
the Sun, which differs from our value even though our IHR
region covers a similar spatial and metallicity range. Nonetheless, the
consensus from all of these studies is that there exists a radially
biased velocity ellipsoid for the local/inner halo.

There is no such general consensus on the more distant outer halo, however.
Our derived value of $\beta$ = 0.383 for the OHR is between the values
reported by previous studies that probe similar halo regions and metallicity ranges.
For instance, \citet{kafle2017} obtain $\beta$ = 0.62 for metal-poor
stars ([Fe/H] $<$ --1.4) in the region of \z\ $>$ 4 kpc and 11 $<$ $r
\leq$ 15 kpc. One possible cause of the difference is that they use only
stellar radial velocities to calculate the velocity components in the
spherical coordinates. \citet{belokurov2018}, however, report a similar
range of $\beta$ = 0.2 -- 0.4 to our OHR value for metal-poor MS stars
with [Fe/H] $<$ --1.7. Interestingly, \citet{kafle2012} suggest
negative values of $\beta$ from BHB stars observed by SDSS/SEGUE in the
region  $r$ = 14 -- 19 kpc, reaching a minimum
value of --1.2 at $r$ = 17 kpc and increasing again. \citet{king2015}
also report $\beta$ values between --2.64 and --0.12 from F-type and
BHB stars in the region of $r \sim$ 12 -- 15 kpc, reaching a minimum
(about --4.0) around $r \sim$ 23 kpc. Their minimum value, however is
much less than that of \citet{kafle2012}, considering both studies
including the BHB stars. In any case, the lower value of $\beta$
reported by all of these studies for the outer halo (the OHR in our
case) compared with the inner halo provides strong evidence that the two
halos have undergone different assembly histories.

\begin{figure*}
\centering
\epsscale{1.15}
\plottwo{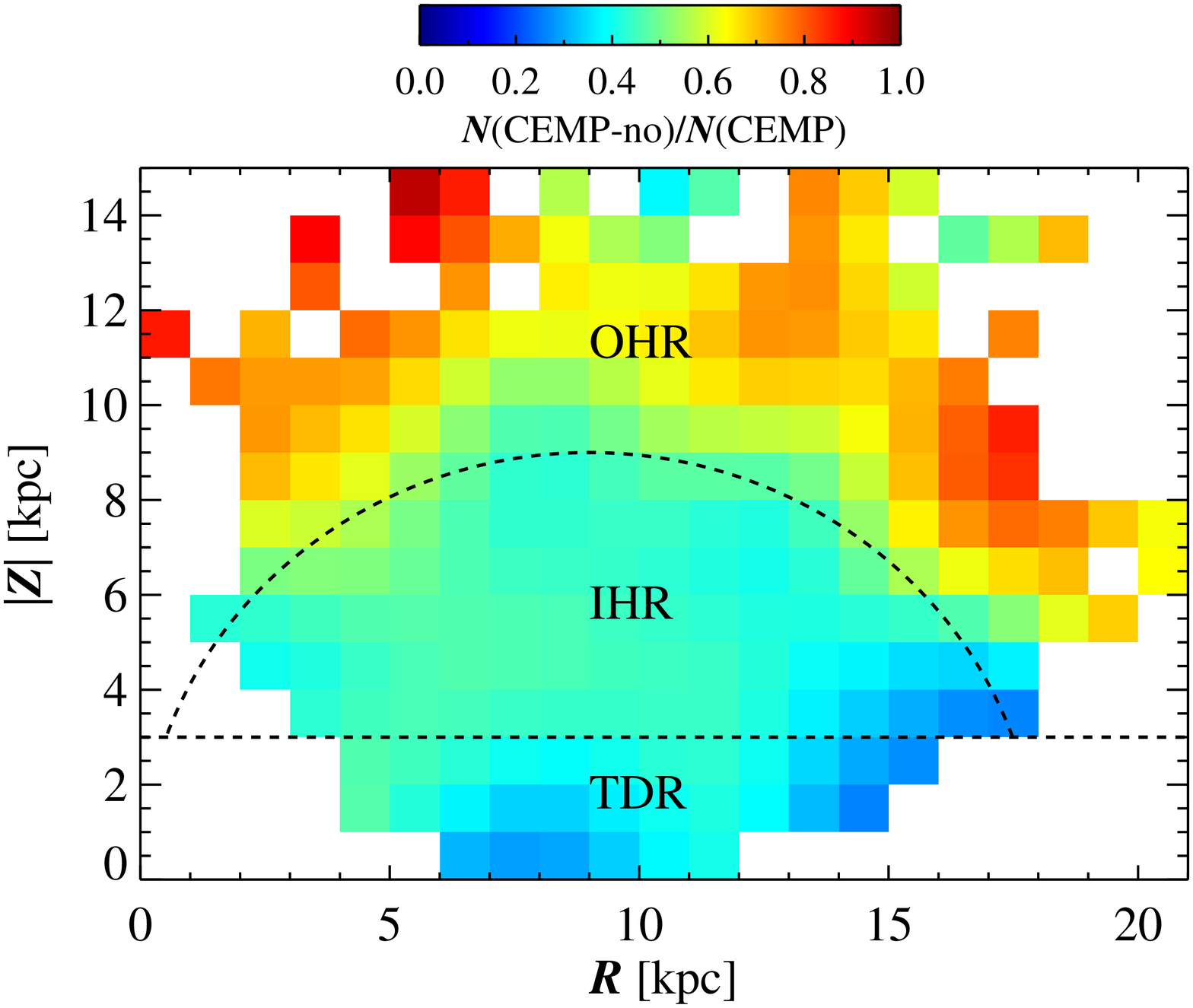}{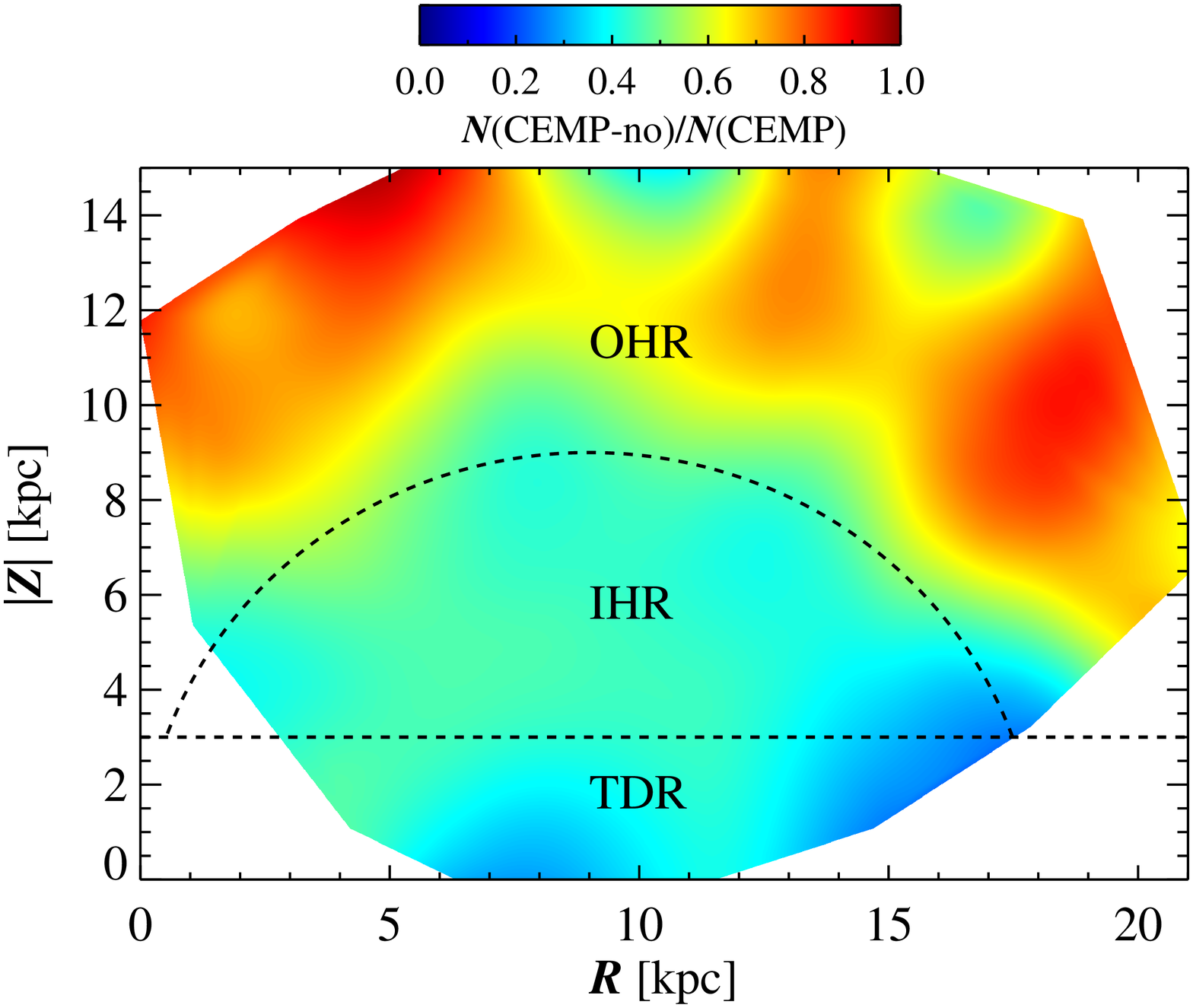}
\caption{$Left~panel$: A map of the ratios of CEMP-no stars relative to the full
sample of CEMP stars in our MSTO sample. Following the definition of
CEMP stars, we only considered stars with [Fe/H] $\le$ --1.0 and [C/Fe]
$\geq$ $+$0.7. The bin size is 1$\times$1 kpc; each pixel contains at
least two stars. A Gaussian kernel has been applied to smooth the
distribution. $Right~panel$: Same as in the left panel, but with a map
produced by interpolating with Quintic polynomials from triangles
produced by Delaunay triangulation.}
\label{fig:sno_map}
\end{figure*}

\begin{figure*}
\centering
\epsscale{1.15}
\plottwo{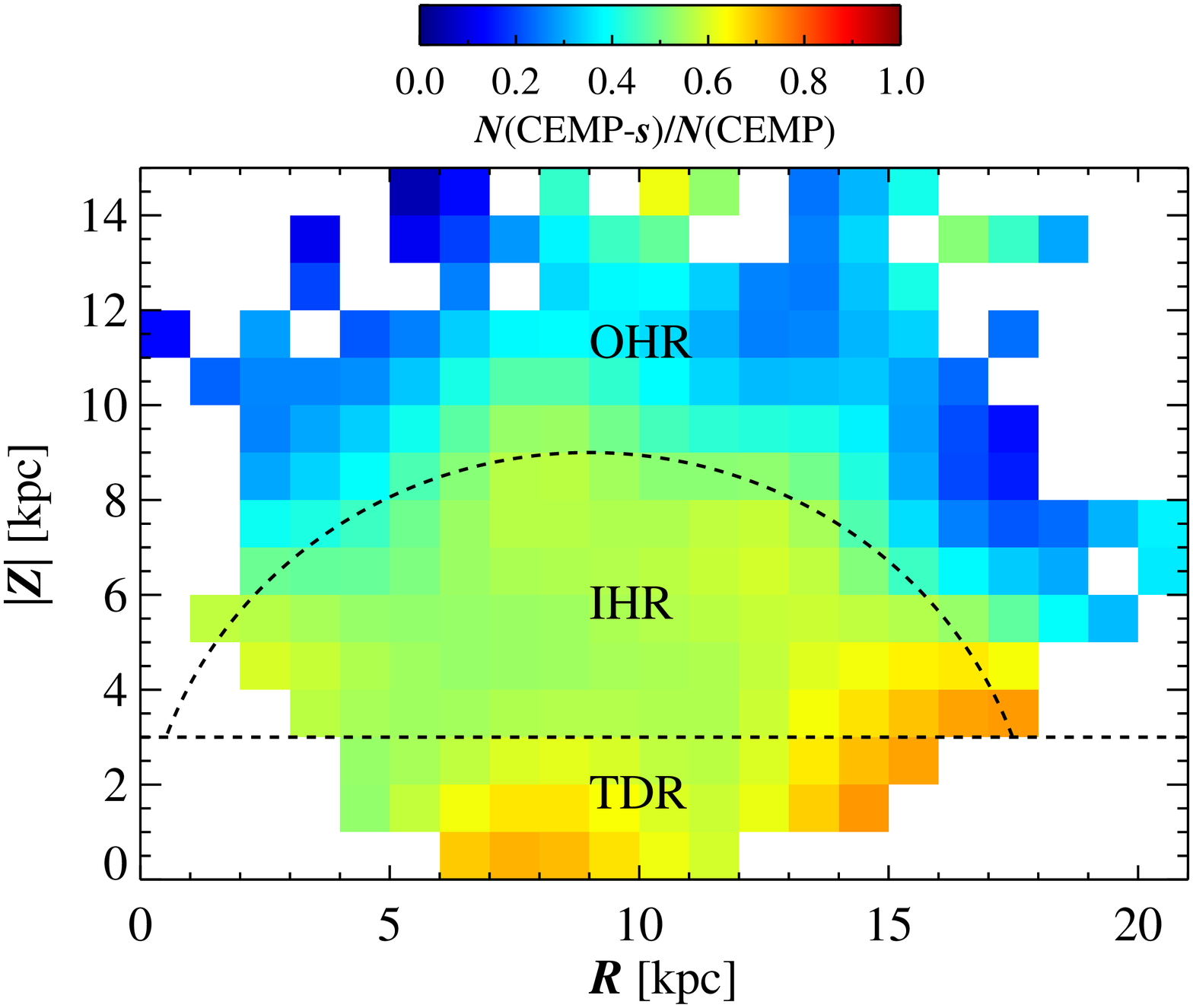}{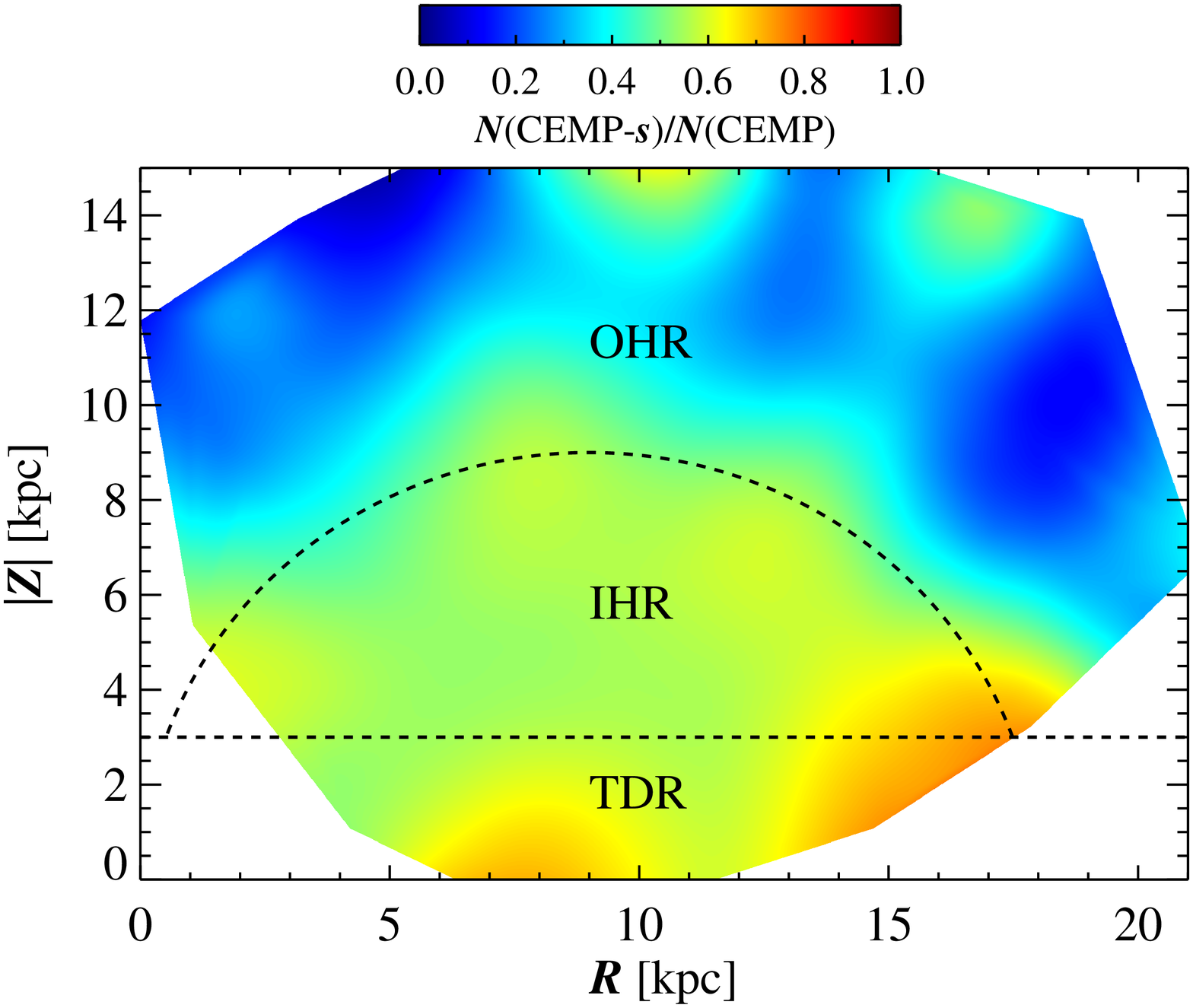}
\caption{Same as in Figure \ref{fig:sno_map}, but for the fraction of
\cemps\ stars relative to the full sample of CEMP stars in our MSTO sample.}
\label{fig:sno_map_s}
\end{figure*}

\subsubsection{The Impact of Observational Errors}

Uncertainties in the observed 
quantities, such as distance, radial velocity, and proper motions in 
our sample of stars can result in the derivation of possibly different 
orbital parameters; hence the structures seen in 
the map of the anisotropy parameter (left panel of Figure \ref{fig:beta}) may be 
distorted by such fluctuations. To assess the impact of the observational 
uncertainties on the anisotropy parameter, we performed a Monte Carlo 
simulation to synthesize 100 different samples of our MSTO stars by 
perturbing the observed distance, radial velocity, and proper motions 
by an uncertainty of 20\% in the distance, and using the quoted uncertainties
in the radial velocity and proper motion. We assumed a normal error 
distribution for our sample.

After generating 100 simulated samples, we computed standard 
deviations of the rescaled anisotropy parameter ($\beta^{*}$), as shown
in the right panel of Figure \ref{fig:beta}. In the figure, it is clear
that the scatter is less than 0.1 in most of the Galactic locations,
even though the deviations at the edge of the map become slightly
larger because the number of stars becomes smaller. Thus, Figure
\ref{fig:beta} confirms that there is no large impact on our results and
subsequent interpretation arising from uncertainties in the various
observed quantities.

\begin{table*}
\renewcommand{\tabcolsep}{2pt}
\begin{center}
\caption{Fractions and Mean Values of [C/Fe] for CEMP, \cemps, and CEMP-no Stars in each Galactic Region}
\label{tab:frac}
\begin{tabular}{cccccccc}
\hline
\hline
 & \multicolumn{3}{c}{Fraction} & & \multicolumn{3}{c}{\cfemean} \\
 \cline{2-4}  \cline{6-8} \\
Region & CEMP & \cemps\  & CEMP-no & & All & \cemps\ & CEMP-no \\
\hline
 TDR  & 0.031 $\pm$ 0.001 & 0.625 $\pm$ 0.024 & 0.375 $\pm$ 0.017 & & $+$0.097 $\pm$ 0.001 & $+$1.205 $\pm$ 0.013 & $+$0.934 $\pm$ 0.009 \\
 IHR  & 0.098 $\pm$ 0.002 & 0.560 $\pm$ 0.014 & 0.440 $\pm$ 0.012 & & $+$0.259 $\pm$ 0.002 & $+$1.175 $\pm$ 0.008 & $+$0.932 $\pm$ 0.005 \\
 OHR  & 0.355 $\pm$ 0.016 & 0.355 $\pm$ 0.027 & 0.645 $\pm$ 0.040 & & $+$0.603 $\pm$ 0.013 & $+$1.427 $\pm$ 0.043 & $+$1.023 $\pm$ 0.014 \\
\hline
\end{tabular}
\tablecomments{The error in the fraction is derived from Poisson statistics, while the errors listed for \cfemean\ are the standard
errors of the mean derived from 100 bootstrapped resamples.}
\end{center}
\end{table*}

\section{Spatial and Kinematic Properties of CEMP-$s$ and CEMP-no Stars}\label{sec:cempsno}

We have identified different kinematic properties of the stellar
populations in the IHR and OHR from our MSTO sample, which suggest that
each halo component likely experienced a different assembly history. We
now examine how each of the primary sub-classes of CEMP stars are
distributed throughout the halo, and how their kinematic properties
differ. The primary reason for this investigation is that the CEMP-$s$ 
and CEMP-no stars are expected to have formed from
very different mechanisms, each associated with different ranges of
progenitor mass. 

It is presently thought that the astrophysical progenitor of CEMP-no
stars differs in stellar mass from that of \cemps\ stars. The most
likely progenitors of CEMP-no stars are ``faint supernovae" (or
``mixing-and-fallback SNe") with a mass range of $\sim$~20 -- 60 \msun\
(e.g., \citealt{umeda2003, umeda2005, nomoto2013, tominaga2014}), or
``spinstars'' with a mass range of $>$ 60 -- 100 \msun\
(\citealt{meynet2006, meynet2010, chiappini2013}). On the other hand,
the accepted mechanism to produce the \cemps\ stars is binary
mass-transfer from low- to intermediate-mass (1 -- 4 \msun) asymptotic
giant branch (AGB) stars, which can efficiently produce carbon and $s$-process
elements. Eventually, the primary AGB star evolves to become a white dwarf, and
its secondary companion is observed as a \cemps\ star (e.g.,
\citealt{suda2004, herwig2005, lucatello2005, komiya2007, bisterzo2011,
hansen2015}). As a result, variations in the radial velocity for the
surviving star is expected. Strong support for this scenario comes from
long-term radial-velocity studies (\citealt{starkenburg2014,
hansen2016a, hansen2016b, jorissen2016}), which report very different
binary fractions between these two sub-classes of CEMP stars.

Thus, distinct patterns in spatial distributions
and/or kinematics of the \cemps\ and CEMP-no stars throughout the halo, 
if they exist, provide insight to the nature of the environments in which 
they formed, and their star-formation histories.

\citet{carollo2014} previously examined the possible association of
different CEMP sub-classes with the different stellar populations of the
halo, using a small sample ($N \sim 90$) of CEMP stars
with available [C/Fe] and [Ba/Fe] estimates obtained from high-resolution
spectroscopy. After dividing CEMP stars into likely membership in the
IHP and OHP by consideration of their orbital energies,
they calculated the fraction of CEMP-no and CEMP-$s$ stars in each
halo population, finding that the fraction of CEMP-no stars in the OHP
is higher, by about a factor of two, than for \cemps\ stars, while
almost equal fractions of \cemps\ and CEMP-no stars were found in the IHP.
\citet{yoon2018} also reported similar fractions of the 
CEMP-no stars in the IHP and OHP, using subgiant and giant stars from 
the AEGIS survey. They classified the CEMP stars into the \cemps\ and 
CEMP-no stars by \ac, as adopted in this study. Establishing the 
validity of these claims requires examination of much larger samples of CEMP 
stars. We accomplish this by making use of the clear split in the typical 
levels of \ac\ associated with the CEMP-$s$ and CEMP-no stars of our MSTO 
sample, as shown in Figure~\ref{fig:ac_feh}.

Below we first consider the spatial distributions of the relative
fractions of CEMP-no to CEMP$s$ stars, and then examine the kinematic
properties of each sub-class within each Galactic region.

\begin{figure}
\centering
\epsscale{1.15}
\plotone{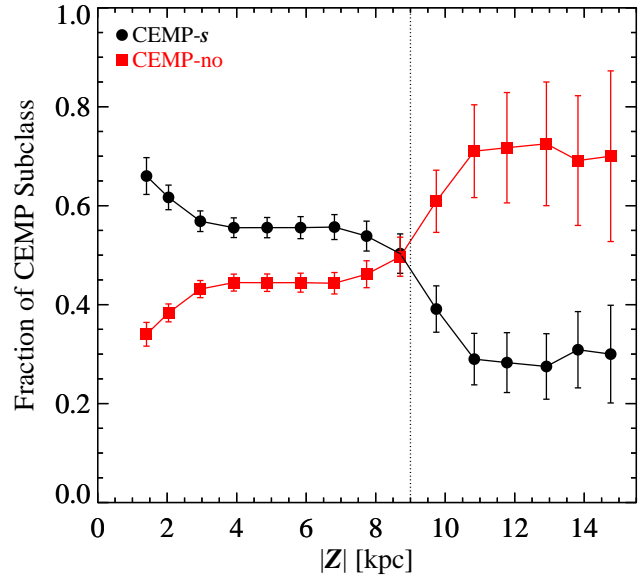}
\caption{Differential fractions of \cemps\ stars (black symbols) 
and CEMP-no stars (red symbols), as a function of \z. The bin size is 2
kpc with 1 kpc overlapped with the next neighboring bin. Error bars are
derived from Poisson statistics. There is an abrupt increase of CEMP-no
stars between \z\ = 8 and 11 kpc. The vertical dotted line at
\z\ = 9 kpc denotes the highest point in the region that separates the
IHR from the OHR in Figure \ref{fig:cmap}.}
\label{fig:sno_frac}
\end{figure}


\begin{table*}
\renewcommand{\tabcolsep}{2pt}
\begin{center}
\caption{Kinematic Properites in Cylindrical Coordinates for \cemps\ and CEMP-no Stars in each Galactic Region}
\label{tab:ks}
\begin{tabular}{llrrrrrrrcc}

\hline
\hline
Region & Subclass & $N_{\rm tot}$ & $V_{\rm R}$~~~~ & $V_{\Phi}$~~~~ & $V_{\rm Z}$~~~~ & $\sigma_{V_{\rm R}}$~~~~ & $\sigma_{V_{\phi}}$~~~~ & $\sigma_{V_{\rm Z}}$~~~~ & $V_{\Phi}$ & Eccentricity \\
                  &          &               & (\kms)~      & (\kms)~         & (\kms)~      & (\kms)~               & (\kms)~                  & (\kms)~              &$p$-value       & $p$-value \\
\hline

TDR & \cemps\ &  1067 &     2.2 $\pm$    4.4 &    14.6 $\pm$    3.5 &    2.1 $\pm$    2.4 &  140.9 $\pm$    3.3 &  103.5 $\pm$    2.4 &   81.6 $\pm$    2.3 &    0.028 &   0.021\\
    & CEMP-no &   640 &   --2.3 $\pm$    5.2 &     0.5 $\pm$    3.9 &    4.0 $\pm$    4.2 &  129.7 $\pm$    3.1 &  102.0 $\pm$    3.3 &  101.5 $\pm$    3.5 & \\
\hline
IHR & \cemps\ &  2403 &   --4.8 $\pm$    2.8 &  --17.4 $\pm$    2.2 &    3.8 $\pm$    2.0 &  141.6 $\pm$    1.9 &  103.0 $\pm$    1.6 &  104.8 $\pm$    1.5 &    0.000 &   0.000\\
    & CEMP-no &  1835 &   --3.8 $\pm$    3.2 &  --26.4 $\pm$    3.0 &    0.3 $\pm$    3.0 &  137.2 $\pm$    2.0 &  115.6 $\pm$    2.2 &  112.1 $\pm$    2.0 & \\
\hline
OHR & \cemps\ &   170 &  --10.6 $\pm$   10.9 &  --45.1 $\pm$    8.0 &    5.4 $\pm$    9.4 &  150.5 $\pm$    7.6 &  120.6 $\pm$    6.9 &  131.2 $\pm$    6.7 &    0.033 &   0.015\\
    & CEMP-no &   195 &  --13.5 $\pm$   10.8 &  --61.5 $\pm$    8.4 &    4.9 $\pm$    8.3 &  139.4 $\pm$    6.8 &  132.6 $\pm$    6.3 &  117.7 $\pm$    5.8 & \\
\hline
\end{tabular}
\tablecomments{$N_{\rm tot}$ is the total number of stars in each subclass of the CEMP stars. The listed uncertainties are derived from 100 bootstrapped resamples. The $p$-values for $V_{\Phi}$
and eccentricity ($e$) are derived by the K-S two sample test on the distributions of $V_{\Phi}$ and eccentricity ($e$) of the \cemps\ and CEMP-no star samples.}

\end{center}
\end{table*}

\begin{table*}
\renewcommand{\tabcolsep}{2pt}
\begin{center}
\caption{Kinematic Properties in Spherical Coordinates for \cemps\ and CEMP-no Stars in each Galactic Region}
\label{tab:beta}
\begin{tabular}{clrrrrrrrrr}
\hline
\hline
Region & Subclass & $N_{\rm tot}$ & $V_{\rm r}$~~~~~ & $V_{\theta}$~~~~~ & $V_{\phi}$~~~~~ & $\sigma_{V_{\rm r}}$~~~~ & $\sigma_{V_{\theta}}$~~~~ & $\sigma_{V_{\phi}}$~~~~ & $\beta$~~~~~~~~ \\
       &          &               &  (\kms)~  &      (\kms)~     & (\kms)~        & (\kms)~              & (\kms)~                    & (\kms)~                  &       \\
\hline
TDR & \cemps\ &  1067 &     2.8 $\pm$    4.8 &   --1.5 $\pm$    2.5 &    14.6 $\pm$    3.5 &  142.7 $\pm$    2.9 &   78.5 $\pm$    2.7 &  103.5 $\pm$    2.4 &    0.586 $\pm$   0.023\\
    & CEMP-no &   640 &   --0.6 $\pm$    5.5 &   --6.4 $\pm$    3.4 &     0.5 $\pm$    3.9 &  132.4 $\pm$    3.8 &   97.8 $\pm$    2.9 &  102.0 $\pm$    2.8 &    0.431 $\pm$   0.040\\
\hline
IHR & \cemps\ &  2403 &   --1.0 $\pm$    2.7 &   --4.3 $\pm$    1.7 &  --17.4 $\pm$    2.2 &  152.4 $\pm$    2.1 &   88.4 $\pm$    1.7 &  103.0 $\pm$    1.9 &    0.603 $\pm$   0.015\\
    & CEMP-no &  1835 &   --0.3 $\pm$    3.5 &   --3.8 $\pm$    2.1 &  --26.4 $\pm$    3.0 &  141.0 $\pm$    2.4 &  107.3 $\pm$    2.1 &  115.6 $\pm$    2.1 &    0.374 $\pm$   0.027\\
\hline
OHR & \cemps\ &   170 &     8.2 $\pm$   12.4 &  --13.0 $\pm$   10.6 &  --45.1 $\pm$    8.0 &  158.8 $\pm$    7.8 &  120.7 $\pm$    8.9 &  120.6 $\pm$    7.5 &    0.423 $\pm$   0.079\\
    & CEMP-no &   195 &   --2.1 $\pm$    9.5 &  --24.0 $\pm$    8.5 &  --61.5 $\pm$    8.4 &  136.7 $\pm$    6.2 &  119.2 $\pm$    5.8 &  132.6 $\pm$    6.4 &    0.150 $\pm$   0.097\\
\hline
\end{tabular}
\tablecomments{$N_{\rm tot}$ is the total number of stars in each subclass of the CEMP stars. The listed uncertainties are derived from 100 bootstrapped resamples. $\beta$
is the anisotropy parameter computed by Equation (1).}
\end{center}
\end{table*}

\subsection{Spatial Distribution of CEMP-$s$ and CEMP-no Stars}

The left panel of Figure~\ref{fig:sno_map} shows the fraction of
CEMP-no stars relative to the full sample of CEMP stars among our MSTO
sample, in the \z\ versus $R$ plane. The right panel is the same as in the
left panel, but for a map constructed after interpolating with Quintic
polynomials from triangles produced by Delaunay triangulation over the
map. This map is for a better visualization of the fractions of the 
CEMP-no stars in the OHR.

Inspection of these figures immediately reveals two clear results: (1)
An increasing fraction of CEMP-no stars as one moves farther into the
OHR -- more than 60\% of the CEMP stars in the OHR are CEMP-no stars,
even though there exist a few substructures with relatively higher and
lower fractions of CEMP-no stars; and (2) The boundary between the IHR
and OHR defined in Figure~\ref{fig:cmap} also clearly divides the two
halo regions in terms of lower and higher fractions of CEMP-no stars. A
similar map of \cemps\ star fractions is shown in
Figure~\ref{fig:sno_map_s}. As in the CEMP-no map, we note regions with
relatively high (and low) fractions of \cemps\ stars, as discussed in
more detail below. It is noteworthy from inspection of Figures
\ref{fig:sno_map} and \ref{fig:sno_map_s} that the blue
(Figure~\ref{fig:sno_map}) and orange (Figure~\ref{fig:sno_map_s}) regions
in the range of $13 < R < 18$ kpc and \z\ $< 6$ kpc, indicating relatively
large fractions of \cemps\ stars, appear to be associated with the
Monoceros Stream (\citealt{newberg2002, ivezic2008}). This suggests that the
parent dwarf responsible for the Monoceros Stream has experienced
prolonged star formation, consistent with a comparatively high mass.

Quantitatively, we obtained 0.440 $\pm$ 0.012 and 0.645 $\pm$ 0.040 for the
ratio of CEMP-no to CEMP stars for the IHR and OHR, respectively, as
listed in Table \ref{tab:frac}. Our computed ratios confirm that the
\cemps\ stars in the IHR are favored over the CEMP-no stars, while the CEMP-no stars
outnumber the \cemps\ by about a factor of two in the
OHR, in excellent agreement with the inferred fractions of CEMP-no stars
-- 0.43 (IHP) and 0.70 (OHP) reported by \citet{carollo2014}, even
though we employed a totally different selection criterion for
the regions expected to be dominated by stars of the IHP and OHP from
theirs. Our derived value of 0.440 for the 
fraction of the CEMP-no stars in the IHR agrees with that (0.47) of 
\citet{yoon2018}, while their derived fraction (0.78) of the CEMP-no stars 
for the OHP is rather higher than ours (0.65). 

Taken as a whole, we obtained CEMP frequencies
of 0.098 $\pm$ 0.002 and 0.355 $\pm$ 0.016 for the IHR and OHR, respectively,
as listed in Table~\ref{tab:frac}. Among the stars of our MSTO sample,
the OHR possesses roughly three times as many CEMP stars as does the IHR,
even higher than the contrast reported previously by \citet{carollo2012}.

We also examined the average carbon-to-iron ratios
($\langle$[C/Fe]$\rangle$) for stars in each Galactic region, as listed
in the last three columns of Table~\ref{tab:frac}. Generally, the OHR
has a higher $\langle$[C/Fe]$\rangle$ than the IHR, but the level of the
enhancement is not as high as that claimed by \citet{carollo2012}
($\langle$[C/Fe]$\rangle$ $\sim +1.0$ to $+2.0$). As listed in the
table, we obtained $\langle$[C/Fe]$\rangle$ = $+$0.259 and $+$0.603 for the
IHR and OHR, respectively. This contrast likely arises due to the fact
that we did not restrict our sample by low metallicity ([Fe/H] $<$
--1.5) and kinematic properties (\zmax\ $>$ 5 kpc), as was used by
Carollo et al., but by the spatial variation of \cfe\ alone.

We also note that the \cfemean\ value of the CEMP-no
stars is somewhat lower than that of the \cemps\ stars for all three
regions we considered, implying that the mechanism responsible for
producing the \cemps\ stars produces more carbon at a given metallicity
than that responsible for the CEMP-no stars, as revealed 
in several previous studies (e.g., \citealt{yong2013, spite2013, bonifacio2015, 
hansen2016a,hansen2016b,yoon2016,yoon2018}).

The clear association of different CEMP sub-classes with different
Galactic halo regions is revealed in Figure~\ref{fig:sno_frac} as
well, which exhibits the differential fractions of CEMP-no (red symbols)
and \cemps\ (black symbols) stars among CEMP stars in our MSTO sample as a function of \z.
Each bin has a size of 2 kpc with 1 kpc overlapped with the next
neighboring bin. Error bars are calculated from Poisson statistics.
As in Figure~\ref{fig:sno_map}, we only consider stars with
[Fe/H] $\leq$ --1.0 and [C/Fe] $\ge$ $+$0.7.

Figure~\ref{fig:sno_frac} indicates that there is a very slow increase
in the CEMP-no fractions up to \z\ = 8 kpc, followed by a steep increase
between \z\ = 8 and 11 kpc, and an essentially flat behavior at higher
\z. This trend is also well-reflected by the boundaries of the Galactic
regions seen in Figures \ref{fig:sno_map} and \ref{fig:sno_map_s}.
Consequently, the results from all three figures are clear evidence that
the underlying stellar population in the OHR does not share the same
parent population as the IHR.

\begin{figure*}
\centering
\plottwo{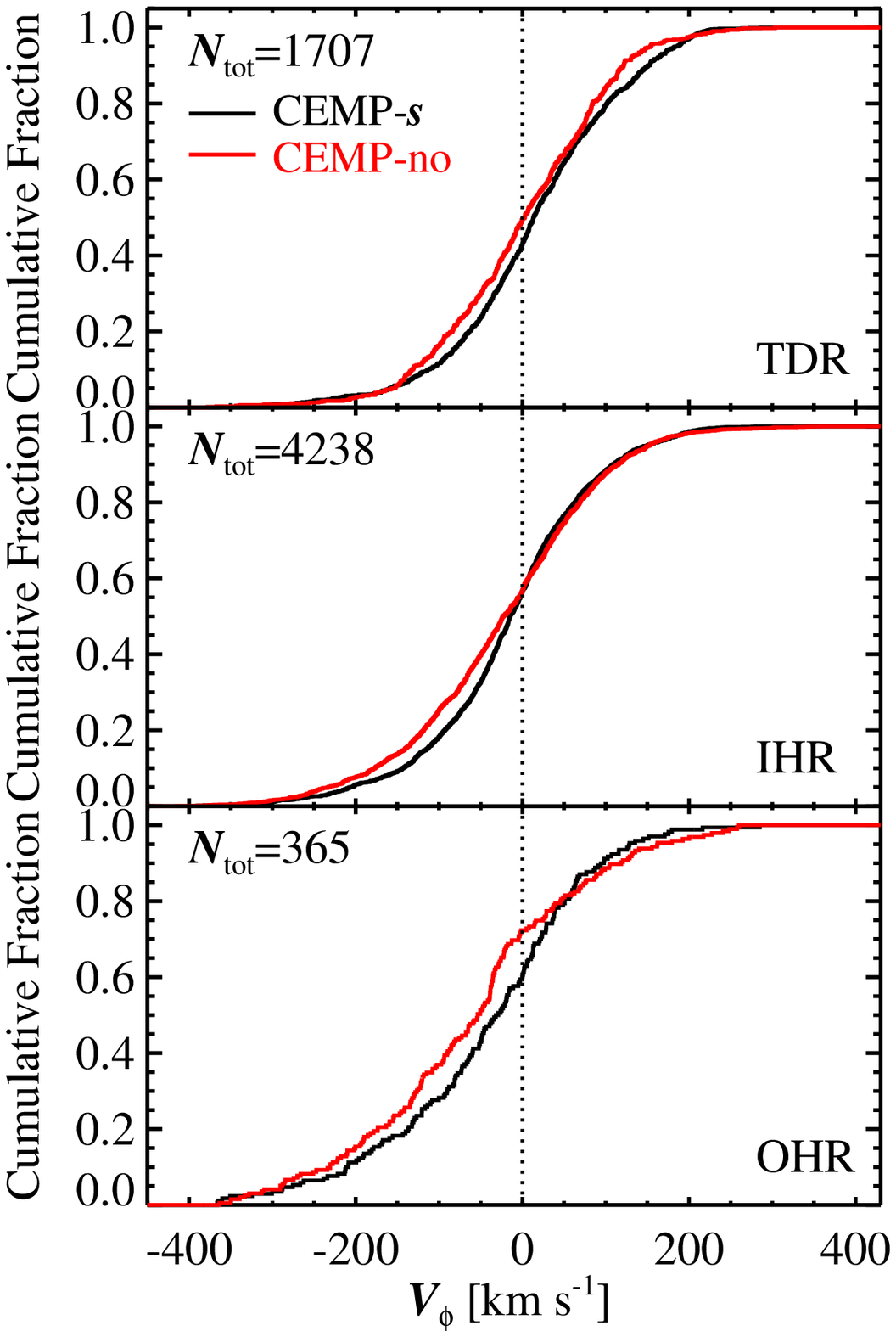}{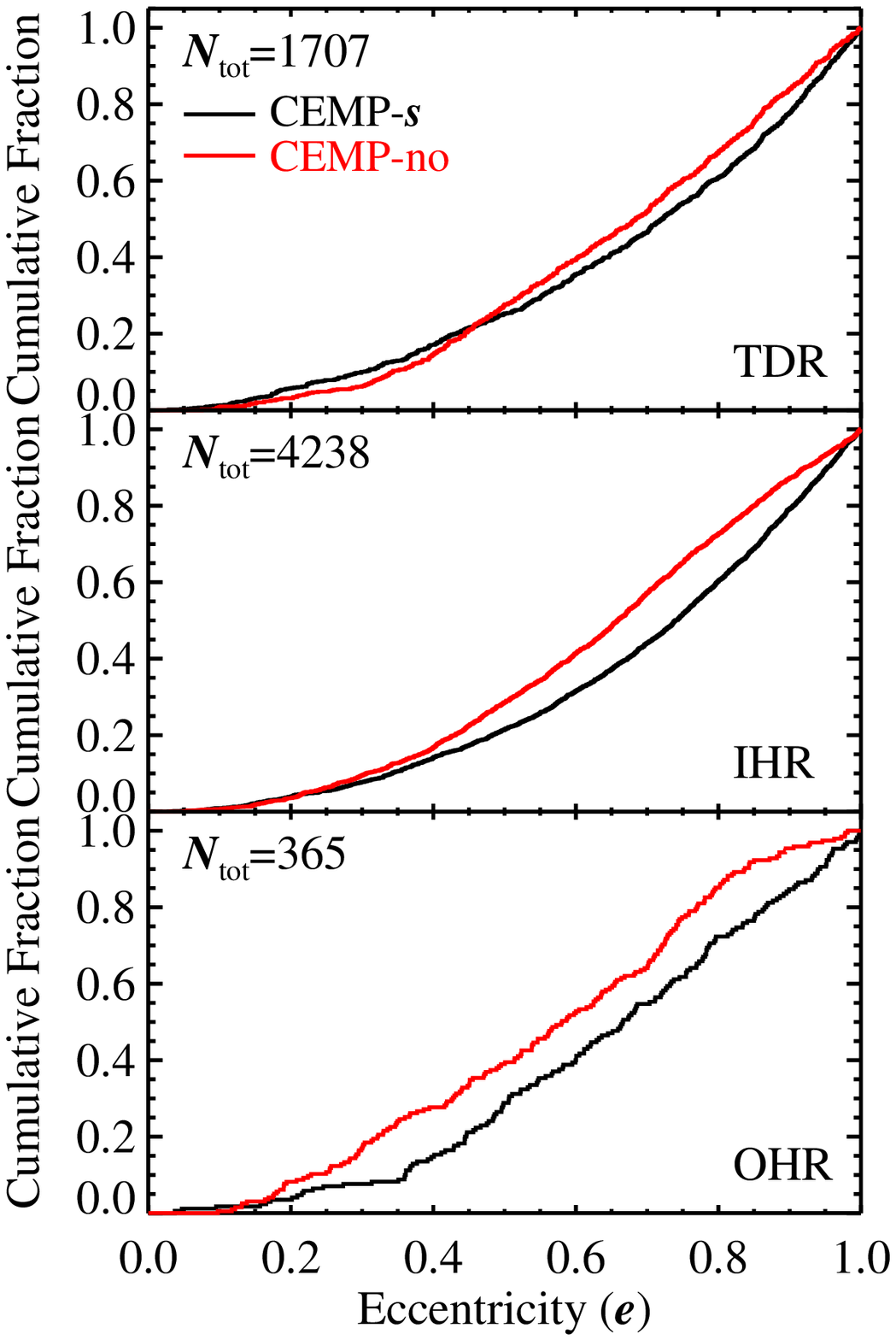}
\caption{$Left~panel$: Cumulative distributions of rotation velocities
for CEMP-$s$ (black line) and CEMP-no (red line) stars in each Galactic component.
The total number of stars considered in each region is denoted by $N_{\rm tot}$.
The vertical dotted line provides a reference at 0 \kms. $Right~panel$: Same as the
left panels, but for orbital eccentricities. It is clear that the cumulative
distributions in the kinematic properties differ between the
\cemps\ and CEMP-no stars in the two halo components, indicating that they do not
share the same parent population.}
\label{fig:vtecc}
\end{figure*}

\subsection{Kinematic Characteristics of \cemps\ and CEMP-no Stars}

We now consider whether or not there exist differences in the
kinematic properties for the sub-classes of CEMP stars.

We first examine the average velocity properties of each CEMP sub-class
in each Galactic region, as summarized in Table~\ref{tab:ks} (and in
Table~\ref{tab:beta} for spherical coordinates). According to the tables,
in the OHR, the average rotation velocity of \cemps\ stars is
$\langle$\vPhi$\rangle$ = --45.1 $\pm$ 8.0 \kms\ and --61.5 $\pm$ 8.4
\kms\ for the CEMP-no stars. For the IHR, the mean rotation velocity of
CEMP-no stars is lower by about 9 \kms\ than that of \cemps\ stars. The
general trend is that CEMP-no stars exhibit larger retrograde motions
than the \cemps\ stars. By comparison with Table \ref{tab:kin}, which
provides kinematic information for the entire sample of our MSTO sample,
CEMP stars exhibit lower rotation velocities than carbon-normal stars.

Figure~\ref{fig:vtecc} shows the cumulative distribution of rotation
velocities (left panels) and orbital eccentricities (right panels) of
CEMP-no (red curve) and \cemps\ (black curve) stars in each region of
the MW. The total number of stars considered in each region is denoted
by $N_{\rm tot}$, shown in the legend of each panel. A non-parametric
two-sample K-S test of the distribution of the rotation velocities for
CEMP-no and \cemps\ stars yields $p$-values of $< 0.001$ and 0.033 for
the IHR and OHR, respectively, as listed in Table \ref{tab:ks}. It is
thus unlikely that the two sub-classes of CEMP stars share the same
parent population. Similarly, the $p$-values of $< 0.001$ and 0.015 for
the eccentricity distribution of CEMP-no and
\cemps\ stars in the IHR and OHR, respectively, reject the hypothesis of a
common parent population.

We also derived the anisotropy parameter for each CEMP sub-class in each
halo component, as listed in the last column of Table \ref{tab:beta}. In
the IHR, we obtained $\beta$ = 0.603 for the \cemps\ stars and 0.374 from
the CEMP-no stars; for the OHR, we found $\beta$ = 0.423 for the \cemps\
stars and 0.150 for the CEMP-no stars. Overall, as $\beta$ = 0.383 for the
entire sample in the OHR (see Table \ref{tab:kin}), the CEMP stars in 
the OHR exhibit a more isotropic distribution of orbits than for the IHR. 
Within a given halo component, the CEMP-no stars exhibit a more isotropic 
velocity ellipsoid than the \cemps\ stars.

\section{Insights into the Formation of the Galactic Halo}\label{sec:insight}

A number of recent cosmological numerical simulations of MW-like galaxies
suggest that the inner halos form from mergers of a few relatively
massive gas-rich mini-halos, which undergo dissipative collapse while
conserving angular momentum (e.g., \citealt{zolotov2009, font2011, mccarthy2012,
tissera2012, tissera2013, tissera2014, cooper2015}). In these mini-halos, stars that are
eventually distributed in the inner halo form continuously to rapidly
enrich the interstellar medium. Consequently, the prolonged star
formation in such systems lead to the formation of moderately metal-poor
stars ([Fe/H] $\sim$ --1.5), having numerous intermediate-mass ($\sim$ 1
-- 4 \msun), which become AGB stars in the late stage of their
evolution. Some of these stars form in a binary system that produce the
CEMP-$s$ stars observed at present.

In this formation scenario, we can easily understand the higher fraction
of \cemps\ in the IHR that we identified in Figure \ref{fig:sno_map_s}
and Table \ref{tab:frac}. Kinematically, this formation mechanism can
also naturally explain the higher value of $\beta$ for the inner halo.
The radial velocity component of the stars, which formed out of
gas-rich sub-fragments during the dissipative collapse will naturally be
enhanced. After merging, these stars would have a flattened
distribution, dominated by eccentric orbits.

According to recent theoretical work (e.g.,
\citealt{debennassuti2014, debennassuti2017, salvadori2016, amorisco2017,
starkenburg2017}), the stellar populations in the outer halos of MW-like
galaxies mostly consist of stars that were born in numerous lower-mass
mini-halos, and later accreted into the Galactic halo via chaotic, dissipationless
mergers. In such low-mass environments, star formation is truncated
before forming large numbers of relatively metal-rich, low-mass stars,
the dominant source to produce \cemps\ stars. As a result, it is
expected that there would be more metal-poor ([Fe/H] $<$ --2.0) and
CEMP-no stars formed than \cemps\ stars in the Galactic outer halo.

The fact that most CEMP-no stars appear in the regime of [Fe/H] $<$
--2.5, and the larger fraction of the CEMP-no stars in the OHR, as shown
in Figure \ref{fig:sno_map}, meet with theoretical expectations, and
suggests that the stars in the OHR were accreted from relatively
low-mass dwarf satellites. In addition, since the accreted stars were
born in low-mass fragments, they could be easily dispersed by the
Galactic tidal force, leading to the formation of a more spherical
outer halo. We might expect that the stars from such systems were
accreted from all directions, have less eccentric orbits (lower value of
the anisotropy parameter), and possibly exhibit counter rotation
relative to the inner halo, which was already in place. Once again,
these results are commensurate with what we have found for the CEMP-no
stars in our MSTO sample. Therefore, from both a chemical and kinematic
perspective, we can infer that the stellar population of the OHR is
dominated by donated stars from numerous low-mass dwarf satellites.

What kind of dwarf satellites can contribute to the buildup of the
Galactic outer halo? The best candidates for such objects are the
ultra-faint dwarf (UFD) galaxies around the MW (e.g.,
\citealt{belokurov2006a, belokurov2006b, zucker2006, simon2007},
and see \citealt{simon2018} for the current demography of UFDs). In these
systems, we expect that star formation occurred early and was stopped or
truncated due to the shallow potential well of the systems (e.g., \citealt{brown2014,
webster2015}). The CEMP-no stars were produced from the nucleosynthetic products of
the first burst of the star formation, but not the \cemps\ stars, because 
the duration of the star formation was not long enough to form the low-mass, 
relatively high metallicity ($>$ --2.5) AGB stars. Consequently, CEMP-no stars
should dominate over \cemps\ stars in such systems. This picture is
consistent with our results summarized on Tables \ref{tab:frac},
\ref{tab:ks}, and \ref{tab:beta}, as well as with the relative dominance of CEMP-no stars
over \cemps\ stars in the UFDs discovered around the MW (e.g.,
\citealt{frebel2014, frebel2015, frebel2016, ji2016}). This inference 
leads us to conclude that a large fraction of the stars in the OHR were 
accreted from systems similar to disrupted UFD galaxies. \citet{yoon2019} 
also reported the same conclusion that the CEMP-no stars in the Galactic 
halo were accreted from very low-mass systems such as UFD galaxies by 
comparing the subgroups of the CEMP-no stars with the CEMP stars 
found in the dwarf satellite galaxies around the MW.

Additionally, our results suggest that, if all the CEMP-no stars in the
outer halo were born in small satellite galaxies such as UFD galaxies
that were disrupted into the Galactic halo, we can estimate a lower
limit on the fraction of the accreted stars in the outer halo from the
the ratio of CEMP-no to all CEMP stars. From Table \ref{tab:frac},
which reports a CEMP-no fraction of 0.645 and a total CEMP fraction of
0.355 for the OHR, we have a fraction of 0.229 (0.645 $\times$ 0.355);
hence at least about 23\% of the outer-halo stars came from systems
such as UFDs. 

\section{Summary and Conclusions}\label{sec:conclusion}

We have examined the chemical and kinematic features of the Galactic
halo, making use of over 100,000 MSTO stars from the SDSS. To
accomplish, we first divided the halo region into IHR and OHR, based on
the varying levels of carbon-to-iron ratios shown in a carbonicity map
in the \z\ versus $R$ plane. On average, the stars in the OHR exhibit
higher \cfe, by about 0.2 dex, than stars in the IHR. We found that the stellar
population in the OHR shows a clear retrograde motion (--49.3 \kms),
with an anisotropy parameter of $\beta = 0.383$, while the population in 
the IHR exhibits almost no net rotation (--2.6 \kms) with $\beta = 0.656$. 
These kinematic differences between the OHR and IHR can be
interpreted as arising from the different assembly histories these
components experienced.

While analyzing our MSTO stars, we have also investigated the impacts of
target-selection bias on our MSTO sample, and found no significant influence on the
identification of the Galactic halo regions. We confirmed that our derived 
distance scale agrees well with that derived from \gaia\ DR2 parallaxes, without any trend
evident with respect to [Fe/H], \logg, and [C/Fe]. The impact of observational 
uncertainties in the distances, radial velocities, and proper motions on the derived 
spatial velocities is also small, thus there 
is no significant influence on the velocity structures revealed by our
results arising from these fluctuations.

We further divided the CEMP stars in the IHR and OHR into high-\ac\ and
low-\ac\ stars, which we associate with \cemps\ and CEMP-no stars,
respectively, based on the absolute carbon abundance,
\ac. First we have mapped the fractions of the \cemps\ and CEMP-no
stars in the \z\ and $R$ plane, and found that there are almost twice
as many CEMP-no stars than \cemps\ stars in the OHR, while there are almost equal
numbers of \cemps\ and CEMP-no stars in the IHR. This information on the
different ratios of CEMP-no to CEMP-$s$ stars in the Galactic halo can
provide valuable constraints on the mass distributions of the mini-halos
from which they formed, as well as their star-formation histories.

We have also examined the kinematic properties for the \cemps\ and
CEMP-no stars, and found that the CEMP-no stars exhibit more
retrograde motions, and a more isotropic distribution of their orbits,
compared to the \cemps\ stars in each halo region. The chemical and
kinematic properties found from the MSTO stars indicate that most of the
stars in the OHR may be accreted from small satellite galaxies such as
the UFD galaxies, as discussed in \citet{yoon2019}. Very conservatively, 
we estimated from our MSTO stars that at least about 23\% of the outer halo 
stars came from systems such as UFDs, based on the CEMP-no star fraction.

Finally, our work demonstrates (Figures \ref{fig:vel_mean}, 
\ref{fig:vel_disp}, \ref{fig:beta}, and Tables \ref{tab:kin}, \ref{tab:ks}, and
\ref{tab:beta}) that, even if we separate the halo components based on the
spatial distribution of carbonicity, we are able to identify
distinct kinematic signatures among the divided halo components, which
can provide clues to the origin of the dichotomy of the Galactic halo.

As a third paper in the ``Chemical Cartogrpahy" series,
we are planning to carry out an analysis of kinematic and chemical properties
of CEMP giants observed with SDSS, SEGUE, and LAMOST, in order 
to probe the more distant Galactic halo, and characterize the
nature of the dwarf galaxies that contributed stars into the Galactic halo.

\acknowledgments

We thank an anonymous referee for his/her careful review of this paper
to improve the clarity of the presentation.

Funding for SDSS-III has been provided by the Alfred P. Sloan
Foundation, the Participating Institutions, the National Science
Foundation, and the U.S. Department of Energy Office of Science. The
SDSS-III Web site is http://www.sdss3.org/.

The authors thank J. Yoon for reviewing this manuscript and suggesting
useful comments to improve this paper. Y.S.L. acknowledges support from
the National Research Foundation (NRF) of Korea grant funded by the
Ministry of Science and ICT (No.2017R1A5A1070354 and
NRF-2018R1A2B6003961). T.C.B. acknowledges partial support for this work
from grant PHY 14-30152; Physics Frontier Center/JINA Center for the
Evolution of the Elements (JINA-CEE), awarded by the US National Science
Foundation, and from the Leverhulme Trust (UK), during his visiting
professorship at the University of Hull, when this paper was finished.

\clearpage

\end{document}